\documentclass[showpacs,twocolumn,prb,eqsecnumbers]{revtex4}
\usepackage{amsmath}
\usepackage{epsf}
\usepackage{graphicx}
\usepackage{dcolumn}
\usepackage{bm}

\newcommand{\nn}{\nonumber}
\newcommand{\beq}{\begin{equation}}
\newcommand{\eeq}{\end{equation}}
\newcommand{\bea}{\begin{eqnarray}}
\newcommand{\eea}{\end{eqnarray}}
\newcommand{\bwt}{\begin{widetext}}
\newcommand{\ewt}{\end{widetext}}
\newcommand{\dz}{\Delta}
\newcommand{\dzz}{{\tilde\Delta}}
\newcommand{\up}{\uparrow}
\newcommand{\down}{\downarrow}
\newcommand{\om}{\Omega_m}
\newcommand{\fs}{f_{s,0}}
\newcommand{\vf}{{\tilde v}_{F}}

\begin{document}

\title{
Nonanalytic paramagnetic response
of itinerant fermions away and near a ferromagnetic quantum phase transition}
\author{Dmitrii L. Maslov$^{1}$ and Andrey V. Chubukov$^{2}$}
\date{\today}
\affiliation{
$^{1}$Department of
Physics, University of Florida, P. O. Box 118440, Gainesville, FL
32611-8440 \\
$^{2}$ Department of Physics, University of Wisconsin-Madison,
1150 University Ave., Madison, WI 53706-1390}

\begin{abstract}
We study nonanalytic paramagnetic response of an interacting Fermi
system both away and in the vicinity of a ferromagnetic quantum
phase transition (QCP). Previous studies found that
 (i) the spin
 susceptibility $\chi$
 scales linearly with either the temperature $T$ or magnetic field $H$ in the weak-coupling regime; (ii)
the interaction in the Cooper channel affects this scaling via
logarithmic renormalization of prefactors of the $T$, $|H|$ terms,
and may even reverse the signs of these terms at low enough
energies.
 We show  that Cooper renormalization becomes effective only at very
  low energies, which get even smaller near a QCP. However,
  even in the absence of such renormalization,
 generic (non-Cooper) higher-order processes  may also inverse the sign
 of $T,|H|$  scaling.
 We  derive
   the thermodynamic potential as a function of
 magnetization and show that it contains, in addition to regular terms,
 a non-analytic $|M|^3$ term,
   which becomes $M^4/T$ at finite $T$.
 We show that
   regular ($M^2, M^4, ...$) terms
 originate from fermions with energies of order of the bandwidth,
 while the non-analytic term comes from  low-energy fermions.
We
  consider the
 vicinity
 of
 a ferromagnetic  QCP
 by generalizing the Eliashberg treatment of the spin-fermion model
to finite magnetic field, and show that the
 $|M|^3$ term crosses over to a non-Fermi-liquid form $|M|^{7/2}$
near a QCP. The prefactor of the $|M|^{7/2}$ term is negative,
which
 indicates that the system undergoes a first-order rather than a
continuous transition to ferromagnetism. We compare two scenarios
of the breakdown of a continuous QCP: a first-order instability
and a spiral phase;
 the latter may arise from
  the nonanalytic dependence of $\chi$ on the momentum.
 In a model with a long-range interaction in the spin channel, we show
that the first-order
 transition occurs  before the spiral
 instability.
\end{abstract}

 \maketitle
\section{Introduction}
The Landau Fermi-liquid (FL) theory postulates that at low enough
energies a system of interacting fermions behaves as a
weakly-interacting gas of
quasiparticles with renormalized parameters: effective mass, Land{\'e} $g$%
-factor, etc. \cite{agd} The thermodynamics of a canonical
 FL is constructed under the
assumption that the fermion-fermion interaction is absorbed
entirely into a set of
 the renormalization factors (Landau parameters), while the
residual interaction between quasiparticles can be neglected. In
this approximation,
  the FL behaves as a Fermi gas of free quasiparticles.
In particular,
 the specific heat coefficient, $\gamma
(T, H)=C(T,H)/T$, and the uniform, static spin susceptibility,
$\chi (T,H)$,
  remain finite
 in the limit of $T,H\rightarrow 0$, while
 their $T$ and $H$ dependences follow the familiar Sommerfeld
expansions in powers of $T^2$ and $H^2$.

It has long been known that neglecting the residual interaction
leaves some important physics behind.
 In particular,
  non-trivial
{\it kinetics}
  of a FL
  is entirely due to the
 residual interaction among quasiparticles.
 The effect of the
 residual
 interaction on {\it thermodynamics} of
 Fermi systems
 has been studied intensively in recent years (for a review, see Refs.
\onlinecite{belitz_rmp,rosch_rmp}). It is
  well established by now
  that  $T$ and $H$ dependences of
$\gamma (T,H)$ and $\chi (T,H)$ are nonanalytic. In two dimensions
(2D), both $\gamma $ and $\chi $ are linear rather then quadratic
in $T$ and $\left| H\right| $
 (Refs.~\onlinecite{coffey,marenko,chitov_millis,chm03,chmgg_prl,chmgg_prb,ch_s,
betouras,chmm,catelani,aleiner_efetov,efetov,finn,we_short,finn_new}).
In addition, the nonuniform spin susceptibility, $\chi \left(
q\right)$, scales linearly with $|q|$
  for $q\ll k_F$ (Refs.~\onlinecite{belitz,chm03}).

The nonanalytic behavior originates from a dynamic, long-range
component of the residual interaction mediated by virtual
particle-hole pairs.
Two regions
  in the space of momentum transfers
 contribute to the long-range dynamics.
  The first one is
the region of small $q$, where the long-range interaction arises due to the $%
\Omega /q$ form of the fermion polarizability (this
 form is also the reason for Landau damping). In real space, this
component of the interaction falls off slowly, e.g., as $\Omega
/r$ in 2D. The second one is the region around $2k_{F}$, where the
Kohn anomaly generates \emph{dynamic} Friedel oscillations falling
off as $\Omega \cos (2k_{F}r)/r^{1/2}$ in 2D. With this in mind,
the non-analytic behavior of the free energy can be obtained by
 the following scaling argument. The range of the interaction via
particle-hole pairs is determined by a characteristic size of the
pair, $L_{\mathrm{ph}}$,
which is large at small energy scales. At finite temperature, $L_{\text{ph}%
}\sim v_{F}/T$  by the uncertainty principle. To second order
in the bare interaction, two quasiparticles
  interact via a
single particle-hole pair. The energy
 of order $T$,
 carried by such a pair,
  is distributed over a volume
 $L^D_{\mathrm{ph}}\propto T^{-D}$. The contribution from such
process to the free energy per unit volume is of order $\delta
F\sim u^{2}T/L_{\text{ph}}^{D}\propto T^{D+1}$, where $u$ is the
dimensionless coupling constant. Consequently, $\gamma(T)=-
\partial^2 F/\partial T^2\propto T^{D-1}$. Likewise,
at $T=0$ but in finite magnetic field, a
characteristic energy scale is the Zeeman splitting $2\mu _{B}|H|$ and $L_{%
\text{ph}}\sim v_{F}/\mu _{B}|H|.$ Hence,
  $\delta F\propto \left|
H\right| ^{D+1}$ and $\chi \left( H\right) \propto |H|^{D-1}$. For
$D=2$, this implies that $\gamma(T)\propto T$ and $\chi(H)\propto
|H|$. For $D=3$, power-counting misses logarithmic factors which
are recovered by
 an explicit calculation.

A perturbation theory indeed shows~ that $\gamma $ and $\chi $
depend linearly on $T$ and $\left| H\right| $ in 2D, and as
$T^{2}\ln T$ and $H^{2}\ln \left| H\right|$ in 3D.
 To second order in the interaction, it has been
found\cite{chm03,chmgg_prl,chmgg_prb,ch_s,betouras} that
\begin{subequations}
\begin{eqnarray}
&&\delta \gamma (T,H=0)=-\frac{9\zeta (3)}{\pi ^{2}}\left[
f_{c}^{2}(\pi )+3f_{s}^{2}(\pi )\right] \frac{T}{\epsilon
_{F}}\gamma _{0}^{2D}
\label{in_1} \\
&&\delta \chi (T=0,H)=f_{s}^{2}(\pi )\frac{\mu _{B}\left| H\right| }{%
\epsilon _{F}}\chi _{0}^{2D}  \label{in_2} \\
&&\delta \chi (T,H=0)=f_{s}^{2}(\pi )\frac{T}{2\epsilon _{F}}\chi
_{0}^{2D}
\label{in_3} \\
&&\delta \chi (T=0,H=0,q)=\frac{4}{3\pi }f_{s}^{2}(\pi )\frac{|q|}{k_{F}}%
\chi _{0}^{2D},  \label{in_4}
\end{eqnarray}
\end{subequations}
 where $f_{c}\left( \pi \right) =(m/\pi )\left[
U\left( 0\right) -U\left( 2k_{F}\right) /2\right]$  and $f_{s}(\pi
)=-(m/2\pi )U(2k_{F})$ are the charge/spin component of the
(first-order) backscattering amplitude $f_{c/s}(\theta=\pi)$,
$\theta$ is the angle between the incoming momenta of two
fermions. Also in Eqs.~(\ref{in_1}-\ref{in_4}), $\gamma
_{0}^{2D}=m\pi /3$ is the specific heat coefficient of a 2D Fermi
gas, $\chi
_{0}^{2D}=\mu _{B}^{2}m/\pi $ is the spin susceptibility of a 2D Fermi gas, $%
\epsilon _{F}$ is the Fermi energy$,\mu _{B}$ is the Bohr
magneton, and all relevant energy scales--$T,\mu _{B}|H|,$ and
$v_{F}\left| q\right| $--are small compared to $\epsilon _{F}.$
(The scaling forms as functions of all three variables can also be
obtained, see Ref. \onlinecite{betouras}.) Scattering processes
contributing to Eqs. (\ref{in_1}-\ref{in_4}) are characterized by
special kinematics (''backscattering''): two fermions move in
almost opposite directions before a collision and then either
continue to move along the same path (momentum transfer $q=0)$ or
scatter back (momentum transfer $2k_{F}).$

The intriguing feature of the perturbative results is that the
spin susceptibility is not only nonanalytic but also an
\emph{increasing }function of all three arguments: $H$, $T,$ and
$q$. Since one should expect the susceptibility to decrease at
least at energies much larger than $\epsilon _{F},$ a natural
conclusion is that $\chi $ has a maximum at intermediate energies.
 If this
 behavior
 survives beyond weak-coupling,
it implies
  non-trivial
 consequences
 for a magnetic phase transition in such a  system.
 Indeed,
  a maximum of $\chi (T,H, q=0)$ at
finite $H$
 gives rise to a local
  minimum in the free energy
  at finite
magnetization $M$. As
 $\chi (M=0)$ increases,
  this minimum
  becomes degenerate in energy
with a non-magnetic state
  implying that
  a ferromagnetic state emerges via
a discontinuous, first-order transition accompanied by a
metamagnetic response away from the critical point. On the other
hand, a maximum of $\chi (T, H=0, q)$ at finite $q$ implies
  that the system may also undergo a
   transition into a spiral rather
  than uniform magnetic state. Both
 scenarios imply a breakdown of the Hertz-Millis-Moriya (HMM)
 model of a continuous,
 quantum,  ferromagnetic
 phase transition~\cite{hertz,millis_qcp,moriya}.
The first-order instability has been discussed in recent
 literature. \cite{belitz_rmp,rosch_rmp}
It is not clear, however,
 which of the two instabilities--the first-order
 or spiral one--occurs first. One of the aims of this paper is to clarify this
 issue.

Experimentally, a linear $T$ dependence of the specific heat
coefficient has been observed in
 monolayers
 of
$^{3}$He (Ref.[~\onlinecite{saunders}]); both the sign and the
magnitude of the effect are consistent with Eq. (\ref{in_1})
(Ref.~\onlinecite {chmgg_prl,chmgg_prb}). For the spin
susceptibility, the experimental situation
 is less clear. A quasi-linear dependence of $\chi $
 on $T$ was observed in a Si-based 2D heterostructure
\cite{reznikov}; however, the slope is opposite in sign to that in
(\ref{in_3}). On the other hand, a number of experiments on
this and other  heterostructures (Si Ref.~\onlinecite{pudalov_gershenson}, $n$%
-GaAs Ref.~\onlinecite{stormer}, and AlAs Ref.~\onlinecite{shayegan}) have found that $\chi $ \emph{%
increases} with magnetization, in agreement with
 Eq. (\ref{in_2}). A
linear temperature dependence of $\chi $ has recently been
observed in the normal phase of  Fe-based pnictides; \cite{pn}
the sign of the slope is consistent with Eq. (\ref{in_3}).

 A linear $\left| q\right| $-dependence of
$\chi (T=0, H=0, q) $ has recently been proposed to influence
 ordering of nuclear spins via a Ruderman-Kittel-Kasuya-Yosida
(RKKY) interaction mediated by \emph{interacting} rather than free
electrons. \cite {loss,loss_long}
 Because of the $|q|$ term,
  the dispersion of
nuclear spin waves in the RKKY-ordered state, $\omega_s(q)\propto
\chi(q)$, is linear rather than quadratic in $q$.  In 2D, this
implies that the nuclear magnetic order is stable with respect to
thermal fluctuations,
  which
   opens a possibility to freeze nuclear
spins at experimentally accessible temperatures
 with
 potential applications
 in quantum computing.

 Conflicting
observations of the temperature
 and magnetic-field
  dependences of $\chi (T,H)$ and
 potential applications
 in quantum computing
call for a detailed theory of the nonanalytic effects in the
 spin response
  of 2D and 3D Fermi systems. In particular, it is
important to understand whether the weak-coupling results can
  be extended into a non-perturbative regime
 near a ferromagnetic transition.

Several groups have recently investigated this issue.
\cite{chmm,catelani,aleiner_efetov,efetov,finn,we_short,finn_new,loss_long,loss_new}
 It turns out that the result
for the specific heat is robust: for $D <3$, \emph{all}
higher-order corrections  can be absorbed into renormalization of
the backscattering amplitudes $f_{c}\left( \pi \right) $ and
$f_{s}\left( \pi \right) $ in the second-order result, Eq.
(\ref{in_1})
(Ref.~\onlinecite{chmm,catelani,aleiner_efetov,efetov}). One
particular consequence of this result, which still awaits for an
experimental verification, is the additional logarithmic
dependence of the specific heat coefficient
  $\gamma (T) \propto T^{D-1}/(\ln {\epsilon _{F}/T})^2$
 resulting from
 renormalizations of $f_{c/s}\left( \pi \right)$ in the Cooper
channel
 ($f_{c/s}\left( \pi \right)\propto \ln ^{-1}(\epsilon _{F}/T)$ in
the limit of $T\rightarrow 0$,  Refs.~\onlinecite{chmm,aleiner_efetov,chm_cooper}).
 In 3D, there are
additional $T^2 \ln {T}$ nerms in $\gamma$, which are not
expressed via backscattering ~\cite{Pethick73_a,chmm}.

 For the spin
susceptibility, the situation is more complex:
 even in 2D, not all
higher-order processes can be absorbed into renormalization of the
backscattering amplitudes in the second-order results.
  The remaining
processes do not have special kinematics: the momenta of incoming
fermions are not correlated and momenta transfers are generic
rather than peaked either near $0$ or near $2k_{F}.$ The
 signs of these extra linear contributions to $\chi (T,H)$ alternate with
order
 of the perturbation theory, which opens a possibility
 for sign of $\chi (T,H)$
 to be reversed  upon resummation.
 In addition, the
backscattering contribution is suppressed by Cooper logarithms,
leaving the non-backscattering processes as the main contributors
to linear in $T$ and $|H|$
 terms in the susceptibility
 at
sufficiently low
  $T$. \cite{finn,finn_new}

In this paper, we develop a general theory of the nonanalytic
behavior of the spin susceptibility in two and three dimensions,
both in the FL-regime and also in the vicinity of a ferromagnetic
quantum critical point (QCP). In Sec. \ref{sec:2D}, we discuss the
2D case. After a brief review of the perturbation theory for $\chi
$ in Sec. \ref{sec:delta}, we construct in Sec. \ref{sec:beyond}
an expansion of the exact susceptibility in skeleton diagrams with
an increasing number of dynamic polarization bubbles. Physically,
such an expansion corresponds to collecting all processes
involving a given number of virtual particle-hole pairs. In Sec.
\ref {sec:two_bubbles}, we show that all diagrams with two dynamic
bubbles
 give effectively second-order
 results (\ref{in_2},\ref{in_3}) but with the exact rather than perturbative backscattering amplitudes.
In Sec.\ref{sec:other}, we consider processes with more than two
dynamic bubbles
 and show that they also give rise to linear $T$ and $|H|$ terms in $\chi (T,H)$.
 We
evaluate the diagrams up to fourth order in dynamic bubbles and
calculate $\chi $ explicitly for a model form of the scattering
amplitude parameterized by the first two harmonics, $f_{s,0}$ and
$f_{s,1}$. In Sec. \ref{sec:sign}, we address an issue of the sign of the $T$%
- and $H$-dependences of $\chi .$ We show that higher-order
process can reverse the sign of backscattering contribution for a
strong enough interaction, even if
  logarithmic renormalizations in the Cooper channel are neglected.
  In the same Section,
 we also analyze  the role of Cooper renormalizations
 for a system with
 a
  short-range interaction
   and
for a 2D electron gas with Coulomb interaction
 in the large $N$ limit,  relevant mostly for valley-degenerate
semiconductor heterostructures. In agreement with
Ref.~\onlinecite{finn_new}, we find
 that the slope of $\chi (T, H)$  in a Coulomb gas changes sign
 below a certain energy;
however, this energy is of order $E^*=\epsilon _{F}\exp \left(
-N^{3/2}/\sqrt{2}\right)$ in the large-$N$ model.
  Already for the case of two valleys ($N=4$),
 $E^*$ is too low for this mechanism to be responsible for the
observed negative sign of the $T$-dependence of $\chi $ in s
Si-MOSFET (Ref.~\onlinecite{reznikov}).

Next, we obtain  a general form of the thermodynamic potential for
a 2D FL with an arbitrary strong interaction  (Sec.
\ref{sec:fl_ren}) and extend the analysis of the magnetic-field
and temperature dependences of $\chi (T,H)$
 to both FL- and non-FL regions near
 a ferromagnetic QCP in 2D (Sec. \ref {sec:qcp}).
  In Sec. \ref{sec:at_qcp},
    we
 neglect Cooper renormalizations and show
  that
  while $\chi (T,H)$ \emph{increases} with $H, T$ in both regimes,
  the $|H|, T$ scaling holds only up to a certain energy which decreases as the QCP is approached.
  At higher energies,
  the magnetic-field and temperature dependences of $\chi (T,H)$
 are $|H|^{3/2}$ and $T \ln T$, respectively.
 The increase of $\chi$ with $H$ signals an imminent
breakdown of the continuous ferromagnetic transition. We discuss
possible scenarios of quantum- and finite-temperature
ferromagnetic phase transitions in Sec.~\ref{sec:pt}.
  In
particular, we show
 that for a large radius of the
  interaction in the spin channel  the first-order transition always preempts the spiral instability.
 Finally, in
Sec. \ref{sec:cooper_qcp}, we show that the increase of $\chi$
with $H$ and $T$ near a QCP is not affected by renormalization in
the Cooper channel, as this renormalization becomes relevant only
below an energy which decreases exponentially
 as the QCP is approached.

In Sec.~\ref{sec:3D}, we consider
 the 3D case.  In Sec. \ref{sec:3DH},
 we show that $\propto H^{2}\ln \left| H\right|$ form of
$\chi(T=0,H)$ in a 3D FL transforms into a weaker, $H^{2}\ln \ln
\left| H\right|$ form near
  a ferromagnetic QCP.
  In Sec.~\ref{sec:3DT}, we analyze the $T$%
-dependence of $\chi (T, H=0)$ in 3D.
 A 3D FL
 is peculiar in a sense that
 $\chi (T)$ scales as $T^{2}$ without an extra logarithmic factor
\cite{bealmonod68,pethick}. We generalize the earlier result
 for the $T^2$ scaling
  by Beal-Monod et al. \cite{bealmonod68} and
show that the prefactor of the $T^{2}$ term is non-universal: its
magnitude and {\it sign} depend
 on details of the fermion dispersion. For the $k^2$ dispersion
 discussed in Ref. \onlinecite{bealmonod68}, the prefactor of the $T^2$ term is negative,
  i.e.,  $\chi (T)$ decreases with $T$.
However,  $\chi $ may increase with $T$ for a  more complex
dispersion.
  A increase
  of $\chi (T)$ with $T$
  has been observed in a number of exchange-enhanced
paramagnetic metals. \cite {co}

Section \ref{sec:concl}
 summarizes our conclusions. Some technical details of the
derivations are given in Appendices \ref{sec:largeN}-
\ref{sec:app3D}. Some of the results presented here were published
in a shorter form in Ref.~\onlinecite{we_short}.

\section{Magnetic response of a 2D Fermi liquid}
\label{sec:2D}
  The spin susceptibility at zero temperature and in zero magnetic field, $\chi (T=0,H=0)$ is
 described
   by the conventional FL theory (Ref.~\onlinecite{agd}).
 The subject of our study
  is
the temperature- and field-dependent part
  of the susceptibility:
 $\delta \chi (T,H) = \chi (T,H) - \chi (0,0)$. The most straightforward way to obtain $\delta \chi (T,H)$
  is to
evaluate the thermodynamic potential
 $\Xi \left(T,H \right)$ and
differentiate it twice with respect to the field. In contrast to
 the linear-response theory, which generates a large number of
diagrams, the number of relevant diagrams for the thermodynamic
potential is rather small.
 The prefactor of the $H^2$ term in the thermodynamic potential
gives the $T$-dependent spin susceptibility, while the nonanalytic
$|H|^3$ term gives the field-dependent (nonlinear) susceptibility.
\subsection{Second-order perturbation theory}
\label{sec:delta}
  To second order in the
interaction
  $\delta \chi (T,H)$ was considered in
 Refs.\onlinecite{chm03,chmgg_prl,ch_s,betouras}, where it
 was found that
\begin{itemize}
\item $\chi (T,|H|)$ is nonanalytic in both arguments and scales
as $\max\{T,|H|\})$; \item the nonanalyticity comes from the
states near the Fermi surface; \item
 only $2k_F$ scattering is relevant, thus
 the prefactors of
 the linear terms in $T$ and in $H$ contain only the $2k_F$ component of the unteraction
 $U(q)$.
 \end{itemize}
 In this Section, we
 overview briefly the
  second-order perturbation theory,
because later we will need to understand
 what replaces $U(2k_F)$ in the interaction vertices beyond the second order.

\begin{figure}[tbp]
\caption{The field-dependent part of the thermodynamic potential at second
order. $k,p,q$ are the four-momenta: $k\equiv (\mathbf{k%
}, \protect\omega _{m}$), etc.}\label{fig:fig1}
\par
\begin{center}
\epsfxsize=0.3\columnwidth\epsffile{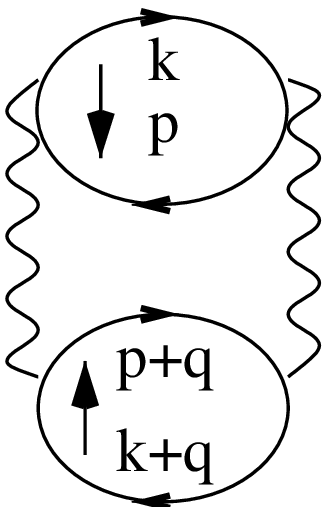}
\end{center}
\end{figure}

At second order in $U(q)$, the field-dependent part of the
thermodynamic potential $\Xi (T,H)$ is given by
 a single
  diagram shown in Fig.~\ref{fig:fig1}.
  In this diagram, the spins of
  fermions in one of the bubbles are opposite to
  those in another bubble.
  The nonanalytic contribution to $\Xi (T,H)$ originates from
 the
  $2k_F$ nonanalyticity of the {\it dynamic}
 polarization bubble in zero field and, hence,
  is proportional
  to
  $U(q=2k_F)$.
 Finite magnetic field cuts off the nonanalyticity,
 but at a price that the derivatives with respect to the field become nonanalytic in $H$.
 With all  four fermions
  near the Fermi surface, the
$2k_F$ diagram necessarily contains two spin-up fermions with
momenta near
 ${\bf k}_F$ and $-{\bf k}_F$ and two spin-down fermions also with
momenta near ${\bf k}_F$ and $-{\bf k}_F$. These four fermions can
 be  re-grouped
  into two up-down bubbles, each
 with a small momentum transfer.
This simplifies the computations substantially because  the
polarization bubble
 has a much simpler form
   for small $q$
 than  for $q$ near $2k_F$.

The magnetic field enters the problem via the Zeeman shifts of
single-fermion energies in the Green's functions
\begin{equation}
G_{\uparrow, \downarrow }\left(\mathbf{k}, \omega_m\right) =\frac{1}{%
i\omega _{m}-\epsilon _{\mathbf{k}} \pm \Delta/2},
\label{e_3}
\end{equation}
 where \beq\dz\equiv 2\mu _{\mathrm{B}}H\eeq is the Zeeman energy.
The up-down polarization bubble is defined as
\begin{equation}
\Pi _{\uparrow \downarrow }\left( q,\Omega _{m}\right) =T\sum_k
G_{\uparrow }\left(\omega_m+\om,{\bf k}+{\bf q}\right)
G_{\downarrow }\left(\omega_m,{\bf k}\right),
\label{pi}\end{equation}
 where $T\sum_k$ is a shorthand for
$T\sum_{\omega_m}\int d^2k/(2\pi)^2$. For $q\ll k_F$, the
 up-down
 bubble can be separated into the static and dynamic parts as
\begin{equation}
\Pi _{\uparrow \downarrow }\left(q, \Omega _{m}\right) =-\nu
\left( 1-P_{\uparrow \downarrow }\right),
\label{def_P}\end{equation}
 where $\nu=m/2\pi$ is the density of
states at the Fermi surface and
\begin{equation}
P_{\uparrow \downarrow }\left(q, \Omega _{m}\right) =\frac{\left| \Omega
_{m}\right| }{\sqrt{\left( \Omega _{m}-i\dz\right) ^{2}+v_{F}^{2}q^{2}}}.
\label{bubbleud2D}
\end{equation}
  Re-expressed in terms of the up-down bubbles, the diagram in
Fig.~\ref{fig:fig1}
 reads
 \begin{equation}
\Xi _{2}\left(T,H\right) =-\frac{U^2(2k_F)}{2}T\sum_q \Pi
_{\uparrow \downarrow }^{2}\left( \Omega _{m},q\right).
\label{second}
\end{equation}
The nonanalytic
 part of $\Xi_{2} (T,H)$
 is obtained by keeping the square of the dynamic
term in Eq.(\ref{def_P}), i.e., replacing $\Pi^2_{\up\down}$ by
 $\nu^2 P_{\up\down}^2$. This gives
\bea
&&\Xi _{2}\left(T,H\right) = - \frac{u^2_{2k_F}}{2} T \sum_q \frac{\Omega_m^2}{(\Omega_m - i \Delta)^2 + (v_F q)^2} \nonumber \\
&& = - \frac{u_{2k_F}^2}{8\pi v^2_F} T \sum_{\Omega_m} \Omega_m^2
\ln{\frac{W^2}{(\Omega_m - i \Delta)^2}}, \label{ya_1} \eea where
\beq u_{2k_F}\equiv \nu U(2k_F)\eeq and $W$ is the high-energy
cutoff
 which, in general, is of order
 of the
bandwidth.

The logarithm in the frequency sum in Eq.~(\ref{ya_1}) originates
from the $\Omega_m/|q|$ form of the polarization bubble at $v_F q
\gg \Omega_m, \Delta$, i.e., from the long-range tail of the
dynamic bubble (in real space, $\Omega_m/|q|$ transforms into
$\Omega_m/r$).
 If not for the logarithm, $\Xi_{2} (T,H)$ would be expandable in
 powers of
 $T^2$ and
$H^2$.
 The logarithm breaks analyticity. Replacing the Matsubara sum by a
contour integral, and subtracting off the field-independent part,
we obtain from Eq.~(\ref{ya_1}) \beq \Xi _{2}\left(T,H\right)
=-\frac{u^2_{2k_F}}{8\pi v_F^2} \int_0^{|\dz|}
d\Omega\Omega^2\coth\left(\frac{\Omega}{2T}\right)
\label{Xi2H2D_a} \eeq The integral in Eq.~(\ref{Xi2H2D_a}) can be
solved exactly (in terms of polylogarithmic functions), but we
actually do not need this solution, as the $T$- and $H$-dependent
spin susceptibility
 can be obtained directly from
Eq.~(\ref{Xi2H2D_a}) by differentiating it twice with respect to
$H$. This yields
\beq \delta \chi_{2} (T,H)=-\frac{\partial^2 \Xi_2}{\partial
H^2}=u^2_{2k_F}\frac{\left| \dz\right|
}{2E_{F}}S\left(\frac{|\dz|}{2T}\right)\chi_{0}^{2D}, \label{chi2}
\eeq where $\chi^{2D}_{0}=\mu_B^2 m/\pi$ is the spin
susceptibility of a free 2D Fermi gas,
  and the scaling function
$S(x)$ is
 \beq S(x)=\coth x-\frac{x}{2\sinh^2x}. \eeq
The asymptotic limits of $S$ are $S(x\to 0)=1/2x$ and
$S(x\to\infty)=1$.
 Substituting these limits into Eq.~(\ref{chi2}), we find that
 the susceptibility {\it increases} linearly with
the largest of the two energy scales, $T$ and $\dz$. Schematically,
\beq
 \delta\chi_2(T,H)=
 u^2_{2k_F}\frac{E}{2\epsilon_F}\chi_{0}^{2D}
 \label{chi2_HT}
\eeq
where $E\equiv \max \left\{T,\left|\dz
 \right|\right\}$.

If the same calculation is performed in real rather than Matsubara
frequencies, the frequency integral contains the product of the
real and imaginary parts of
the retarded, dynamic bubble: $u^2_{2k_F} {\rm Re
}P^{R}_{\up\down}{\rm Im}P^{R}_{\up\down}$.
 This allows for a transparent physical interpretation of the two-bubble
 diagram. \cite{chmm} Indeed,
 Im${\rm \Pi} ^{R}$ can be thought of as the spectral density of particle-hole pairs, while $%
u_{2k_F}^{2} $Re${\rm \Pi} ^{R}$ as of the dynamic interaction
between the two fermions in the particle-hole pair. The product
 $d\Omega d^{2}qu_{2k_F}^{2}$Re${\rm \Pi} ^{R}$Im${\rm \Pi}
^{R}\propto d\Omega d^{2}q\left( \Omega /q\right) ^{2}$ is then
the potential energy of a single
 particle-hole pair excited above the ground state.
 In this language, an increase of $\delta \chi_{2} (T,H)$ with
 both $H$ and  $T$
can be understood as the consequence of the fact that the
 magnetic field gaps out
 soft particle-hole pairs,
 suppressing their contribution to the thermodynamic potential.

\begin{figure}[tbp]
\caption{Diagrams for the thermodynamic potential beyond second
order.} \label{fig:fig2}
\begin{center}
\epsfxsize=0.8\columnwidth
\epsffile{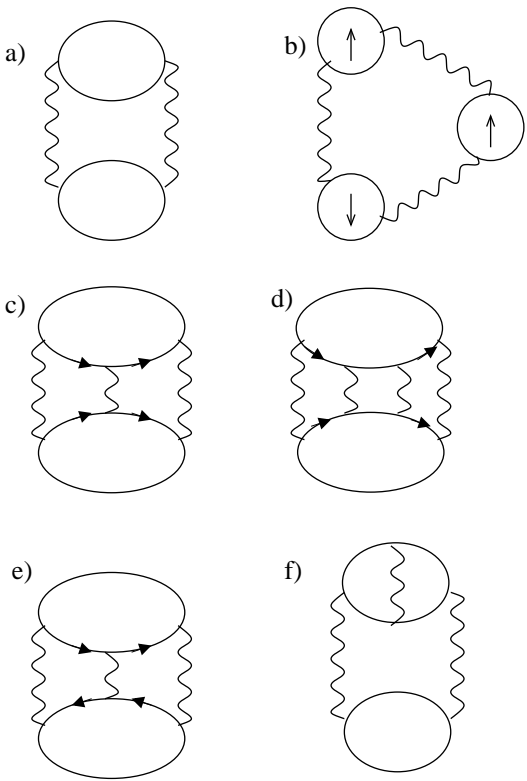}
\end{center}
\end{figure}

\subsection{Beyond second order}
\label{sec:beyond}
 Higher-order diagrams for $\Xi (T,H)$ can be
divided into two groups. The first group is formed by diagrams
 in which a nonanalyticity is produced in the same way as at second
order: by extracting a product of only two dynamic bubble from the
whole diagram. The rest of the diagram goes into dressing up of
the fermion propagators and renormalization of the $2k_F$
interaction lines into full {\it static} vertices. In
real-frequency language, these diagrams describe higher-order
corrections to the effective static interaction in a single-pair
process.\cite{chmm} The second group is formed by diagrams in
which
 a nonanalyticity is produced by combining more than two dynamic
bubbles.

These two groups of diagrams describe two distinct physical
processes. As we will show in this Section, the first group
corresponds to scattering events in which fermions, moving in
almost opposite directions before the collision, reverse their
respective directions of motion. We dub this process as
"backscattering". The second group describes scattering events
with no correlation between initial directions of motion.

Third-order diagrams {\it b},{\it c}, and {\it f} in
Fig.~\ref{fig:fig2} belong to the first group. Nonanalytic
contributions to $\chi$ from these diagrams are obtained by
selecting two dynamic
 up-down
  bubbles  and setting $q=\Omega_m=0$ in the rest of the diagram.
As an example, we consider diagram {\it b}. Fermions from any of
the two bubbles with opposite spins can be re-grouped into two
 up-down
  bubbles in the same way as in the second-order
diagram in Fig.~\ref{fig:fig1}.
 Retaining only
 the dynamic part
 of these two bubbles, we obtain the same nonanalyticity
 as at
   second order.
 The remaining, third bubble can then be evaluated
 at zero external frequency, which means that it
 renormalizes the {\it static} vertex.
\begin{figure}[tbp]
\caption{Skeleton backscattering diagram.}\label{fig:fig3}
\par
\begin{center}
\epsfxsize=0.5\columnwidth\epsffile{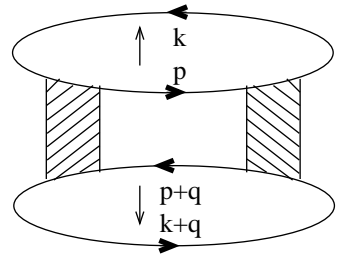}
\end{center}
\end{figure}
 Diagrams of the first type to all orders
 can be cast into a single skeleton diagram, shown
in Fig.~\ref{fig:fig3}. The fermion Green's functions in this
diagram are
  of a
 FL form
\begin{equation}
G_{\up,\down} = \frac{Z}{i \omega_m - \vf (k-k_F) \pm \dzz/2},
\label{n_1}
\end{equation}
where $\vf$ is the renormalized Fermi velocity, $Z$ is the
quasiparticle residue, and $\dzz = 2 {\tilde \mu}_B H$ with
${\tilde \mu}_B$ being the effective Bohr magneton, which we
discuss below. A hatched block
   in Fig.~\ref{fig:fig3}
  is
 the spin component
 of the renormalized static vertex,
$\Gamma_s(\mathbf{k},\mathbf{p};q)$,
 obtained from the dynamic one in the limit of $\Omega_m/\vf q\to
0$.
 We will follow a standard
procedure  \cite{agd} and absorb
  factors of $Z$
into $\Gamma_s$.
 In the low-energy limit
($T,
\dzz\ll \epsilon_F$), the fermion momenta ${\bf k}=\mathbf{n}_k k$
and ${\bf p}=\mathbf{n}_p p$ are confined to the Fermi surface, so
that $\Gamma_s$ depends on the angle
 between $\mathbf{n}_k$ and $\mathbf{n}_p$:
 (as well as on $q$) :
$\Gamma_s=\Gamma_s\left(\mathbf{n}_k\cdot\mathbf{n}_p; q
\right)$.
 At first order in $U$, the vertex reduces to
$\Gamma_s (\mathbf{k},\mathbf{p};q) = -U(|{\bf k} - {\bf p}|)$,
and only $U(2k_F)$ contributes to the nonanalyticity.
 Beyond the lowest order, however, more complicated angular
averages of the interaction
 occur, and it is not a'priori clear what
  the prefactor of
the nonanalytic term is.
 We now
show, using the same procedure as in Ref.[~\onlinecite{chmm}], that
this prefactor is precisely the square of the spin component of
the backscattering amplitude:
  $\nu^2
 \Gamma_s^2 (\mathbf{n}_k\cdot\mathbf{n}_p=-1,q=0)$.

\subsubsection{Contribution to the susceptibility
from diagrams with two dynamic bubbles} \label{sec:two_bubbles}
The nonanalytic contribution
 of
 the skeleton diagram in Fig.~\ref{fig:fig3} is given
by~\cite{chmm}\bwt \bea \Xi _{2s} (T, H)=-\frac{1}{2}T\sum_q\int
d\mathbf {n}_k\int d\mathbf{n}_{p} \times \left[\nu \Gamma
_{s}\left( \mathbf{n}_k\cdot\mathbf
{n}_p;q\right) \right] ^{2}\mathcal{P}%
_{\uparrow \downarrow }\left( \Omega
_{m},q;\mathbf{n}_{k}\right) \mathcal{P%
}_{\uparrow \downarrow }\left( \Omega _{m},q;\mathbf{n}_{p}\right)
, \label{xis1}\eea\ewt where
\beq
\mathcal{P}_{\uparrow \downarrow }\left( \Omega
_{m},q;\mathbf{n}_{k}\right) =\frac{1}{2\pi }\frac{i\Omega
_{m}/\vf q}{\left( i\Omega
_{m}+\dzz\right) /\vf q-\mathbf{n}%
_{k}\cdot \mathbf{n}_{q}}
\end{equation}
is the propagator of a particle-hole pair
 moving with in the direction of $\mathbf{n}_k$ with small energy
 $\Omega_m$ and
momentum $\mathbf {q}$. [$P_{\up\down}$ in Eq.~(\ref{def_P}) is
obtained from $\mathcal{P}_{\up\down}$ by averaging over
$\mathbf{n}_k$:
 $P_{\up\down}(\Omega_m,q)=\int d\mathbf{n}_k \mathcal{P}%
_{\uparrow \downarrow }\left( \Omega
_{m},q;\mathbf{n}_{k}\right)$.] The vertex $\Gamma_s$ can be
 expanded in angular harmonics as
\begin{equation}
\Gamma_s\left(\mathbf{n}_k\cdot\mathbf{n}_p;q\right)
=\sum_{l=0}^{\infty }\Gamma%
_{s,l}(q)\cos\left(l\theta_{kp}\right),
\end{equation}
where $\theta_{kp} = \cos^{-1} \left({{\bf n}_k {\bf
n}_p}\right)$. Substituting this expansion into (\ref{xis1}), we
obtain \bea \Xi_{2s} (T,H) &=&-\frac{1}{2}\nu^2 T\sum_q \left(
\frac{\Omega _{m}}{\vf q}\right)
^{2}\nn\\
&&\times\sum_{l,l'}\Gamma_{s,l}(q) \Gamma_{s,l^{\prime}}(q)
A_{ll^{\prime }}, \eea where \bea A_{ll^{\prime
}}&=&\frac{1}{\left( 2\pi \right) ^{2}} \int\int d\mathbf{n}
_{k}d\mathbf{n}
_{p}\cos(l\theta_{kp})\cos(l'\theta_{kp})\nn\\
&&\frac{1}{%
iz-\mathbf{n}_k\cdot\mathbf{n}_q}\frac{1}{iz-\mathbf{n}_p\cdot\mathbf{n}_q}
\eea
and
\begin{equation}
z=\frac{\Omega _{m}-i\dzz}{\vf q}.
\end{%
equation}
Using the identities
\begin{eqnarray}
(ia-b)^{-1} &=&-i\text{sgn(Re}a)\int_{0}^{\infty }d\lambda
e^{-\lambda \text{%
sgn(Re}a)(a-ib)} \nn\\
i^lJ_l(\lambda)&=&\int^{\pi}_0\frac
{d\theta}{\pi}e^{i\lambda\cos\theta}\cos(l\theta)
\end{eqnarray} and
\begin{equation}
\int_{0}^{\infty }d\lambda J_{l}\left(
\lambda \right) e^{-a\lambda }=\frac{%
\left( \sqrt{a^{2}+1}-a\right) ^{l}}{%
\sqrt{a^{2}+1}}
\end{equation}
(valid for $\text{Re}a>0$), we
 obtain for $A_{ll^{\prime }}$ \beq
A_{ll'}=\frac{(-)^{l+l'}}{2}\frac{\left[\left(\sqrt{z^2+1}-z\right)^{l+l'}
+\left(\sqrt{z^2+1}-z\right)^{l-l'}\right]}{z^2+1}.\eeq The
expression for $\Xi_{2s} (T,H)$ reduces then to
 \bea
\Xi_{2s}(T, H)=-\frac{1}{2}T\sum_qB(z)\frac{\Omega^2_m}{(\Omega_m
-i\dzz)^{2}+(\vf q)^{2}}\label{xis2}\eea where \bea B(z)
&=&\sum_{l,l^{\prime }=0}^{\infty
}\left[\frac{\left(\sqrt{z^{2}+1}-z\right)^{l+l'}+\left(\sqrt{z^{2}+1}-z\right)^{l-l'}}{2}\right]
\nn\\
&&\times (-)^{l+l^{\prime
}}\nu^2\Gamma_{s,l}(q)\Gamma_{s,l^{\prime }}(q).
\label{xis3} \eea
 As it was the case
 for the second-order diagram, the nonanalyticity in $\Xi_{2s} (T,H)$
 is associated with the logarithmic divergence of the integral over $q$
[see Eq. (\ref{ya_1})].
 Because  the logarithm comes from the ``tails'' of the integrand,
  typical $\vf q$ are much lager than
 both  $\Omega_m$ and $\dzz$, i.e., typical $z$ are small.
  Therefore, one can safely put $z=0$ in the factor in square brackets of Eq.~(\ref{xis3}),
   upon which it reduces to unity.
 On the other hand, since $q$ is still smaller than the momentum
cutoff of the interaction, one can set $q=0$ in the vertex.
   Next, we recall
 that the small-momentum limit of
 $\nu \Gamma_s ({\bf k},{\bf p};q)$ is the scattering amplitude $f_s (\theta
 _{kp})$ ( Ref.\onlinecite{agd}). Therefore,
 \begin{equation}
B\left( 0\right) =\sum_{l,l^{\prime }=0 }^{\infty
}(-)^{l+l^{\prime }}f_{s,l}f_{s,l'}=\left[
\sum_{l=0}^{\infty}(-)^lf_{s,l}\right]^2 =\left[ f _{s}\left( \pi
\right) \right] ^{2},
\end{equation}
 which is a square of the {\it exact} backscattering amplitude.

The rest of the integral in Eq.~(\ref{xis3}) is evaluated in the
same way as
 it was done at second order and yields the same scaling form as in
  Eq. (\ref{chi2}), with $v_F \rightarrow \vf$ and $\dz \rightarrow \dzz$.
 Therefore, the contribution of
 the  skeleton diagram in Fig.
 \ref{fig:fig3}
to the spin susceptibility
  is given by
 \beq
\delta \chi _{2s} \left(T,H\right)
=\left[f_s(\pi)\right]^2\left(\frac{{\tilde\mu_B}}{\mu_B}\right)^2\frac{|\dzz|}{2{\tilde
\epsilon}_F}S \left(\frac{|\dzz|}{2T}\right)\chi_{0}^{2D},
\label{chi_gen}\eeq where ${\tilde \epsilon}_F=\vf k_F/2$ is the
renormalized Fermi energy.

The renormalized Fermi velocity $\vf$ and Bohr magneton ${\tilde
\mu_B}$ can be expressed in terms of the Landau parameters
$g_{c,l}$ and $g_{s,l}$
 (Ref.~\onlinecite{agd}), where $c$ and $s$ stand for
charge and spin.
 The Fermi velocity is given by
$\vf =v_F/
 (1 + g_{c,1})$,
  while
 renormalization of the Bohr magneton follows
from the requirement that
  the Zeeman energy of a spin
in the magnetic field is
 $2 {\tilde \mu}_B H = 2\mu_B  H/ \left(1+g_{s,0}\right)$.
 Then, \beq {\tilde \mu}_B =  \frac{\mu_B}{1 + g_{s,0}}.
 \label{n_9}
\eeq

We also recall that harmonics of the  Landau interaction function
 are related to harmonics of the scattering amplitude~\cite{agd}.
 In 2D, this relation is given by
 \beq
f_{a,n}=\frac{g_{a,n}}{1+\frac{g_{a,n}}{2-\delta{n,0}}},
\label{yep}
 \eeq
 where $a = c,s$.
 To first order in the
 interaction,
\bea g_{c,n} &=& f_{c,n}= \nu \left[2U(0)\delta_{n,0} - \int_0^\pi
\frac{d \theta}{\pi}  \cos n \theta U\left(2k_F
\sin{\theta/2}\right) \right] \nonumber \\ g_{s,n} &=& f_{s,n} =-
\nu \int_0^\pi \frac{d \theta}{\pi} \cos n \theta U\left(2k_F
\sin{\theta/2}\right). \eea In general, $g_{a,l} = Z^2 (v_F/\vf)
\nu \Gamma_{a,l}^\Omega$, $f_{a,l} = Z^2 (v_F/\vf) \nu
\Gamma_{a,l}^q$, where $\Gamma^\Omega$ and $\Gamma^q$ are
renormalized vertices
 at small momentum and frequency transfers, in the limits $\vf q/\Omega_{m}
 \to 0$
and $\Omega_m/\vf q \to 0$, respectively.

It is convenient to re-express $\delta\chi(T,H)$ in terms of the
 actual (renormalized)
spin susceptibility at $T=H=0$, rather than of the
 susceptibility of a Fermi gas
$\chi^{2D}_0$.
  The renormalized spin susceptibility at $H=T=0$
is given by~\cite{agd}
\bea \chi (0,0)&=&
  \frac{m}{\pi} \frac{v_F}{\vf} \frac{\mu^2_B}{1+g_{s,0}} =
  \frac{v_F}{\vf} \left(1-f_{s,0}\right) \chi^{2D}_0.
\label{n_8} \eea Expressing $\chi^{2D}_0$ via $\chi (0,0)$ and
substituting the result back into Eq.~(\ref{chi_gen}), we obtain
\beq \delta \chi _{2s}\left(T,H\right) =\chi
(0,0)~\left[f_s(\pi)\right]^2\frac{{\tilde\mu_B}}{\mu_B}\frac{|\dzz|}{2
\epsilon_F}S \left(\frac{|\dzz|}{2T}\right).
 \label{chi_gen_1}\eeq

 Before concluding this Section, we note that
 the exact backscattering amplitude $f_s (\pi)$
  depends
logarithmically on $T$ and $H$
  due to singular
renormalizations in the Cooper channel.
\cite{chmm,efetov,finn,finn_new,chm_cooper} To see this, one needs
to recall that $f_s (\pi)$ is equal (up to a prefactor) to the
Cooper vertex for scattering from the states with momenta ${\bf
k}$ and $-{\bf k}$ into the states with momenta $-{\bf k}$ and
${\bf k}$, respectively.
  We will discuss this special feature of the backscattering amplitude
 in Secs. \ref{sec:Cooper} and \ref{sec:coulomb}, but for a
moment
 continue with
 the consideration of
 higher-order contributions to $\Xi (T,H)$.

\subsubsection{Contributions to susceptibility from more
 than three dynamic bubbles}\label{sec:other}

  There are other diagrams at third and higher orders, which do not belong to
 the skeleton diagram in Fig. \ref{fig:fig3}. In zero magnetic field, these additional diagrams yield only
analytic contributions to $\Xi (T, H=0)$
(Refs.~\onlinecite{chmgg_prl,chmgg_prb,chmm,chm_cooper}). This is
not so in the presence of the magnetic field, as we are now going
to demonstrate.

 At third order, there is only one diagram which cannot be fully
 absorbed into  Fig. (\ref{fig:fig3})--diagram {\it e} in
 Fig.~(\ref{fig:fig2}). For a local interaction ($U(q)=\mathrm{const}$),
 this diagram contains a cube of the up-down bubble
\bea &&\Xi_{3e}(T,H)=- \frac{u^{3}}{3}T\sum_q \left(1 -
P_{\uparrow \downarrow }
\right)^{3}\notag\\
&&=- \frac{u^{3}}{3}T\sum_q\left(1-3 P_{\up\down}+
3P^2_{\up\down}-P^3_{\up\down}\right). \label{xi3e} \eea
 As one can readily verify, the first two terms do not give rise to
non-analyticities, while
 the $P^2_{\uparrow \downarrow }$ term
  has already been accounted for in the skeleton diagram of
Fig.~(\ref{fig:fig3}).
 The new contribution comes from the $P^3_{\uparrow \downarrow }$
term.  Keeping only this term and
 integrating over $q$, we obtain
\bwt
\begin{equation}
\Xi _{3e}\left(T,H\right)= \frac{u^{3}}{6\pi}T\sum_{\om}\int_{0}^{\infty }dqq\frac{\Omega _{m}^{3}}{%
\left[ \left( \Omega _{m}-i\dz\right) ^{2}+v_{F}^{2}q^{2}\right]
^{3/2}}
=\frac{u^3}{3\pi v_F^2}T\sum_{\om>0}\frac{\om^4}{\om^2+\dz^2}.
\label{n_2}
\end{equation}
\ewt
 Subtracting off the ultraviolet contribution
   and summing over $\Omega_m$, we find
  \beq
  \Xi_{3e}(T,H)=\frac{u^3}{12\pi
  v_F^2}|\dz|^3\coth\frac{|\dz|}{2T}.
  \label{xi3e_1}
  \eeq
 Differentiating twice
 with respect to the field,
  we obtain the new contribution to the susceptibility
\begin{equation}
\delta \chi _{3e}(T,H)
=-u^3\frac{|\dz|}{\epsilon_F}R\left(\frac{|\dz|}{2T}\right)
\chi^{2D}_0, \label{chi3e}\end{equation} where \beq
R(x)= \coth
x-\frac{x}{\sinh^2x}+\frac{x^2}{3\sinh^3x}. \eeq In the two
limits,
 $R(x\to\infty)=1$ and
$R(x\to 0)=1/3x$.  We see that  $\delta\chi_{3e}$ has the same nonanalytic dependence
on $T$ and $H$ as the second-order diagram: it scales linearly
with the largest of the two energy scales \beq \delta \chi _{
3e}(T,H)=-u^3\frac{\max\{|\dz|,2T/3\}}{\epsilon_F}. \eeq

There is one essential difference between the second- and
third-order contributions: the nonanalyticity in $\delta \chi
_{3e}(T,H)$
 does \emph{not} arise from a
logarithmically divergent integral over $q$. Indeed, the momentum
integral in Eq.~(\ref{n_2}) is convergent and comes from the
region
 $q\sim |\Omega _{m}|/v_{F}\sim \left|\dz\right|/v_F
$.  This means that Eq.~(\ref{chi3e})
 cannot be obtained by replacing the
dynamic part of the bubble by its asymptotic form at large
$v_Fq/\Omega_m$, which was the case for the backscattering
contribution.

Notice that the {\it sign} of the third-order non-backscattering
contribution is opposite to the second order result. This opens a
possibility of inverting the sign of $\delta\chi$ in the
non-perturbative regime (see Sec.~\ref{sec:sign} for a more
detailed discussion).

To go beyond the  perturbation theory
 for this new type of
processes, we apply the same procedure as for backscattering.
  Namely, we combine all diagrams with three dynamic bubbles into a
``third-order'' skeleton diagram by replacing the bare
interactions in Fig.~\ref{fig:fig3}{\it e} by the renormalized
vertices evaluated in the limit of $\Omega_m/v_F q \rightarrow 0$.
 This limit ensures that we obtain contributions with no more than
 three dynamic bubbles. The renormalized
vertices
 are then again the spin components of the scattering amplitude
 $f_s$,
 and
 the  third-order skeleton diagram
  reduces to
 \bwt
\bea \Xi _{{\rm 3s},3}=- \frac{1}{3}\sum_q\int d\mathbf{n}%
_{k}\int d\mathbf{n}_{p}\int
d\mathbf{n}_{s}f_{s}\left(\mathbf{n}_k\cdot\mathbf{n} _{l}\right)
f_{s}\left( \mathbf{n}_l\cdot\mathbf{n}_p\right) f _{s}\left(
\mathbf{n}_l\cdot\mathbf{n}_p\right)
\mathcal{P}
_{\uparrow \downarrow }\left( \Omega _{m},q;%
\mathbf{n}_{k}\right)
\mathcal{P}_{\uparrow \downarrow }\left( \Omega _{m},q;%
\mathbf{n}%
_{l}\right) \mathcal{P}_{\uparrow \downarrow }\left( \Omega _{m},q;
\mathbf{n}_{p}\right). \label{xis3a}\eea \ewt
 As it was done for the second-order skeleton diagram, we
 replace
 the bare $v_F$ and $\dz$ by their renormalized values and
 absorb
 the quasiparticle residue $Z$ into $f_s$.
  We now show that the three vertices
in Eq.~(\ref{xis3a}) do {\it not} form the cube of the
 backscattering amplitude, i.e.,
 that
  the nonanalyticity in $\delta \chi _{3s}$ comes
   from scattering of fermions
  with uncorrelated directions of the initial momenta.
    To demonstrate this, we  adopt a simplified
model, in which the angular dependence of $f_s \left(
\mathbf{n}_l\cdot\mathbf{n}_p\right)$  is approximated
 by
 the
  first two harmonics:
 \beq
f_s(\mathbf{n}_k\cdot\mathbf{n}_p)=f_{s,0}+
\mathbf{n}_k\cdot\mathbf{n}_pf_{s,1}. \label{2harm}\eeq In this
model, the backscattering  amplitude is equal
 to
\beq
f_s(\pi)=f_{s,0}-f_{s,1}=f_{s,0}\left(1-\frac{f_{s,1}}{f_{s,0}}\right).
 \label{back2harm}\eeq
Substituting Eq.~(\ref{2harm}) into Eq.~(\ref{xis3a}), performing
straightforward angular integrations,
 and differentiating twice with respect to the magnetic field, we
obtain for the
 three-bubble contribution to the spin susceptibility:
 \begin{subequations}
 \bwt
\begin{eqnarray}
&& \delta\chi_{3s}(T=0,H)= \left(f_{s,0}\right)^3
F_3\left(\frac{f_{s,1}}{f_{s,0}}\right)\left(\frac{{\tilde\mu}_B}{\mu_B}\right)^2
\frac{|\dzz|}{{\tilde \epsilon}_F} \chi^{2D}_0
 \label{any3H} \\
&&\delta\chi_{3s}(H=0,T)= \left(f_{s,0}\right)^3
F_3\left(\frac{f_{s,1}}{f_{s,0}}\right)\left(\frac{{\tilde\mu}_B}{\mu_B}\right)^2
\frac{2T}{3{\tilde \epsilon}_F} \chi^{2D}_0\label{any3T}, \eea\ewt
\end{subequations} where \bea
F_3(x)&=&1-3\left( 2\ln 2-1\right)x+3(3\ln 2-2)x^2\nn\\
&&+\left( 5/2 -3\ln 2\right)x^3. \label{F3} \eea
 Obviously, the
product $\left(f_{s,0}\right)^3F_3\left(f_{s,1}/f_{s,0}\right)$
 in the prefactors of Eqs.~(\ref{any3H},\ref{any3T}) does not
reduce to the cube of $f_s (\pi)$ from Eq.~(\ref{back2harm}).

A similar consideration can be extended to higher orders. At
fourth order, we get an additional nonanalytic contribution to
$\chi$ from processes with four
 dynamic particle-hole bubbles, at fifth order--from five dynamic bubbles,
and so on. Each of these contributions can
  be converted into a
skeleton diagram by dressing up fermion Green's functions
and interaction lines, and neither of them is expressed
 solely
  via the
backscattering amplitude. For example, approximating
$f_s(\mathbf{n}_k\cdot\mathbf{n}_p)$ as in  Eq.~(\ref{2harm}), we
obtain for the fourth-order skeleton contribution
\begin{subequations}
\bwt
\begin{eqnarray}
&&\delta\chi_{4s}(T=0,H)=\left(f_{s,0}\right)^4
F_4\left(\frac{f_{s,1}}{f_{s,0}}\right)\left(\frac{{\tilde\mu}_B}{\mu_B}\right)^2
\frac{3|\dzz|}{2{\tilde \epsilon}_F} \chi^{2D}_0\label{any4H}\\
&&\delta\chi_{4s}(H=0,T)=\left(f_{s,0}\right)^4
F_4\left(\frac{f_{s,1}}{f_{s,0}}\right)\left(\frac{{\tilde\mu}_B}{\mu_B}\right)^2
\frac{3T}{4{\tilde \epsilon}_F} \chi^{2D}_0,\label{any4T}
\end{eqnarray}
\ewt
\end{subequations}
 where \bea F_4(x)&=&1-4(3-4\ln 2)x-(50-6\ln 2)x^2
\nn\\
&&-(50-72\ln 2)x^3+\left(20\ln 2-\frac{41}{3}\right)x^4.
\label{F4} \eea

\subsubsection{Isotropic scattering}

If one further neglects $f_{s,1}$ compared to $f_{s,0}$, i.e.,
approximates the scattering amplitude by a constant, the
contributions to the  thermodynamic potential from skeleton
diagrams from all
 orders form geometric series and can be summed up.
  Doing so, we obtain
  \bea &&
  \Xi(T,H) = - \frac{|\dzz|^3}{24\pi (\vf)^2}
~\left(\fs^2 + 2\fs^3+3\fs^4+\ldots\right)\nn\\
&&= -\frac{|\dzz|^3}{24\pi (\vf)^2} \left(\frac{f_{s,0}}{1-
f_{s,0}}\right)^2 =  -\frac{ g^2_{s,0} |\dzz|^3}{24\pi (\vf)^2}
\label{fs0H_a} \eea for $|\dzz| \gg T$, and \bea &&
 \Xi(T,H) = - ~\frac{T \dzz^2}{4\pi (\vf)^2}
 \left(\frac{1}{2}\fs^2 + \frac{2}{3}\fs^3+\frac{3}{4}\fs^4+\ldots\right)\nn\\
&&= -\frac{T \dzz^2}{4\pi (\vf)^2}~
 \left[\ln\left(1 - f_{s,0}\right) + \frac{f_{s,0}}{1- f_{s,0}}\right] \nonumber \\
&& = -\frac{T \dzz^2}{4\pi (\vf)^2}~
 \left[\ln\left(1 + g_{s,0}\right)^{-1} +  g_{s,0}\right]
\label{fs0T_a} \eea for  $|\dzz|\ll T$.

Differentiating $\Xi(T,H)$ with respect to the field, we obtain
\bea &&
 \delta\chi(T=0,H) =\left(\frac{f_{s,0}}{1-
f_{s,0}}\right)^2\left(\frac{{\tilde\mu}_B}{\mu_B}\right)^2\frac{|\dzz|}{2{\tilde
\epsilon}_F} \chi^{2D}_0 \nonumber \\
&& =
g^2_{s,0}\left(\frac{{\tilde\mu}_B}{\mu_B}\right)^2\frac{|\dzz|}{2{\tilde
\epsilon}_F} \chi^{2D}_0 \label{fs0H} \eea and \bea
&&\delta\chi(T,H=0) =
 \left[\ln\left(1 - f_{s,0}\right) + \frac{f_{s,0}}{1- f_{s,0}}\right]\left(\frac{{\tilde\mu}_B}{\mu_B}\right)^2 \frac{T}{{\tilde \epsilon}_F} \chi^{2D}_0 \nonumber \\
&&= \left[\ln\left(1 + g_{s,0}\right)^{-1} + g_{s,0}\right]
\left(\frac{{\tilde\mu}_B}{\mu_B}\right)^2 \frac{T}{{\tilde
\epsilon}_F} \chi^{2D}_0. \label{fs0T} \eea
 Equation
(\ref{fs0T}), without
 FL renormalization of the Fermi energy,
 was derived earlier in Refs.~\onlinecite{we_short,finn,finn_new}.

In what follows, we will also need a full
expression for the spin susceptibility for the case in which the
angular dependence of the scattering amplitude is approximated by
first two harmonics, as in Eq. (\ref{2harm}). Such
 an
 expression
can be obtained for the case when $f_{s,1}\ll f_{s,0}$
 while  $f_{s,0}$ is arbitrary. The calculation of $\delta \chi$
 is tedious but straightforward. We present the result only for
$\delta \chi (T, H=0)$
 \beq \delta\chi(T,H=0) = \left[
F_0 (f_{s,0}) - 2 \frac{f_{s,1}}{f_{s,0}}
F_1 (f_{s,0})\right]
\left(\frac{{\tilde\mu}_B}{\mu_B}\right)^2\frac{T}{{\tilde
\epsilon}_F} \chi^{2D}_0,
 \label{yepp1} \eeq where \bwt\bea
&&
F_0 (x) = \ln\left(1 - x\right) + \frac{x}{1- x} \label{yepp1_1}\nonumber \\
&& F_1 (x) = \left(\frac{x}{1-x}\right)^2 - x^3 \left[\frac{4 \ln
2}{(x+1)^3} - \frac{4 \ln {(1-x)}}{(x+1)^3} - \frac{2}{(x+1)^2
(1-x)} -\frac{2 (x^2 +1)}{(x^2-1)^2}\right]. \label{yepp2} \eea
\ewt In the two limits, $F_1 (x \ll 1) \approx x^2$ and $F_1 (x
\gg 1) \approx 2 x$.

\subsection{The sign of the temperature and magnetic-field
dependences of the spin susceptibility} \label{sec:sign}

In the previous Section, we calculated the third and fourth order
skeleton diagrams for a model form of $f_s(\theta)$ given by
Eq.~(\ref{2harm}). Beyond weak coupling, the expansion in skeleton
diagrams does not have a natural small parameter. Still, it is
worthwhile to analyze the result for not too strong interaction.
To fourth order in the scattering amplitude, the field dependence
of $\delta\chi (T=0,H)$ is given by
 the
sum of Eqs.~(\ref{chi_gen})
[taken
 in the limit of $T\to 0$],
 (\ref{any3H}), and
(\ref{any4H}).
 Explicitly, \beq \chi(T=0,H)=\frac{|\dzz|}{{\tilde
\epsilon}_F}\left(\frac{{\tilde\mu}_B}{\mu_B}\right)^2\chi_0^{2D}S_H\left(f_{s0},f_{s,1}\right),
\label{slopeH} \eeq where \bwt\beq
S_H\left(f_{s0},f_{s,1}\right)=\frac{1}{2}\left(f_{s,0}-f_{s,1}\right)^2+f_{s,0}^3F_{3}\left(\frac{f_{s,1}}{f_{s,0}}\right)
+\frac{3}{2}f_{s,0}^4F_{4}\left(\frac{f_{s,1}}{f_{s,0}}\right)\label{SH}\eeq\ewt
with functions $F_{3}(x)$ and $F_4(x)$ defined in Eqs.~(\ref{F3})
and (\ref{F4}), respectively. The first term in Eq.~(\ref{SH}) is
the square of the backscattering amplitude in the two-harmonic
approximation [cf. Eq.~(\ref{back2harm})].

 Likewise, the $T$ dependence
 of  $\delta\chi (T,H=0)$
 is
given by
 the
 sum of Eqs.~(\ref{chi_gen}) [taken in the limit of
$H\to 0 $], (\ref{any3T}), and (\ref{any4H}): \beq
\chi(T,H=0)=\frac{T}{{\tilde
\epsilon}_F}\left(\frac{{\tilde\mu}_B}{\mu_B}\right)^2\chi_0^{2D}S_T\left(f_{s0},f_{s,1}\right),
\label{slopeT} \eeq where \bwt\beq
S_T\left(f_{s0},f_{s,1}\right)=\frac{1}{2}\left(f_{s,0}-f_{s,1}\right)^2+
\frac{2}{3}f_{s,0}^3F_{3}\left(\frac{f_{s,1}}{f_{s,0}}\right)
+\frac{3}{4}f_{s,0}^4F_{4}\left(\frac{f_{s,1}}{f_{s,0}}\right).\label{ST}\eeq\ewt

 In
Figs.~\ref{fig:fig4} and \ref{fig:fig5}, we plot
$S_H(f_{s,0},f_{s,1})$ and $S_T(f_{s,0},f_{s,1})$,
correspondingly, as functions of $f_{s,1}/f_{s,0}$ for a range of
$f_{s,0}$. We see that if both $|f_{s,0}|$ and
$\left|f_{s,1}/f_{s,0}\right|$ are sufficiently large (but still
less than one), the signs of the slopes are opposite to those of
the backscattering contribution, i.e., $\chi$ decreases with $T$
and $|H|$.

To get an idea about the numerical values of $f_{s,0}$ and
$f_{s,1}$, we use
 available data for
  Landau parameters.
 A system of fermions with repulsive interaction is expected to
exhibit enhanced ferromagnetic fluctuations, which corresponds to
a negative value of $g_{s,0}$. Indeed, the
 Landau parameter $g_{s,0}$ is negative in He$^3$ (in both bulk
\cite{greywall} and film \cite{greywall90} forms), 2D gases in
semiconductor heterostructures, \cite{pudalov_gershenson,stormer,
shayegan} and many other fermion systems.
  In bulk He$3$,
 $g_{s,0}=-0.70$ and $g_{s,1}=-0.55$ at ambient pressure.
\cite{greywall}  In Si MOSFETs, $g_{s,0}$ is also close to $-0.7$
in a wide interval of densities. \cite{pudalov_gershenson} The
first harmonic of the spin Landau function, $g_{s,1}$ has not been
measured in 2D gases.
 Taking the bulk He$^{3}$ values
  as rough estimates for the 2D case as well,
we obtain with the help of Eq.~(\ref{yep}): $f_{s,0}=-2.3$,
$f_{s,1}=-0.76$,
 and $f_{s,1}/f_{s,0}=0.33$.
  Although the magnitude of $f_{s,0}$
is probably too large for our truncated perturbation theory to be
 accurate,
  Figs.~\ref{fig:fig4} and \ref{fig:fig5} indicate that both
$\delta \chi (T,0)$ and $\delta \chi (0,H)$ are already negative
  for this value of $f_{s,1}/f_{s,0}$.

We thus see  that the field and temperature dependences of $\chi$
are non-universal: while the slopes are positive at weak coupling,
they well may become negative at sufficiently strong coupling.
[Later on, however, we will show that in the vicinity of a
ferromagnetic QCP the
 sign of $\delta \chi (T,H)$ is definitely positive.]

\subsubsection{Cooper renormalization}
\label{sec:Cooper}
 References \onlinecite{finn,finn_new}
considered a more subtle mechanism for changing the sign of
$\delta\chi$ as compared to the second-order result, namely,
renormalization of the backscattering amplitude in the Cooper
channel.  As we have already said, the backscattering amplitude is
special in that it is given by
  a
 fully renormalized vertex with zero total
incoming momentum
 and momentum transfer of $2k_F$.
 Therefore, it can be
 expressed
 via angular harmonics of the irreducible Cooper amplitude,
$\gamma_C$, as
 \beq
 f_{s}(\pi)=\sum_l(-)^l\frac{\gamma_{C,l}}{1+\gamma_{C,l}\ln(W/E)},
 \label{fs_harm}\eeq
where $E =\max\{T,\dzz\} $ is an appropriate energy scale. This gives rise
to two effects. First, if at least one of $\gamma_{C,l}$ is
negative, i.e., $\gamma_{C,l_0} <0$, the system undergoes a
superconducting transition of the Kohn-Luttinger type \cite{kl}
into a state
 with orbital momentum $l_0$ at  $
 E_{{\rm
KL}}=W\exp\left(-|\gamma_{C,l_0}|\right)$.
 The backscattering contribution to the
 spin susceptibility $\delta\chi_{2s}\propto
E/\left(1+\gamma_{C,l_0}\ln(W/E)\right)^2$
 diverges at $
 E_{{\rm
KL}}$ as well.
 Above $E_{KL}$, $\delta\chi_{2s}$ is non-monotonic: it decreases
with $E$ for $E_{KL}<E<\mathrm{e}^2E_{KL}\approx 7.39 E_{KL}$ and
increases with $E$ for $E>\mathrm{e}^2E_{KL}$
(Ref.~\onlinecite{finn}). For $E\gg \mathrm{e}^2E_{KL}$, Cooper
renormalization is weak and $\delta\chi_{2s}$ becomes linear in
$E$. Second, if all $\gamma_{c,l}$ are positive,
 $f_s(\pi)$ scales down to zero as $1/\ln(W/E)$ for $E\to 0$.
 Consequently, the backscattering contribution to $\chi$ is reduced by a
 factor of $1/\ln^2(W/E)$.
  In this situation,
the dominant contribution to $\delta \chi$ comes from
non--backscattering terms,\cite{finn_new} which do not contain
 singular
Cooper renormalizations.
 In Ref.~\onlinecite{finn_new}, this effect was accounted for in
 a model of isotropic scattering amplitude,
 $f_{s}(\theta)=f_{s,0}$, by subtracting off
the backscattering contribution from Eqs.~(\ref{fs0H}) and
(\ref{fs0T}). This gives \bwt \bea \delta\chi(T=0, H\to 0)
\rightarrow
\left[\left(\frac{f_{s,0}}{1-f_{s,0}}\right)^2-\fs^2\right]\left(\frac{{\tilde\mu}_B}{\mu_B}\right)^2
\frac{|\dzz|}{2{\tilde \epsilon}_F} \chi^{2D}_0
=\left[2\fs^3+3\fs^4+\ldots\right]\left(\frac{{\tilde\mu}_B}{\mu_B}\right)^2\frac{|\dzz|}{2{\tilde
\epsilon}_F} \chi^{2D}_0 \label{fs0H0} \eea \ewt and \bwt \bea
\delta\chi(T\to 0, H=0)
 \rightarrow
\left[\ln\left(1-f_{s,0}\right)+
\frac{f_{s,0}}{1-f_{s,0}}-\frac{1}{2}\fs^2\right]\left(\frac{{\tilde\mu}_B}
{\mu_B}\right)^2\frac{T}{{\tilde\epsilon}_F}
\chi^{2D}_0=\left[\frac{2}{3}
\fs^3+\frac{3}{4}\fs^4+\ldots\right]\left(\frac{{\tilde\mu}_B}{\mu_B}\right)^2\frac{T}{{\tilde\epsilon}_F} \chi^{2D}_0.\nn\\
\label{fs0T0} \eea \ewt
 The signs of the $H$ and $T$ dependences in
Eqs.~(\ref{fs0H0}) and (\ref{fs0T0}) now coincide with the sign of
$f_{s,0}$;
 for negative $f_{s,0}$, expected for a repulsive
interaction, they are opposite to the second-order result. This
mechanism was proposed in Ref.~\onlinecite{finn_new} as an
explanation of the negative sign of the slope of $\delta \chi (T,
H=0) \propto T$ observed in Ref.~\onlinecite{reznikov}.
\begin{figure}[tbp]
\caption{(color online) Function
$S_{H}\left(f_{s,0},f_{s,1}\right)$, which determines the sign of
the magnetic-field dependence of the spin susceptibility
 calculated
 to fourth
order in the skeleton interaction
 [cf. Eq.~(\ref{SH})],
 plotted as a function of $f_{s,1}/f_{s,0}$ for
$f_{s,0}=-0.3\;\mathrm{(dashed)},-0.5\;\mathrm{(dotted)},-0.7\;
\mathrm{(dashed)}$. The curve for $f_{s,0}=-0.3$ was multiplied by
a factor of 10 for clarity.}
\begin{center}
\epsfxsize=1.0\columnwidth\epsffile{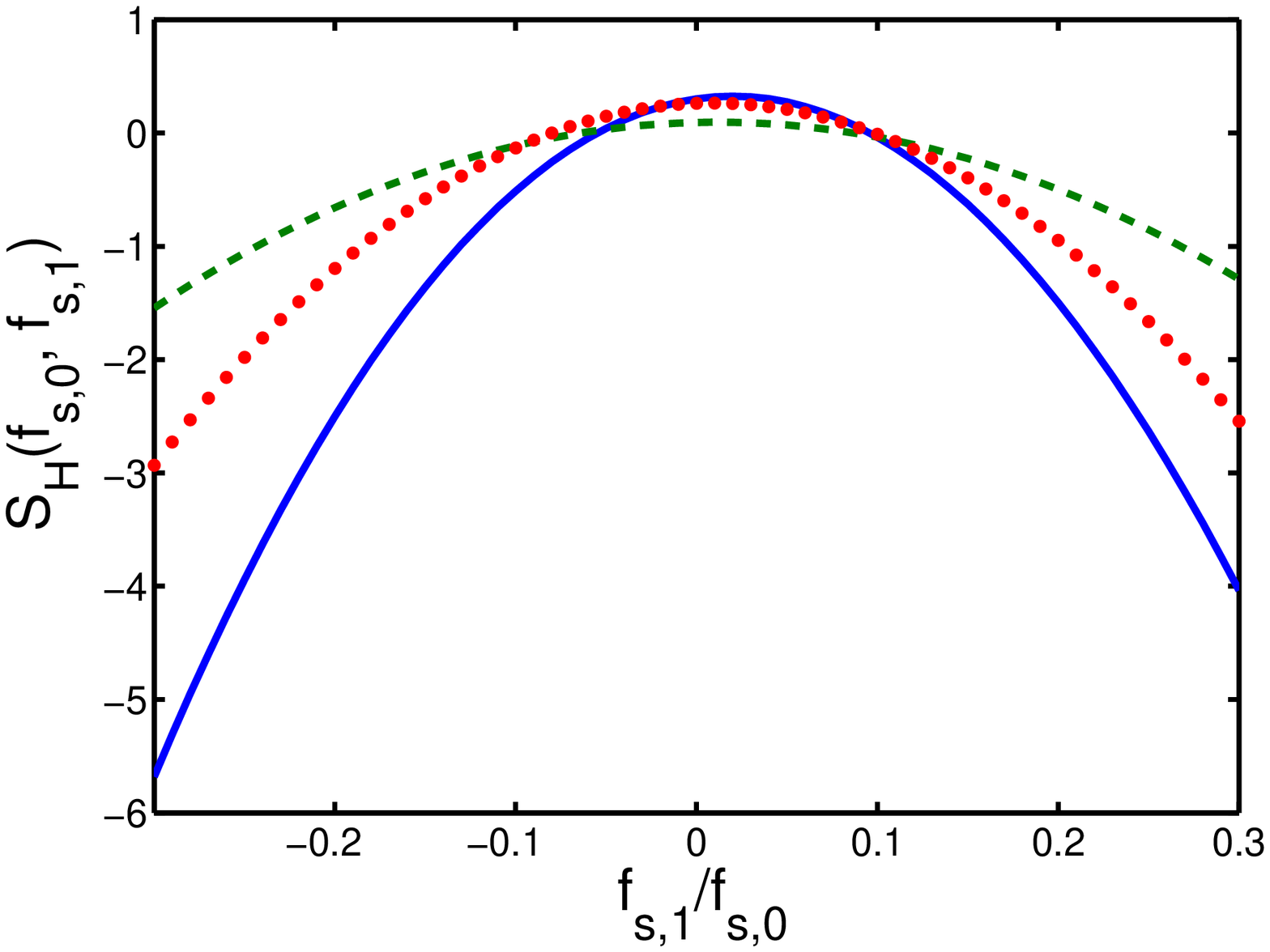}
\end{center}
\label{fig:fig4}
\end{figure}

\begin{figure}[tbp]
\caption{(color online) Function
$S_{T}\left(f_{s,0},f_{s,1}\right)$, which determines the sign of
the magnetic-field dependence of the spin susceptibility
 calculated
 to fourth
order in the skeleton interaction
 [cf. Eq.~(\ref{ST})],
  plotted as a function of
$f_{s,1}/f_{s,0}$ for $f_{s,0}=-0.3\; \mathrm{(dashed)},-0.5\;
\mathrm{(dotted)},-0.7\; \mathrm{(solid)}$. The curve for
$f_{s,0}=-0.3$ was multiplied by a factor of 10 for clarity.}
\begin{center}
\epsfxsize=1.0\columnwidth\epsffile{ 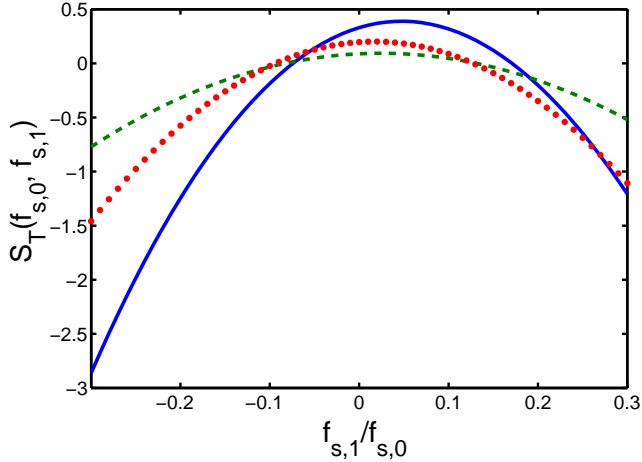}
\end{center}
\label{fig:fig5}
\end{figure}

A simple way to estimate the validity of the approximation used in
Eqs.~(\ref{fs0H0}) and (\ref{fs0T0}) is to consider the scattering
amplitude
 with two rather than one
 components:
 \beq f_s (\phi) = f_{s,0} + f_{s,1} \cos
\phi. \label{yepp3} \eeq The vanishing of $f_s (\pi)$ at $T \to 0$
implies that $f_{s,0} = f_{s,1}$ at $T=0$. In the approximation
used
 to derive
  (\ref{fs0T0}),  this
relation was accounted for
in the quadratic
 but not in higher-order terms. Substituting $f_{s,0} =
f_{s,1}$ into
 the  third and fourth order terms in $f_{s}$, we
 obtain, instead of Eq.~(\ref{fs0T0}),
\bea \delta\chi(T, H=0)&=& \left[(0.02)\times\frac{2}{3} \fs^3
+(-45.65)\times\frac{3}{4}\fs^4+ \ldots\right]
\nn\\
&&\times\left(\frac{{\tilde\mu}_B}{\mu_B}\right)^2\frac{T}{{\tilde\epsilon}_F}
\chi^{2D}_0
 \label{yepp4} \eea
 Comparing (\ref{yepp4}) to (\ref{fs0T0}), we see
that the prefactors differ substantially, making
 it difficult to draw a general conclusion.
  At the same time, the signs of both terms in (\ref{yepp4}) are negative for $f_{s,0} <0$;
  hence, to this order, $\delta\chi(T, H=0) <0$, which is consistent with Ref.~\onlinecite{finn_new}.

\subsubsection{Coulomb interaction in the large N limit}
\label{sec:coulomb}
 Another issue is that the results of
 Refs.~\onlinecite{finn,finn_new},
   as well as Eq. (\ref{yepp4}),
 are valid only below a
characteristic energy scale at which Cooper renormalizations of $f_{s,0}$
become significant. For a weak interaction, this scale is
exponentially small. To estimate this scale beyond the
weak-coupling regime, we consider
 the effect of
 Cooper
renormalization of the backscattering amplitude
  on the spin susceptibility
 in
  a large $N$ model for
  the
Coulomb interaction, developed
 earlier in Refs.~\onlinecite{takada,iordanskii,suhas_dm}. To be
specific, we assume that there are $N_v$ degenerate electron
valleys, so that the total (spin$\times$valley) degeneracy is
$N=2N_v$. While this model is especially relevant to Si- and
AlAs-based heterostructures, which have at least two valleys
($N=4$), it can also provide a useful insight even for
single-valley system ($N=2$).

 In an $N$-fold degenerate 2D Fermi gas, the Fermi
momentum is scaled down by a factor of $\sqrt{N}$:
$k_F=n/\sqrt{4\pi N}$, where $n$ is the number density of
electrons. On the other hand, the inverse screening radius
$\kappa$, which is proportional to the density of states, is
scaled up by a factor of $N$. Their ratio, \beq
g_N=\kappa/k_F=r_sN^{3/2}/2\label{gc}\eeq with $r_s=me^2/\sqrt{\pi
n}$, defines an effective coupling constant. We still need to
assume that $r_s\ll 1$; only then the mean-field, random-phase
approximation (RPA) is valid. If $N\sim 1$, then $g_N\sim r_s\ll
1$, which implies that there is only a weak-coupling regime. If
$N\gg 1$, there are two regimes: weak coupling ($g_N\ll 1$) and
strong coupling ($g_N\gg 1$); the latter is of the most interest
for us.

The details of the calculation are given in Appendix
\ref{sec:largeN}.
 Here
 we present only
the result
 for the field-dependent part of the spin susceptibility
 \beq\delta\chi
=\frac{|\Delta|}{8\epsilon_F}\left(\frac{N^2}{L_{C}^2}-\frac{2}{N}\right)
\chi^{2D}_0, \label{175a}\end{equation} where $L_C
=\ln\left(\epsilon_F/|\dz|\right)$ is the Cooper logarithm and, as
before, $\Delta = 2 \mu_B H$. Equation (\ref{175a}) is valid for
$L_C\gg 1$, i.e., for $\dz\ll E_C\equiv
\epsilon_F\exp\left(-N\right)$. The first term in Eq.~(\ref{175a})
is the backscattering contribution, which is the leading term in
the $1/N$ expansion. The second term is the contribution from
other processes, which is the next-to-leading term in this
expansion. As expected, the backscattering contribution is scaled
down by a factor of $1/L_C^2$. The change of sign occurs at
$L_C=N^{3/2}/\sqrt{2}$ or \beq\Delta=E^*=
\epsilon_F\exp\left(-N^{3/2}/\sqrt{2}\right)\ll E_C.\eeq

For an estimate,  we consider a 2D electron gas in the $
(001)$ plane of a Si MOSFET, where $N=4$; correspondingly,
$E_C=0.018\epsilon_F$ and $E^*=0.0035\epsilon_F$. In the
experiment
 of Ref.~\onlinecite{reznikov}, the
 highest Fermi energy in this measurement is about
$40$K. Then, $E_C= 0.70$ K and $E^*=0.14$ K. Both
 energies
 are smaller than the disorder broadening in these samples.
This shows
 that the mechanism of the sign reversal of $
\delta\chi$ due to Cooper renormalization does not have room to
develop until disorder becomes important.
 $\delta \chi$ in Ref.~ \onlinecite{reznikov}.

Notice also that $E_C$ is still
  larger than the energy scale
of the Kohn-Luttinger superconducting instability. Indeed, in 2D
the Kohn-Luttinger effect starts only at third order in the
interaction, \cite{chubukov_kl} which implies that $E_{KL}\sim
\exp(-N^3)\epsilon_F\ll E_C$.

The above estimates  are based on additional
assumptions, such as $N\gg 1$, and therefore cannot give rigorous
results regarding the $T$
and $H$ dependences measured in 2D heterostructures. The
 crucial experimental check for the many-body nature of
these effects is the $T/H$ scaling of the susceptibility which, to
the best of our knowledge, has not been performed in detail.

We should also point out that while the $T$-dependence of $\chi$
in Si MOSFET was obtained in a thermodynamic measurement (via the
magnetocapacitance), \cite{reznikov} the $H$-dependence of $\chi$
is all heterostructures \cite{pudalov_gershenson,stormer,shayegan}
was extracted from Shubnikov-de Haas oscillations. While it is
known that Shubnikov-de Haas oscillations contain a renormalized
spin susceptibility of a FL, $\chi(0,0)$, it remains to be
verified that the field-dependence of the susceptibility can also
be extracted from such a measurement. \cite{loss_comment}

\subsection{Thermodynamic potential of a 2D Fermi liquid as a function of magnetization}
Preparing the ground
 for the analysis of a ferromagnetic QCP in
Sec.~\ref{sec:qcp}, it is
 convenient to obtain the thermodynamic potential in terms of
magnetization rather than of magnetic field.
 In this formulation, the susceptibility is defined as \beq
\chi^{-1}=\frac{1}{\mu_B^2}\frac{\partial^2 \Xi}{\partial M^2},
\label{chu1} \eeq where
 $M=n_{\up}-n_{\down}$ and $n_{\up/\down}$ is the number density
of spin-up/down fermions.

In the RPA, which neglects
  FL
renormalizations ($Z = \vf/v_F =1$), the recipe for finding the
free energy was given
 in Ref.~\onlinecite{bealmonod68}. For a
 local
 interaction $U=u/\nu$, \beq
\Xi(\Delta,M,T)=-\frac{\nu\Delta^2}{4}+\frac{M\Delta}{2}-\frac{uM^2}{4\nu}+\delta\Xi,\label{bm}\eeq
 where $\delta
\Xi(T,\Delta)$ is the sum of the RPA
(ladder)
series \beq \delta \Xi =
T \sum_{q}\left( \ln\left[1
 +(u/\nu)\Pi_{\uparrow
\downarrow}\right]
 -(u/\nu)\Pi_{\up\down}\right)
. \label{n_5a} \eeq (The second term in Eq.~(\ref{n_5a})
compensates for the first-order contribution not present in the
ladder series.) The relation between $M$ and $\Delta$ is found
from the condition $\left(\partial \Xi/\partial
\Delta\right)\big|_{T,M}=0$, which gives $M=\nu \dz$. Neglecting
the RPA term, one obtains the Stoner-like spin susceptibility
$\chi(0,0) = 2\mu^2_B\nu/(1-u)$, which is consistent with
(\ref{n_8}) for $Z = \vf/v_F =1$ and $f_{0,s} = -u/(1-u)$
 (or, equivalently, $g_{s,0} = -u$).
Evaluating further $\chi
(T,\Delta)$ with the RPA term
 included,
   and using the relation
between $\Delta$ and $M$, one reproduces Eqs.~(\ref{fs0H}) and
(\ref{fs0T}) without FL renormalizations.

\subsubsection{ Fermi-liquid renormalizations}
\label{sec:fl_ren}
 Equation~(\ref{bm})  can be generalized to a FL.
 In this Section, we assume that the system is
 away from the immediate vicinity of a QCP,
 and the effective interaction
can be considered as static.
 We discuss specific conditions below.
  For a static interaction
   the
 relevant fermion  self-energy
   depends on $k$ but
not on $\omega_m$: \beq
  \Sigma^{{\mathrm F}}(\omega_m,k) \approx
  - \left(\vf-v_F\right)(k-k_F).
\eeq In this case, the quasiparticle residue
  $Z=\left(1-i\partial\Sigma/\partial\omega_m\right)^{-1}$
 is equal to unity and the
fermion Green's function is given by
 \beq
G_{\up,\down}(\omega_m,k)=\frac{1}{i\omega_m-\vf(k-k_F)\pm\dzz/2},
\label{gff}\eeq where, as before, ${\tilde \Delta} = 2 {\tilde
\mu}_B H = 2 \mu_B H/(1 + g_{s,0})$. We will see later in
  Sec.~\ref{sec:qcp} that
 the self-energy becomes  predominantly
   $\omega_m$- but not $k$-
   dependent in the immediate vicinity of a QCP.

Since our primary interest is the spin susceptibility at small momenta, we
  focus on the Pomeranchuk instability towards
a ferromagnetic state.
 In the  FL theory,
  this
  instability occurs when $g_{s,0}$
approaches $-1$.
  All other
 partial components of
 the spin and charge scattering amplitudes are assumed to remain finite at criticality and, without loss of generality,
 can be taken to be small.

 The primary goal of the present Section is
 to demonstrate that
  some terms in the thermodynamic potential of Eq.~(\ref{bm}) are renormalized
   at energies comparable to the bandwidth, where
   $\Sigma$ is
   static, while others are renormalized
   at much smaller energies,
 of  order $\max \{T, M/(v_F \nu/\vf)\}$,
 where $\Sigma^F$
 is dynamic.

Consider first the $\Delta^2$ term in Eq.~(\ref{bm}), which
  is
 the thermodynamic potential of free fermions.
 For a FL, this term can be
 calculated using the renormalized Green's function (\ref{gff}).
Applying the Luttinger-Ward formula \cite{LW} for the
thermodynamic potential of free
 fermions and expanding $\Xi$ to order $\dzz^2$, we obtain at $T=0$
 \bea
\Xi^{(0)}&=&-\sum_{\sigma=\up,\down}\int\frac{d\omega_m}{2\pi}\nu\int^{W}_{-W}
v_Fd(k-k_F)\ln\left(-G^{-1}_{\sigma}\right)\nonumber\\
&&=\frac{\nu\dzz^2}{4}\int\frac{d\omega_m}{2\pi}\int^{W}_{-W}
\frac{v_F
d(k-k_F)}{\left(\vf(k-k_F)-i\omega_m\right)^2}.\label{dm_1}\eea
  The integrals in Eq.~(\ref{dm_1}) are controlled by large momenta and
energies, of order
   $\omega_m\sim
v_F|k-k_F|\sim W $.
 Integrating first over $k$ and then over $\omega_m$, we
 find
  that
 \beq
\Xi^{(0)} =-\frac{v_F}{\vf}\frac{\nu\dzz^2}{4}.
\label{t_1}\eeq

To evaluate the second term in (\ref{bm}), we
need the
  relation between the magnetization and Zeeman energy,
 which
  can be found by expressing the number densities of spin-up and
spin-down fermions in terms of the Green's functions \bea
M &=& n_{\up}-n_{\down}\nonumber\\
&=&\nu\int\frac{d\omega_m}{2\pi}\int^{W}_{-W} v_F
d(k-k_F)\left[G_{\up}-G_{\down}\right]. \label{mfl}\eea Using the
Green's functions from Eq.~(\ref{gff}), we again find that the
integrals in Eq.~(\ref{mfl}) are controlled by  energies of order
$W$. Performing the integrations, we obtain \beq M=\nu
\frac{v_F}{\vf} \dzz. \label{yep_1} \eeq
 Using (\ref{t_1}) and
  recalling that the relation (\ref{yep_1}) must follow from
 the condition
$\partial \Xi
/\partial {\tilde \Delta} =0$, we find
 that the second term in (\ref{bm}) retains its form, but $\Delta$ changes to
${\tilde \Delta}$.

In a similar way, the Hubbard $U$ in
 the Hartree term in Eq.~(\ref{bm}) is replaced by the Landau parameter
 \beq U = \frac{u}{\nu} \rightarrow - \frac{\vf}{v_F} ~\frac{g_{s,0}}{\nu}, \eeq so that the Hartree term
becomes \beq \frac{\vf g_{s,0}}{4v_F\nu}~ M^2. \eeq

 As a result, we have \beq \Xi (T, M,{\tilde \Delta}) =
-\frac{v_F}{4\vf} \nu {\tilde \Delta}^2 + \frac{1}{2}M {\tilde
\Delta}  + \frac{\vf g_{s,0}}{4v_F\nu} M^2 + \delta \Xi (T,{\tilde
\Delta}). \label{n_4} \eeq
 So far,
  all renormalizations in Eq.~(\ref{n_4}) are from energies of order
$W$.

Consider next the RPA term, $\delta \Xi (T,\dzz)$.
Re-calculating it with $G(k,\omega_m)$ from
Eq.~(\ref{gff}), we find that
 it retains
 the same form as in Eq. (\ref{n_5a}),
 except
  that $U$ is replaced  again by the Landau parameter
 and the polarization bubble
$\Pi_{\uparrow \downarrow}$ contains the renormalized Fermi
velocity:
 \beq \delta \Xi (T,{\tilde
\Delta}) = T\sum_q
\left( \ln\left[1
 -\frac{g_{s,0} \vf}{v_F \nu}~ \Pi_{\uparrow \downarrow}
  \right]+\frac{g_{s,0} \vf}{v_F \nu}\Pi_{\up\down}\right). \label{n_5} \eeq

There are both analytic and nonanalytic terms in $\delta \Xi$.
 The analytic, ${\tilde \Delta}^2$ contribution
comes from energies $O(W)$ and can be
 absorbed into the $M^2$ term in Eq.~(\ref{n_4}), once
 the prefactor is expressed in terms of the Landau parameter
$g_{s,0}$ rather than of the bare interaction.  The leading
nonanalytic term has the same $|{\tilde \Delta}|^3$ form, as
 in the perturbation theory,
  but the prefactor now contains the
 FL parameters
$g_{s,0}$
 and $v_F/\vf$. At $T=0$,
 \beq \delta \Xi (0,{\tilde
\Delta}) \rightarrow -
\frac{ |{\tilde \Delta}|^3}{24 \pi {\vf}^2}
 \left(\frac{f_{s,0}}{1- f_{s,0}}\right)^2.
\label{yep_2} \eeq
The key point
 is that the ${\tilde \Delta}^3$
term in Eq.~(\ref{yep_2}) comes from
 small energies: $\omega_m\sim \vf (k-k_F)\sim \dzz\ll W$.
 Therefore,
 $\vf$ in Eq.~(\ref{yep_2})  is the Fermi velocity on
  the small-energy scale.

The equilibrium condition  $\left(\partial \Xi/\partial
\Delta\right)\big|_{T,M}=0$ now gives $ M = {\tilde \Delta} \nu (
v_F/ \vf) + O({\tilde \Delta}^2)$, consistent with
Eq.~(\ref{yep_1}), and the spin susceptibility
 at $T = M=0$
is obtained from
Eq.~(\ref{chu1})
 \beq \chi
(0,0) = \mu^2_B \frac{m}{\pi} ~ \frac{v_F}{\vf}~\frac{1}{1 +
g_{s,0}} =
 \mu^2_B \frac{m}{\pi}~ \frac{v_F}{\vf}~ \left(1-f_{s,0}\right).
\label{n_7}\eeq As
 expected,  this result coincides with
the general expression for the renormalized spin susceptibility in
a FL (Ref.~\onlinecite{agd}).

Using Eqs.~(\ref{n_4}) and (\ref{n_5}), we can now
 construct
 an
 expansion of the
 thermodynamic potential in powers of magnetization. To order $|M|^3$ and at $T=0$, we obtain
\begin{widetext}
\bea \Xi(T=0,M) &=& \frac{1}{4 \nu (1 -f_{s,0})}
\frac{\vf}{v_F}M^2   -
\frac{1}{24 \pi \nu^3 v^2_F} \frac{\vf}{v_F}
 \left(\frac{f_{s,0}}{1- f_{s,0}}\right)^2|M|^3 + b M^4 + ... \nonumber \\
&&=  \frac{1}{4 \nu} \frac{\vf}{v_F} (1 +g_{s,0})M^2 - \frac{
g^2_{s,0}}{24 \pi\nu^3 v^2_F} \frac{\vf}{v_F}|M|^3
 + b M^4 + ...
\label{n_10}
\eea
\end{widetext}
 For completeness, we added a regular $M^4$ term to $\Xi$.

Evaluating $\chi (T=0,M)$
 with the help of Eq.~(\ref{chu1}) and
 using the relation between ${\tilde M}$ and equilibrium
${\tilde \Delta}$, we reproduce the linear in $H$ term in the spin
susceptibility, Eq. (\ref{fs0H}).

We emphasize again  that the $M^2$ and $|M|^3$ terms in
Eq.~(\ref{n_10}) come from different energy scales. The $M^2$ term
comes from high energies of order $W$, and $\vf$
 in this term is the renormalized Fermi velocity at energies of
order $W$.
  The $|M|^3$ term
 comes from fermions with energies of order $M/(v_F\nu/\vf)={\tilde \Delta}$,
 and $\vf$ in the $|M|^3$ term
 is the Fermi velocity on that scale.

At finite
 temperature, the expression for the thermodynamic potential becomes more involved. We present
only the result and show the details of the derivation later, in
Sec.~(\ref{sec:away}), where we
 compute $\Xi (T,M)$ in the spin-fermion model.
To logarithmic accuracy, we obtain
\begin{widetext}
\bea \Xi (T,M) &=& \frac{M^2}{4 \nu} \frac{\vf}{v_F} \left[(1
-f_{s,0})^{-1} -
 \left(\frac{f_{s,0}}{1- f_{s,0}} + \ln(1 - f_{s,0}) \right)\frac{\vf}{v_F}\frac{T}{\epsilon_F}\right]
-\left(\frac{f_{s,0}}{1- f_{s,0}}\right)^2 \frac{M^4}{576 \pi T
\nu^4 v^2_F} \left(\frac{\vf}{v_F}\right)^2 +
  b M^4 + ... \nonumber \\
&&=\frac{M^2}{4 \nu} \frac{\vf}{v_F}  \left[1 +g _{s,0} -
 \left(g_{s,0} + \ln(1+ g_{s,0})^{-1}\right)\frac{\vf}{v_F}\frac{T}{\epsilon_F}\right] -
 \frac{g^2_{s,0} M^4}{576 \pi T \nu^4 v^2_F} \left(\frac{\vf}{v_F}\right)^2 +
 b M^4 +...
\label{n_10a} \eea
\end{widetext}
The key result here is that for $T\gg M/(v_F\nu/\vf)$ the
expansion of $\Xi$ in powers of $M$ becomes analytic: the $|M|^3$
term is replaced by an
 analytic
$M^4/T$ term.
 Simultaneously, the prefactor for the  quadratic term acquires a linear in $T$
 correction,
 which is
 just a nonanalytic temperature dependence of the
 spin
 susceptibility.
   Indeed, evaluating  $\chi (T,0)$,   from (\ref{n_10a}), we reproduce Eq.~(\ref{fs0T}).

\section{Magnetic response near a 2D ferromagnetic quantum critical point}
\label{sec:qcp}

\subsection{General considerations}\label{sec:qcp_gen}
We now consider
 the immediate vicinity of a ferromagnetic QCP. In what follows,
 we first assume that a continuous ferromagnetic
transition
 does
 exists and obtain the thermodynamic potential $\Xi (M,T)$
along a continuous second-order transition line
by
 extending Eqs.~(\ref{n_10}) and (\ref{n_10a}) to energies below
  the scale where
 the self-energy crosses from static to dynamic forms.
 (\onlinecite{chub_cross,cgy}).
 Next, we show that this line becomes unstable at low
enough temperatures because of
  nonanalyticities
 which survive even in the
vicinity of the QCP. We will argue that the
 instability may occur in two ways: i)
 the second-order phase transition into a uniform ferromagnetic
phase becomes first order or ii)
 the transition occurs via an intermediate magnetic phase
with a spiral magnetic order.
 More specific predictions are possible within more
specific models. One of such models is a model with
 a large radius
 of the interaction in the spin channel.~\cite{dzero} We will show
that in this model the first-order instability occurs before the
spiral one.

 A tendency towards the first-order transition can be seen
already from Eq. (\ref{n_10}). Indeed, the cubic term in $M$ in
$\Xi (M, T=0)$ is negative, which implies that a state with {\it
finite} magnetization is energetically favorable. Close to the
 critical
 point,
 the $M^2$ term is small
and the thermodynamic potential (\ref{n_10}) is negative over some
range of $M$.
 This means that the first-order phase transition
 preempts the second-order
one  at $T=0$.
 Indeed, for the thermodynamic potential of the form
 \beq
 \Xi(M)=a M^2-c |M|^3+b M^4\label{tp_gen}
 \eeq
 with $c>0$,
 the magnetization jumps to
    finite value of
 $M_{0}=2a/c$ already
 for $a =c^2/4b
 >0$, i.e., before the second-order transition
  takes place.

At
 finite $T$, the $|M|^3$ term in the thermodynamic potential
 crosses over into
   a $-M^4/T$ one, which is still negative. If $T$ is
low
 enough, this negative term
  is larger than  the regular, $M^4$
term, and the transition remains first order until the $-M^4/T$
term becomes smaller than the regular $bM^4$ term. At higher $T$,
the transition becomes second-order.

This analysis is, however, incomplete
 because it is based on the
 result
  for $\Xi (T.M)$
 derived under assumptions that the quasiparticle residue $Z=1$ and
the effective Fermi velocity $\vf = k_F/m^*$ is finite.
  As we have
already mentioned, this is true only if
 the self-energy is static.  Since the first-order jump in
magnetization, $M_0$, is proportional to the critical parameter
$1+g_{s,0}$, the corresponding energy scale $M_0/\nu$ is also
small and
 falls into the regime where
 the self-energy is dynamic
 and
 Eq.~(\ref{n_10}) is no longer valid.

If the self-energy is dynamic,
 the $Z$ factor and $\vf/v_F$ are
both given by
    $\left(1-i\partial \Sigma (\omega_m)/\partial \omega_m\right)^{-1}$
(so that the product $Z v_F/\vf$ remains intact.)
 This would not lead to substantial changes if $Z$ remained finite at a QCP.
However, it is well established by now that
  $\partial \Sigma (\omega_m)/\partial \omega_m$ diverges at a ferromagnetic QCP in 2D; hence,
  both $Z$ and $\vf/v_F$ vanish. \cite{aim,2/3,chub_cross,rosch_rmp,pepin_prl,pepin_prb}
 One then might be
tempted to conclude that
 the nonanalytic
 term in Eq.~(\ref{n_10}) vanishes,
 as it
 is proportional
  as an overall factor
 to the renormalized velocity $\vf$ evaluated at
  low energies. We will show, however, that
 the nonanalytic term in
the free energy  survives even at the QCP, albeit in a weaker form
($|M|^{3}$ is replaced by $|M|^{7/2}$).

Before proceeding further, we mention two
 paradoxes with the
vanishing of $Z$ and $\vf/v_F$ at
 a ferromagnetic QCP. First, there
seems to be a contradiction with the Stoner criterion which says
 that a ferromagnetic transition occurs
 at some critical, {\it
finite} interaction strength. If we formally use the FL relation
$g_{s,0} = Z (Z v_F/\vf) \Gamma^\Omega$ (Ref. \onlinecite{agd})
 with $Z \rightarrow 0$ but  $Z v_F/\vf = \text {const}$, we find
that the condition $g_{s,0} =-1$
 can be satisfied only
 if $\Gamma^\Omega\to \infty$.
 Second, in
 the  FL theory,
 the velocity renormalization
  is determined by the $l=1$ harmonic
 of the Landau function in the {\it charge} sector: $v/\vf = {\tilde m}/m = 1 +
g_{c,1}$. Hence, the vanishing of $\vf$ implies that $g_{c,1} =
\infty$. Meanwhile,  the very idea of a Pomeranchuk instability is
that it occurs only in one particular channel, e.g., in the spin
channel with the angular momentum $l=0$ for a ferromagnetic QCP.
 All other channels, including the charge channel with $l=1$,
 remain uncritical,  which
seems to be inconsistent with the condition $g_{c,1} = \infty$.

We make a few
 general remarks about these two paradoxes first and show
specific results later.
\begin{itemize}
\item
The assumption of
the conventional FL theory about a single relevant $l=0$ spin
channel near a
 ferromagnetic QCP is valid
  if
 there is a wide range of energies below the
 cutoff
$W$,
 where the fermion self-energy  is static. Within this
range,
 $Z =1$ and $ v_F/\vf$
differs from
 its  bare value only because of a
 non-singular interaction in the $l=1$ channel.
Then $g_{s,0} = Z^2 (v_F/\vf) \nu \Gamma^\Omega$ is of the same
order as $\nu\Gamma^\Omega$, and a critical value of $g_{s,0}=-1$
is approached already at
 finite interaction strength.
 As we have already demonstrated,
 the Stoner
 enhancement of the spin susceptibility
 comes from fermions with energies of order
 $W$,  hence, at energies below $W$, the susceptibility
 is already enhanced by the Stoner factor
  $(1+g_{s,0})^{-1} > 1$.

 \item
 At some energy scale, $\Lambda < W$,
the self-energy undergoes a crossover between static and dynamic
forms.
  Accordingly,
 $Z$ and  $v_F/\vf$
begin to vary below $\Lambda$ and eventually flow to zero at the QCP.
In this regime, the conventional FL theory based on the static
approximation
 is no longer valid and
 has to be replaced by a
  ``new''
   low-energy
FL theory, in which the ``bare'' fermions are the ones on the scale of
$\Lambda$, the ``bare'' interaction between fermions is in the spin channel,
 and the interaction potential $U$ is replaced by the effective
 interaction,
 which scales as $1/(1+g_{s,0})$.
This low-energy FL theory is a spin-fermion model. The Landau
parameters for
 the
 low-energy FL differ from
those of the conventional FL.
 In particular,
  all
 harmonics
 $g_{c,l}$, including $g_{c,1} = v_F/\vf -1$
diverge at a 2D QCP.~\cite{we_new}

\item The spin-fermion model is valid only if
 the crossover between static and dynamic forms of the self-energy
 occurs on a scale  much smaller than $W$. Otherwise, one cannot
 consider only the $l=0$ spin channel. As we will see,
 the condition  $\Lambda \ll W$ can be
 satisfied if the
 interaction is
 sufficiently long-range or, else, if the model is extended
to $N\gg 1$ fermion flavors. We will assume below that at least
one of these two conditions is satisfied.
\end{itemize}

 Because both $Z$ and $v_F/\vf$ on the scale of $\Lambda$
 are
just  constants,
 we will absorb $Z$ into the effective spin interaction, and
measure the velocity renormalization below the cutoff
 with respect to its value at the cutoff.  In other words,
   we
assume that the bare fermion propagator is $G^{-1} (k, \omega_m)
= i\omega_m - v_F (k-k_F)$, and
  use symbols $Z$ and $\vf$
 to describe renormalizations at energies smaller than $\Lambda$.

\subsection{Spin-fermion model in zero magnetic field}
\label{h=0} We now consider in
 detail the
 low-energy  effective theory
   near a ferromagnetic QCP: the spin-fermion model  We first
 review  briefly
 the properties of this model
 in zero magnetic field \cite{pepin_prb}
  and then show how the model
 is modified in the presence of the field.

The spin-fermion model
 includes low-energy fermions with a bare
propagator $G (k, \omega_m)$,
 collective
  spin excitations
 ${\mathbf S}_{{\mathbf q}}$, whose
  bare propagator is  the static spin susceptibility
 $\chi (q)$, and the spin-fermion interaction, described by the
Hamiltonian \beq H_{int} = \frac{g}
 {{\cal N}}
\sum_{\mathbf{k},\mathbf{q}, \alpha, \alpha'}
c^{\dagger}_{\mathbf{k}, \alpha}
 {\vec\sigma}_{\alpha, \alpha'}\cdot
\mathbf{S}_{\mathbf{q}}c_{\mathbf{k}+\mathbf{q} , \alpha'},
\label{n_11} \eeq where ${\cal N}$ is the number of lattice sites.
 The spin-fermion coupling $g$ is related to the Landau parameter
$g_{s,0}$
  as $g = (\pi/m) \sqrt{-g_{s,0}}$.
 Near the QCP, $g_{s,0}\approx -1$ and
   $g\approx
  \pi/m$. The  bare boson propagator
$\chi (q)$ is proportional to
 $1/(1 +g_{s,0})$
 for $q\to 0$.
 We assume that the $q$ dependence of $\chi$ at small but finite
  $q$ is described by the standard Ornstein-Zernike formula
  \beq \chi
 (q) = \frac{m}{\pi} ~\frac{1}{1 + g_{s,0} + (aq)^2} = \frac{m}{\pi a^2}~\frac{1}{\xi^{-2} + q^2}
\label{n_12}, \eeq where \beq \xi^{-2} = (1 + g_{s,0}) a^{-2}.
\eeq
  Similar to the $1+g_{s,0}$ term, the analytic $q^2$ term in
 $\chi (q)$  comes from  fermions with energies
 comparable to  $W$.
 This term can be obtained in the RPA scheme, but
  one has to assume either that the dispersion is different from a free-fermion one,
  i.e., from $k^2/2m$, or that the exchange
  interaction is momentum-dependent; otherwise the
particle-hole polarization bubble does not depend on $q$ for
$q\leq 2k_F$ in 2D.
  If the $q^2$ term is
 comes from
 the momentum dependence of the interaction,
   the length $a$ is
 the radius of the interaction.
 In this case, the RPA is justified for a sufficiently
long-ranged interaction, i.e., for $a k_F \gg 1$
(Ref.~\onlinecite{dzero}.)

The spin-fermion interaction affects both fermion and boson propagators.
  Collective spin excitations acquire a
   self-energy
 $\Sigma^{{\mathrm B}} (q, \Omega)$,
while fermions acquire a self-energy $\Sigma^{{\rm F}}(k, \omega_m)$ which
gives rise to renormalizations of $Z$ and of the Fermi
velocity: \bea
&&G (k, \omega_m) = \frac{1}{i\omega_m  - v_F (k-k_F)+ \Sigma^{{\mathrm F}} (k, \omega_m)} \nonumber \\
&& \chi(q, \Omega_m) = \frac{m}{\pi} ~\frac{1}{1 + g_{s,0} +
(aq)^2 +
 \Sigma^{{\mathrm B}}
 (q, \Omega_m)}.
\label{n_14}
\eea

To one-loop order, the
 $T=0$ self-energies
  behave  as
 \bwt
 \bea
&&\Sigma^{{\mathrm B}} = {\tilde g}^2 \frac{\Omega_m}{v_F q} \nonumber \\
&&\Sigma^{{\mathrm F}}(k =k_F, \omega_m) = \left\{
\begin{array}{lcl}
i\lambda ~\omega_m, & \mbox{~for~} & \omega_m \ll \omega_0/\lambda^3, \nonumber \\
i\omega_0^{1/3} \omega_m^{2/3}, & \mbox{~for~} &
  \omega_0/\lambda^3\ll \omega_m \ll
 \omega_{\max} \sim \omega^{1/4}_0 \ \epsilon^{3/4} _F
\end{array}
\right. \nonumber \\
&& \Sigma^{{\mathrm F}}(k, \omega_m =0)
  \sim
 v_F(k-k_F) \left(\frac{\omega_0}{\varepsilon_F}\right)^{1/4}
\times {\cal S}\left(\frac{\lambda}{\tilde g}\right) \label{n_15}
\eea where \beq \omega_0 =   \varepsilon_F \frac{3\sqrt{3}}{4}
\left(\frac{{\tilde g}}{ak_F}\right)^4,\;~~ \lambda = \frac{3
{\tilde g}^2}{4 a k_F}~\frac{1}{\sqrt{1 + g_{s,0}}}, ~~
 {\tilde g} = 2 g \nu \approx \frac{m g}{\pi}, \label{n_23}
\eeq \ewt
 and ${\cal S}(x \ll 1) \sim x, ~~{\cal S}(x \gg 1) = O(1)$. To
simplify the formulas, we assume that $\omega_m>0$.
 Eq.~(\ref{n_15}) is valid
 for $\omega_0/\lambda^3\ll\omega_{\max}$, i.e., for $\delta \ll 1$.

In the RPA, a ferromagnetic transition occurs
 at
 ${\tilde g} =1$, but
  the critical value of ${\tilde g}$
 may differ from one in a more general
 model.
  For $ak_F \gg 1$, $\omega_0$ is parametrically small compared to
$\epsilon_F$, i.e., the
 $k-$dependent part of the self-energy is always smaller than
 $v_F(k-k_F)$.

To compare
the frequency- and momentum-dependences of the self-energy,
 we
 consider the Green's function near the renormalized mass shell:
  $v_F(k-k_F)=i\omega_m+\Sigma_{{\mathrm F}}(k,\omega_m)$.
 For $\omega_m \ll \omega_0$,
 $i\omega_m\ll \Sigma_{{\mathrm F}}$ and
 $
v_F|k-k_F| \sim \omega^{2/3} \omega_0^{1/3}$
 near the mass shell.
 Consequently,
 the $k$-dependent part of the self-energy is smaller than the
 $\omega_m$-dependent part.
  For $\omega_m \gg \omega_0$, $\Sigma^{{\mathrm F}}
  (\omega_m) \ll i\omega_m$ and
  $\epsilon_k
  v_F|k-k_F| \sim \omega_m$
    near the mass shell.
  Comparing
     $\Sigma^{{\mathrm F}} (k, \omega_m =0)$  and
 $\Sigma^{{\mathrm F}} (k =k_F, \omega_m)$, we find that the two
become comparable at $\omega_m \sim \omega_{\max} =
\omega_0\left(\epsilon_F/\omega_0\right)^{3/4}$.
 For $\omega_m \ll \omega_{\max}$, $\Sigma^{{\mathrm F}} (k,
\omega_m)$ depends predominantly on $\omega_m$ while for $\omega_m
\gg \omega_{\max}$ it depends predominantly on $k$. Note that
$\omega_{\max}$ is also the upper cutoff for $\omega_m^{2/3}$
scaling of the fermion self-energy (see Eq.~(\ref{n_15})]. At
larger $\omega_m$, $\Sigma^F$ scales as $\ln \omega_m$.

The scale $\omega_{\max}$ is larger than $\omega_0$ but still
parametrically
 smaller than $\epsilon_F$ [indeed, $\omega_{\max} \sim \omega_0 (ak_F)^3 \sim \epsilon_F/(ak_F)$].
  The upper limit for the low-energy theory, $\Lambda$,
 can then be set somewhere in
 between $\omega_{\max}$ and $W \sim \epsilon_F$;
 its precise location being irrelevant
  as long as $\omega_{\max} \ll \epsilon_F$.

In what follows, we will also need the fermion self-energy at
finite temperatures.  At finite $T$, the fermions interact both
with classical ($\Omega_m=0$) and quantum ($\Omega_m\neq 0$) spin
fluctuations. The quantum contribution to the self-energy is a
scaling function of $\omega_m/T$:
\beq\Sigma^{\mathrm{F}}_{\mathrm{Q}}(\omega_m,T)=i\omega_0^{1/3}\omega_m^{2/3}Q(\omega_m/T),\label{quant}\eeq
where the scaling function $Q(x)$ is such that $Q(x\gg 1)=1$ and
$Q(x\ll 1)\sim 1/x^{2/3}$. The classical contribution contains a
static propagator that diverges at QCP as $q^{-2}$.
 This divergence can be regularized in two ways: by accounting for
a thermally generated mass of spin fluctuations due to mode-mode
coupling (not present in the spin-fermion model)
\cite{millis_qcp,delanna} or by resumming the self-consistent Born
series.\cite{abanov_ch} In the first approach, the
zero-temperature correlation length in Eq.~(\ref{n_12})
$\xi(0)=a/\sqrt{1+g_{s,0}}$ is replaced by  $\xi(T)\sim
a/\left[\left(T/\Lambda\right)\ln\left(\Lambda/T\right)\right]^{1/2}$
(Ref.~\onlinecite{millis_qcp}). The classical part of the
self-energy $\Sigma^{\mathrm{F}}_{\mathrm{C}}\equiv
\Sigma^{\mathrm{F}}\left(\omega_m=\pi T,T\right)$ then becomes
 \beq \Sigma^{\mathrm{F}}_{\mathrm{C}}=i\frac{3}{4}{\tilde
 g}T\frac{\xi(T)}{k_Fa^2}\propto i\left(\frac{T}{\ln\left(\Lambda/T\right)}\right)^{1/2}\label{class}.\eeq
In the second approach, one obtains
  a self-consistent equation for $\Sigma$
   which yields  a similar $T$
dependence of
 $\Sigma^{\mathrm{F}}_T$.

The quantum and classical contributions become comparable at a
characteristic temperature \beq T_{\mathrm{QC}}={\tilde g}^6
\left(ak_F\right)^2\frac{\Lambda^3}{\epsilon_F^2}.\label{TQC}\eeq
For $T\ll T_{\mathrm{QC}}$ the classical contribution dominates
over the quantum one, and vice versa.

\subsubsection{Eliashberg theory}

We now focus on the low-energy region $\omega_m\ll \omega_{\max}$,
where $\Sigma^{{\mathrm F}} (k, \omega_m)$
 is
 predominantly dynamic.
  Within this region,
 there exists another scale, $\omega_0  \sim \omega_{\max}/(ak_F)^3
\ll \omega_{\max}$, at which
 the fermion self-energy
 becomes comparable to $\omega_m$. Below $\omega_0$,
$\Sigma^{{\mathrm F}} (\omega_m) > \omega_m$, i.e., the system
 is
 in a
 strong-coupling regime.
 Close enough to the QCP, i.e., for
 $\lambda\gg 1$,
 the strong-coupling regime,
 on
 its
  turn, is divided into
two more
   subregimes:
  i) $\omega_0/\lambda^3 < \omega_m < \omega_0$, where the
self-energy has a non-FL, $\omega_m^{2/3}$, form and ii)
 $\omega_m <
\omega_0/\lambda^3$, where
 the FL behavior is restored, i.e.,
 $\mathrm {Re} \Sigma^{{\mathrm F}}
(\omega) \propto \omega$ and $\mathrm {Im} \Sigma^{{\mathrm F}}
(\omega) \propto \omega^2$.

Since $\Sigma^{{\mathrm F}} (\omega_m) > \omega_m$
at $\omega_m <
\omega_0$, the
 accuracy of the one-loop approximation
 for
  the self-energy becomes
  an issue.
Previous work \cite{aim,pepin_prb}
  demonstrated that
 the
 self-consistent
 one-loop approximation (the Eliashberg theory) cannot be
 controlled just by a large value of the parameter
  $ak_F$, as higher-order diagrams
 in the strong-coupling regime
  are of the same order in $1/ak_F$ as the one-loop diagram.
  To put the theory
  firmly
   under control, one needs to
 extend it formally to $N\gg 1$ fermion flavors;
   then higher order terms in the self-energy terms are
   small
   in $(\ln N)/N^2$.
  In what follows,
  we neglect this subtlety and assume that
  the
  Eliashberg theory is valid.

The frequency-dependent self-energy $\Sigma^{{\mathrm F}}
(\omega_m)$ from Eq.~(\ref{n_15}) leads to the renormalization of
the $Z$-factor (equal to the inverse velocity renormalization
factor). Right at the QCP, both $Z$ and $v_F/\vf$ depend on
$\omega_m$ as
\beq Z = \frac{\vf}{v_F}=\left(1
  -i \frac{
\partial\Sigma^F(\omega_m)}{
\partial \omega_m}\right)^{-1}
  = \left( 1 +
\left(\frac{\omega_0}{\omega_m}\right)^{1/3}\right)^{-1}.
\label{n_18b} \eeq

The boson self-energy is generated by inserting the dynamic
 fermion  bubbles,
 made out
  full propagators, into the bare spin-fermion interaction.
 Summing up the RPA series for the
renormalized spin-fermion interaction, we obtain
\begin{subequations} \bea
&&\Sigma^{{\mathrm B}} (q, \Omega_m) = {\tilde g}^2~{\tilde P} (q, \Omega_m), \label{n_12b} \\
&&{\tilde P}(q, \Omega_m)  = \frac{|\Omega_m|}
 {\left[\left(\Omega_m -i c_\Omega \Sigma^{{\mathrm F}} (\Omega_m)\right)^2 +
  v^2_F q^2\right]^{1/2}}, \label{n_12a}
\eea\end{subequations} where $c_\Omega$ is a slowly varying
function of $\Omega_m$, which interpolates
 between
 two limits:
  $c_0 =1$ for $\Omega_m\ll \omega_0/\lambda^3$
and $c_\Omega \approx 1.2$ for $\omega_0/\lambda^3 \ll \Omega_m
\ll \omega_0$ (Refs. \onlinecite{pepin_prb,we_short,bec}).
 For free fermions, ${\tilde P}$ is the same as $P$ introduced in
Eq.~(\ref{def_P}).

In the limit of small frequencies, ${\tilde P} (q, \Omega_m)$
 reduces to
 the
 Landau damping form $\Omega_m/v_F q$. The static boson self-energy is
 small in $1/(ak_F)$ and non-singular, and we neglect it.

Equation (\ref{n_12a}) has to be
 treated
  with caution
 because
   ${\tilde P}$ is the dynamic part of
the particle-hole bubble
 made of
  dressed fermions but without vertex
corrections. The latter are irrelevant for $v_F q \gg \Omega$ but
are important for $v_F q \ll \Omega$, as they are
 necessary for the Ward identities to be satisfied.
  This problem is generic to all
  models
in which the effective interaction is peaked at zero momentum
transfer. \cite{chub_ward} Fortunately,
 this complication does not arise in
 the study of non-analyticities
 in the thermodynamic potential and in the
spin susceptibility because, as it will be shown later, we will
only need to know ${\tilde P} (q, \Omega_m)$ for $v_F q \gg
\Omega_m, \Sigma (\Omega_m)$. Therefore, we will be using Eq.
 (\ref{n_12a})
 in what follows.

The thermodynamic potential of the spin-fermion model in the
Eliashberg approximation
 was obtained in Ref.~\onlinecite{chmgg_prb} (see also Sec.~\ref{field}):
 \begin{equation}
\Xi \left(T \right) =\Xi_{{\rm F}}\left(T\right) +\frac{3}{2}%
T \sum_{\Omega_m} \int \frac{d^2q}{4\pi^2}~ \ln\left[
\chi^{-1}\left(q, \Omega_m\right)\right], \label{n_16}
\end{equation}
 where $\Xi_{{\rm F}}\left(T\right) =- \nu T \sum_{\omega
_{m}}|\omega _{m}| =-\pi^2 \nu T^2/3$
 is the
 $T$-dependent part of the thermodynamic potential of a free Fermi
gas and $\chi (q, \Omega_m)$ is given by Eq.~(\ref{n_14}).
 Differentiating $\Xi (T)$ with respect to temperature, one obtains
the specific heat
 $C(T)$, which
  behaves as  $(1 + \lambda)T$ away from the QCP and as
  $T^{2/3}$    at the QCP.

\subsection{Spin-fermion model in
 a magnetic field} \label{field}

We now return to our main discussion and consider the spin-fermion
model in the presence of a magnetic field. First, we derive a
general expression for the thermodynamic potential in a magnetic
field, and then analyze
 the structure of the nonanalytic terms in the vicinity of a
ferromagnetic QCP.
For reasons already explained in Sec.\ref{sec:qcp_gen},
 we take the bare fermion propagator as
\beq G^{0}_{\up,\down}(k, \omega_m) = \frac{1}{i\omega_m - v_F
(k-k_F) \pm {\tilde \Delta}/2}, \label{su_3} \eeq where  ${\tilde
\Delta} = 2{\tilde \mu}_B H$, and ${\tilde \mu}_B = \mu_B/(1 +
g_{s,0})$.\\

\subsubsection{The Luttinger-Ward functional in a magnetic field}

We first  derive the  thermodynamic potential for the spin-fermion
model in finite magnetic field,
 starting from the Luttinger-Ward functional
  \cite{LW} and
 making use of
  the Eliashberg approximation, which neglects vertex
corrections.~\cite{eliashberg,bardeen,prange,chubukov_eliash,chmgg_prb}

The Luttinger-Ward functional\cite{LW} contains
 four terms
\begin{equation}
\Xi (M, {\tilde \Delta}, T) =\Xi_{{\rm F}}+\Xi_{{\rm B}}+\Xi_{{\rm
FB}} + \Xi_M, \label{lw_1}\end{equation} where $\Xi_{{\rm F}}$ is
the potential
 of fermions dressed by
 the interaction with bosons, $\Xi_{{\rm B}}$ is
 the  potential of bosons dressed by  the interaction with fermions,
$\Xi_{{\rm FB}}$ is
the skeleton part which   describes explicitly
 the
 fermion-boson interaction
 at low-energies, and $\Xi_M = g_{s,0} M^2/(4\nu) + (1/2) M {\tilde \Delta}$
 is an extra $M$-dependent high-energy contribution, same as in
 (\ref{n_4}).
 (There is no double counting, as one can verify explicitly.)
For $H=0$, the  Eliashberg form of the Luttinger-Ward
 functional was derived in
 Refs.~\onlinecite{eliashberg,bardeen,chubukov_eliash,chmgg_prb}.
Extending the derivation to the  case of finite spin polarization,
 we obtain
\begin{eqnarray}
\Xi_{{\rm F}} &=&-
 T\sum_{k,\sigma} \left[ \ln \left\{ -G_{\sigma }^{-1}\left(
k\right)/W \right\} -
 \Sigma^{{\mathrm F}}_{\sigma }\left( k\right)
G_{\sigma }\left( k\right) \right]\nonumber \\
\Xi_{{\rm B}}
&=&
  \frac{T}{4}\sum_{q,\sigma,\sigma'}
\left( 1+\delta _{\sigma ,\sigma ^{\prime }}\right) \left[ \ln
\left\{D_{\sigma ,\sigma ^{\prime }}^{-1}\left( q\right)\right\} -
\Sigma^{{\mathrm B}}_{\sigma \sigma ^{\prime }}\left(
q\right) D_{\sigma \sigma ^{\prime }}\left( q\right) \right]\nonumber \\
\Xi_{{\rm FB}} &=& \frac{{\tilde g}^2}{4\nu}T^2 \sum_{kk^{\prime
},\sigma,\sigma'}\left( 1+\delta _{\sigma ,\sigma ^{\prime
}}\right) G_{\sigma }\left( k\right) D_{\sigma \sigma ^{\prime
}}\left( k-k^{\prime }\right) G_{\sigma ^{\prime }}\left(
k^{\prime }\right),
  \label{lw}
\end{eqnarray}
where ${\tilde g}$ is the dimensionless coupling constant defined
in Eq.~ (\ref{n_23}), $k\equiv \left(\mathbf{k}, \omega
_{m}\right) ,q\equiv \left(\mathbf{q}, \Omega _{m}\right)$,
$\sigma,\sigma'=\up,\down$, and
 as before,
 summation over $k$ and $q$ implies
summation over Matsubara frequencies and integration over momenta.
The functions
 $G_{\up,\down}$ are the exact fermion Green's function
\begin{equation}
G _{\uparrow ,\downarrow }=\left(i\omega _{m}+\Sigma^{{\mathrm
F}}_{\uparrow ,\downarrow }\left( \omega_m\right) -\epsilon
_{k}\pm {\tilde \Delta} /2\right)^{-1} \label{su_4}
\end{equation}
and $D_{\sigma \sigma ^{\prime
}}\left( q\right) $ is the propagator of spin fluctuations
\begin{equation}
D^{-1}_{\sigma, \sigma ^{\prime }}\left( q\right) = 1 + g_{s,0} +
(aq)^2 + \Sigma^{{\mathrm B}}_{\sigma \sigma ^{\prime
}}\left(q\right), \label{su_2}
\end{equation}
 where $\Sigma^{{\mathrm F}}_{\sigma}$ and  $\Sigma^{{\mathrm B}}_{\sigma\sigma'}$ are
the fermion and boson self-energies, correspondingly.
 Notice that the fermion self-energy does not contain a constant part evaluated at $\omega_m=0$ and $k=k_F$--
this part has been absorbed into the renormalized Zeeman energy,
$\dzz$.

By construction, the Luttinger-Ward functional is stationary with
respect to variations of the fermion and boson self-energies.
 The stationarity conditions
\begin{equation}
\frac{\delta \Xi }{\delta \Sigma _{\sigma
}^{\mathrm{F}}}=\frac{\delta \Xi }{\delta \Sigma^{{\mathrm
B}}_{\sigma,\sigma'}}
 =0
 \label{n_26}
\end{equation}
 yield \bwt\begin{subequations}
\begin{eqnarray}
\Sigma^{{\mathrm B}}_{\sigma\sigma'} \left( q\right) &=&
\frac{{\tilde g}^2}{\nu} T\sum_{k}G_ {\sigma}
 \left( k+q\right) G_
{\sigma'} \left( k\right)
,  \label{boson} \\
\Sigma^{{\mathrm F}} _{\sigma }\left( k\right) &=&- \frac{{\tilde
g}^2}{\nu} T\sum_{q}D_{\sigma,\sigma} \left( q\right) G_{\sigma
}\left( k+q\right) - \frac{{\tilde g}^2}{4\nu} T\sum_{q}\left[D
_{\sigma,-\sigma}(q)+D_{-\sigma,\sigma}(q)\right]
 G_{-\sigma }\left( k+q\right).
\label{fermion}
\end{eqnarray}
\end{subequations}
\ewt
 In the presence of a magnetic field,
  there are two different  boson self-energies: $\Sigma^{{\mathrm
B}}_{\up\up}=\Sigma^{{\mathrm B}}_{\down\down}$ and
$\Sigma^{{\mathrm B}}_{\up\down}=\Sigma^{{\mathrm B}}_{\down\up}$.
 $\Sigma^{{\mathrm B}}_{\up\up}$ is composed of fermions of the
same spin ($\Sigma^{{\mathrm B}}_{\up\up}=\Sigma^{{\mathrm
B}}_{\down\down}$). It depends on the magnetization only via a
shift of the chemical potential. This is a regular, analytic
dependence which does not lead to nonanalyticities in the
thermodynamic potential. Neglecting this dependence, we set
$\Sigma^{{\mathrm B}}_{\up\up}=\Sigma^{{\mathrm
B}}_{\down\down}\equiv\Sigma^{{\mathrm B}}$, where
$\Sigma^{{\mathrm B}}$ is the boson self-energy in zero field,
given by Eq.~(\ref{n_12b}).
 On the other hand, $ \Sigma^{{\mathrm B}}_{\up\down}$ is composed
of fermions with opposite spins
  and depends strongly
on the magnetization via the Zeeman term:
 $\Sigma^{{\mathrm
B}}_{\up\down}=\Sigma_{\down\up}^B = {\tilde g}^2{\tilde
P}_{\up\down}$,
 where
\begin{equation}
{\tilde P}_{\uparrow \downarrow} \left(q, \Omega _{m}\right) =
\frac{\left| \Omega _{m}\right| }{\sqrt{\left(\Omega _{m} - i
c_\Omega \Sigma^{{\mathrm F}} (\Omega_m) -i{\tilde \Delta}\right)
^{2}+v_{F}^{2}q^{2}}}. \label{su_5}
\end{equation}
 This also implies that
$D_{\up,\up}(q)=D_{\down,\down}(q)\equiv D(q)$ and
$D_{\up,\down}(q)=D_{\down,\up}(q)$. Expressions for
$\Xi_{\mathrm{B}}$ and $\Xi_{\mathrm{FB}}$ can be then simplified
to
 \bwt
\begin{eqnarray}
\Xi_{{\rm B}} &=&
 T\sum_{q}\left[
\ln\left\{ D^{-1}\left( q\right)\right\} -\Sigma^{{\mathrm
B}}\left( q\right) D\left( q\right) \right]
+\frac{T}{2}\sum_{q}\left[ \ln \left\{D_{\uparrow \downarrow
}^{-1}\left( q\right)\right\} -\Sigma^{{\mathrm B}}_{\uparrow
\downarrow }\left( q\right) D_{\uparrow \downarrow }\left(
q\right) \right]
\\
\Xi_{{\rm FB}} &=&
 \frac{{\tilde g}^2}{2\nu}T^2\sum_{kk^{\prime }}\left[ \sum_{\sigma
}D\left( k-k^{\prime }\right) G_{\sigma }\left( k\right) G_{\sigma
}\left( k^{\prime }\right) +D_{\uparrow \downarrow }\left(
k-k^{\prime }\right) G_{\uparrow }\left( k\right) G_{\downarrow
}\left( k^{\prime }\right) \right],
 \end{eqnarray}
\ewt
 while the fermion self-energy changes to \bwt \beq \Sigma^{{\mathrm F}}
_{\sigma }\left( k\right) =- \frac{{\tilde g}^2}{\nu}
T\sum_{q}D\left( q\right) G_{\sigma }\left( k+q\right) -
\frac{{\tilde g}^2}{2\nu} T\sum_{q}D _{\uparrow\downarrow}(q)
 G_{-\sigma }\left( k+q\right).
\label{fermion_a} \eeq \ewt

The field-dependent part of the thermodynamic potential involves only
$D_{\uparrow \downarrow }\left( q\right)$ \bwt
\begin{eqnarray}
 \Xi
   &=&-T\sum_{k,\sigma}
 \left[ \ln \left\{ -G_{\sigma }^{-1}\left(
k\right)/W \right\} -
 \Sigma^{{\mathrm F}}
_{\sigma }\left( k\right) G_{\sigma }\left( k\right) \right] +\frac{T}{2}%
\sum_{q}\left[ \ln D_{_{\uparrow \downarrow }}^{-1}\left( q\right)
-\Sigma^{{\mathrm B}}_{\uparrow \downarrow }\left( q\right)
D_{\uparrow \downarrow }\left(
q\right) \right]  \label{eliash_delta} \\
 &&+\frac{{\tilde g}^{2}}{2\nu}T^2\sum_{kk^{\prime }}D_{\uparrow
\downarrow }\left( k-k^{\prime }\right) G_{\uparrow }\left(
k\right) G_{\downarrow }\left( k^{\prime }\right)
+g_{s,0}\frac{M^2}{4\nu} + \frac{1}{2} M {\tilde \Delta}.  \notag
\end{eqnarray}
\ewt With the help of Eq.~(\ref{boson}), we simplify
Eq.~(\ref{eliash_delta}) to
\bwt
\begin{equation}
\Xi
  =
+g_{s,0}\frac{M^2}{4\nu} + \frac{1}{2} M {\tilde \Delta}
 -T\sum_{k,\sigma}
 \left[ \ln \left\{ -G_{\sigma }^{-1}\left(
k\right)/W \right\} - \Sigma^{{\mathrm F}}
_{\sigma }\left( k\right) G_{\sigma }\left( k\right) \right] +\frac{T}{2}%
\sum_{q}\ln D_{_{\uparrow \downarrow }}^{-1}\left( q\right) .
\label{eliash_delta_2}
\end{equation}
\ewt
The first two terms in Eq.(\ref{eliash_delta_2}) need to be expanded to order $\tilde{\Delta}%
^{2}.$
  In the first, logarithmic term, we proceed in
the same way as in Eq.~(\ref{dm_1}).  Keeping only the
 ${\tilde \Delta}^2$  term, we obtain \bwt
\begin{eqnarray}
-T\sum_{k,\sigma }\ln \left( -G_{\sigma }^{-1}\left( k\right)
/W\right) = \frac{\nu \tilde{\Delta}^{2}}{4}\int \frac{d\omega_m
}{2\pi }\int_{-\Lambda }^{\Lambda }\frac{v_{F}d\left(
k-k_{F}\right) }{\left( v_{F}\left( k-k_{F}\right)
-i\tilde{\omega}_m\right) ^{2}} =-\frac{\nu
\tilde{\Delta}^{2}}{2}\int \frac{d\omega_m }{2\pi }\frac{\Lambda
}{\tilde{\omega}_m^{2}+\Lambda ^{2}},
\end{eqnarray}
\ewt where $\tilde{\omega}_m=\omega_m -i\Sigma^{{\mathrm F}}\left(
\omega_m \right) .$ The integral is controlled by
  frequencies $O\left(\Lambda\right)$,
   where the self-energy is
small. Neglecting the self-energy, we arrive at the
free--fermion-like result
\begin{equation}
-T\sum_{k,\sigma }\ln \left( -G_{\sigma }^{-1}\left( k\right) /W\right) =-%
\frac{\nu \tilde{\Delta}^{2}}{4}.
\end{equation}
Expanding the second, $\Sigma^{{\mathrm F}}G$ term
 in Eq.~(\ref{eliash_delta_2}) to order $\dzz^2$ and integrating
over $k-k_{F},$ we obtain
\begin{equation}
T\sum_{k,\sigma }\Sigma^{{\mathrm F}}G_{\sigma }=\frac{\nu
\tilde{\Delta}^{2}}{2}\int
\frac{d\omega_m }{2\pi }\Sigma^{{\mathrm F}}\left( \omega_m \right) \frac{i\tilde{\omega}_m%
\Lambda }{\left( \tilde{\omega}_m^{2}+\Lambda ^{2}\right) ^{2}}.
\end{equation}
The frequency integral is of order $\Sigma^{{\mathrm F}}\left(
\omega_m \sim \Lambda \right) /\Lambda ,$ which vanishes in the
limit $\Lambda \rightarrow \infty . $

We see, therefore, that $\Xi $ reduces to the sum of the
free-fermion--like contribution (up to a renormalization of the
Zeeman energy), the Hartree interaction, and the $\ln
D_{\up\down}$ term from the boson part $\Xi_{\rm B}$,
\begin{eqnarray}
\Xi &=& -\frac{\nu\dzz^2}{4}+\frac{1}{2}M\dzz+g_{s,0}\frac{M^2}{4\nu}+ \Xi_{\log} \nonumber\\
&& \Xi_{\log} = T\sum_{q}\ln D^{-1}_{\uparrow \downarrow } \left(
q\right). \label{su_1}
\end{eqnarray}
 Comparing this expression with (\ref{n_4}) and (\ref{n_5}), we see
that the fermionic self-energy $\Sigma^F (\omega)$ does not affect
${\tilde \Delta}^2$, $M {\tilde \Delta}$ and $M^2$ terms  in
$\Xi$. This agrees with our earlier result that FL
renormalizations of these three terms come from from energies of
order $W$,
  where $\Sigma^F = \Sigma^F (k)$. However, the low-energy
 $\Sigma^F (\omega)$ is present in the last, $\ln D^{-1}_{\uparrow \downarrow }$ term in (\ref{su_1}),
  which
 gives a nonanalytic contribution to
 to $\Xi$.

Minimizing Eq.~(\ref{su_1}) with respect to $\dzz$, and
differentiating with respect to $M$,
 we obtain the spin
susceptibility as a function of $T$ and $H$.
 Note in passing that the
   approach
 based on differentiation of the thermodynamic potential is completely equivalent to the diagrammatic
  evaluation of
   the linear
   susceptibility $\chi (T,H=0)$, used in earlier work,~\cite{chm03,pepin_prl,pepin_prb}
    and also generates diagrams for the nonlinear susceptibility $\chi(T,H)$.
 We illustrate this point in Appendix \ref{sec:equiv}.\\

\subsection{Nonanalytic terms in the thermodynamic potential}

In this Section, we use Eq.~(\ref{su_1}) to derive the nonanalytic
terms in the thermodynamic potential  in the vicinity of the critical point.
  We will see how the
nonanalytic terms change in the non-FL regime.\\

\subsubsection{Away from criticality}
\label{sec:away} First, we discuss the FL regime, where
$\Sigma^{{\mathrm F}}_{\sigma}=i\omega_m\lambda$. The logarithmic
term in Eq.(\ref{su_1}) reads \bwt
\begin{eqnarray}
\Xi_{\log} =T\sum_{\Omega _{m}}\int \frac{dqq}{2\pi }\ln
\left(\delta+(a q)^2 + \frac{{\tilde g}^2 \left| \Omega
_{m}\right| }{\sqrt{\left( {\tilde \Omega}-i{\tilde \Delta}\right)
^{2}+v_{F}^{2}q^{2}}}\right), \label{eliash}
\end{eqnarray}
\ewt where ${\tilde \Omega}=\Omega _{m} (1 + \lambda)$ and
$\delta=1+g_{s,0}>0$.

Expanding the integrand of Eq.(\ref{eliash}) in $1/q$ and
evaluating the $q-$integral
 to logarithmic accuracy, we find
that the $(aq)^2$ term under the logarithm can be neglected, so
that Eq.~(\ref{eliash}) reduces to
\begin{equation}
\Xi_{\log} = -\frac{1}{2\pi v^2_F }T\sum_{\Omega _{m}}\left(
{\tilde \Omega}-i\dzz\right) ^{2}\ln \frac{\delta \Omega_{0}+
{\tilde g}^2 \left| \Omega _{m}\right| }{\Omega_{0}},
\end{equation}
where $\Omega_{0}=\sqrt{\left({\tilde \Omega}-i{\tilde
\Delta}\right) ^{2}}$.
 Converting the Matsubara sum into a contour integral, we obtain
\begin{equation*}
\Xi_{\log} =\frac{1}{8\pi ^{2} v^2_F}\int d\Omega \coth \left( \frac{\Omega }{%
2T}\right) \left( \text{Im}f_{1}^{R}-\text{Im}f_{2}^{R}\right) ,
\end{equation*}
where $f_{1}^{R}$ and $f_{2}^{R}$ are  retarded functions of
frequency, obtained via analytic continuation of the Matsubara
functions
\begin{eqnarray}
f_{1} &=&\left( {\tilde\Omega}-i{\tilde \Delta}\right) ^{2}\ln \sqrt{\left( {\tilde\Omega}%
-i{\tilde \Delta}\right) ^{2}} \nonumber\\
f_{2} &=&\left( {\tilde\Omega}-i{\tilde \Delta}\right) ^{2}\ln
\left[\delta\sqrt{\left( {\tilde\Omega}-i{\tilde \Delta}\right)
^{2}}+{\tilde g}^2 \left| \Omega _{m}\right| \right].
\end{eqnarray}
Performing the analytic continuation, we find
\begin{eqnarray}
\text{Im}f_{1}^{R} &=&\frac{\pi }{2}\left( {\tilde \Omega} +{\tilde \Delta}\right) ^{2}%
\text{sgn}\left( {\tilde \Omega} +{\tilde \Delta}\right)\nonumber \\
\text{Im}f_{2}^{R} &=&\frac{\pi }{2}\left({\tilde \Omega} +{\tilde \Delta}\right) ^{2}%
\text{sgn}\left[\delta\left({\tilde \Omega} +{\tilde \Delta}\right) +{\tilde g}^2 \Omega %
\right].
\end{eqnarray}
 Assembling the two parts,
 we obtain \bwt
\begin{equation}
\Xi_{\log} =\frac{1}{16\pi v^2_F}\int d\Omega \coth \left(
\frac{\Omega }{2T}\right) \left({\tilde\Omega} +{\tilde \Delta}\right) ^{2}\left[ \text{%
sgn}\left({\tilde \Omega} +{\tilde \Delta}\right)
-\text{sgn}\left[\delta\left({\tilde \Omega} +{\tilde
\Delta}\right) +{\tilde g}^2 \Omega \right] \right].  \label{Xi1}
\end{equation} \ewt In what follows, we choose $\dzz>0$ without
a loss of generality. Because of the sign-functions, the integral
in Eq.~(\ref{Xi1}) is confined to the interval
 $1/\lambda_1 > |{\tilde \Omega}| > \dzz$, where $
\lambda_1=1+{\tilde g}^2/(\lambda\delta)$. Since $\lambda\propto
1/\sqrt{\delta}$, we have $\lambda_1\gg 1$ close enough to the
QCP.

The thermodynamic potential $\Xi_{\log} (T, {\bar \Delta})$ at
finite ${\tilde \Delta}$ and $T$ is then given by
\begin{equation}
\Xi_{\log} \left( T,{\tilde \Delta}\right) =-\frac{\left| {\tilde
\Delta}\right| ^{3}}{8\pi \lambda v_{F}^{2}}\int_{\lambda
_{1}^{-1}}^{1}dx\coth \left( x\frac{|{\tilde \Delta}|
}{2\lambda T}\right) \left( x-1\right) ^{2}.  \label{Xi2}
\end{equation}
This equation parameterizes $\Xi_{\log} (T, {\tilde \Delta})$ as a
 scaling function of ${\tilde \Delta}/2\lambda T.$  At low temperatures, $%
T\ll {\tilde \Delta}/2\lambda ,$ one
 can replace
  $\coth(x)$ by $1+2\exp (-2x)$ and extends the lower
limit of the integral to zero. This gives
\begin{equation}
\Xi_{\log} \left( T\ll{\tilde \Delta}/\lambda\right)
=-\frac{|{\tilde \Delta}|
 ^{3}}{24\pi \vf^2\lambda }+O\left( \exp \left( -\frac{| {\tilde \Delta}| }{\lambda T}%
\right) \right).  \label{Xilow}
\end{equation}
In the opposite limit of high temperatures, $T\gg {\tilde \Delta}/2\lambda$, one expands $%
\coth x$ in series as $\coth x=x^{-1}+x/3-x^{3}/45+\ldots $. The
integrand is now logarithmically divergent, and the lower limit of
the integral is relevant. Performing elementary integration, we
obtain \bwt
\begin{equation}
\Xi_{\log} \left( T\gg{\tilde \Delta}/\lambda\right) =-\frac{\ln \lambda _{1}%
}{4\pi v_{F}^{2}}{\tilde \Delta}^{2}T-\frac{1}{576\pi }\frac{{\tilde \Delta}^{4}}{%
v_{F}^{2}\lambda ^{2}T}+\frac{1}{172800\pi }\frac{{\tilde \Delta}^{6}}{%
v_{F}^{2}\lambda ^{4}T^{3}}+\ldots  \label{Xihigh}
\end{equation}
\ewt

Next, we obtain the thermodynamic potential $\Xi$ as a function of
magnetization. Substituting (\ref{Xilow}) and (\ref{Xihigh}) into
(\ref{su_1}) and expressing $\dzz$ in terms of $M$
 with the help of the relation $\partial \Xi (M, \dzz, T)/\partial
\dzz =0$, we obtain
\begin{subequations}
\begin{widetext}
\bea &&\Xi (T\ll M/\nu\lambda) = \frac{M^2}{4 \nu} (1 + g_{s,0}) -
\frac{
 |M|^3}{24 \pi \nu^3 v^2_F}~\frac{1}{\lambda}
 + b M^4 + ... \label{mo_3} \\
 &&\Xi (T\gg M/\nu\lambda) = \frac{M^2}{4 \nu} \left[(1 +g _{s,0})
-
 \frac{
 T |\ln{(1 + g_{s,0})}|}
 {\epsilon_F}
 \right] -
 \frac{
  M^4}{576 \pi T \nu^4 v^2_F}~\frac{1}{\lambda^2} +
 b M^4 + ...
\label{mo_4}
\eea
\end{widetext}
\end{subequations}
The $M^2$ term in Eq.~(\ref{mo_4}) determines the temperature
dependence of the
  susceptibility
\begin{equation}
  \chi^{-1} \left( T\right) =\frac{1}{\mu^2_B} ~{\frac{\partial
^{2}\Xi }{\partial {M}^{2}}}= \frac{1}{2 \mu_B^2 \nu} \left(1 +
g_{s,0} -
 \frac{
  T |\ln{(1 + g_{s,0})} |}
{\epsilon_F}
   \right).
\label{mo_5}
\end{equation}
   We can now compare Eqs. (\ref{mo_3}) and
(\ref{mo_4}) with Eqs. (\ref{n_10}) and (\ref{n_10a}),
 keeping in mind that one should set $\vf = v_F$ in regular terms
in Eqs.~(\ref{n_10},\ref{n_10a}) as we measure the velocity
renormalization with respect to its value at the upper cutoff for
the spin-fermion model.
We see that
 Eq. ~(\ref{mo_3}) differs from Eq. ~(\ref{n_10}) by a factor of
$1/\lambda$ in the $|M|^3$ term, while Eq. ~(\ref{mo_4}) differs
from Eq. ~(\ref{n_10a}) by a factor of $1/\lambda^2$ in the $M^4$
term.
 The factors
 are
 precisely $\vf/v_F$ and $(\vf/v_F)^2$, respectively, in the spin-fermion model.
This confirms our
  assertion that the nonanalytic $M^3$ term and its finite $T$ equivalent $M^4/T$ are renormalized by fermions with the energies of order $\dzz= M/\nu$.

Equations (\ref{mo_3},\ref{mo_4}) are valid in the FL
 regime,
  where  $\max\{T,M/\nu\lambda\}\ll \omega_0/\lambda^3$. At higher
energies, the non-FL renormalization of the effective mass affects
the functional form of the nonanalytic terms. This regime is
considered in the next Section.

\subsubsection{At criticality}
\label{sec:at_qcp}
The main difference between the non-FL--
 and  FL regimes
is the form of the self-energy,
  which enters
the propagator of spin fluctuations [see Eq.~(\ref{su_5}]. Also,
 for energies
 in
  between $\omega_0/\lambda^3$ and $\omega_0$,
  one can neglect the bare boson frequency compared
to the self-energy, so that
 the logarithmic term in the expression for the thermodynamic
potential
  becomes \beq \Xi_{\log} =T\sum_{\Omega
_{m}}\int \frac{dqq}{2\pi }\ln \left((aq)^2+\frac{{\tilde g}^2
\left| \Omega _{m}\right|}{\sqrt{\left( {\tilde \Omega}-i{\tilde
\Delta}\right)^{2}+v_{F}^{2}q^{2}}}\right), \label{mo_6} \eeq
where
 ${\tilde \Omega}=-ic\Sigma^{\mathrm{F}}(\Omega_m,T)$, $c\approx
1.2$ (see [see Eq.~\ref{n_12a}], and the fermion self-energy is
the sum of the quantum and classical parts, given by
Eqs.~(\ref{quant}) and (\ref{class}), respectively.

\paragraph{$T=0$.}

    We consider first the $T=0$ case, when the
    Matsubara sum
    can be replaced by
 an integral
 and the self-energy is purely quantum: ${\tilde \Omega}=c \omega^{1/3}_0\mathrm
{sign}(\Omega_m)|\Omega _{m}|^{2/3}$.
  Introducing
   new variables
 \beq
 x=\left(v_Fq/\dzz\right)^2,\;y=c^{3/2}\omega^{1/2}_0\Omega_m/|\dzz|^{3/2},
 \eeq
 we rewrite Eq.~(\ref{mo_6}) as
 \bwt
 \beq \Xi_{\log} =\frac{c^{3/2}|\dzz|^{7/2}}{4\pi^2\omega_0^{1/2}v_F^2}\int^{\infty}_0 dx\int^{\infty}_0 dy\mathrm{Re}\ln
 \left(x\left(\frac{a\dzz}{v_F}\right)^2+{\tilde g}^2
c^{3/2}\left(\frac{\dzz}{\omega_0}\right)^{1/2}\frac{y}{\sqrt{(y^{2/3}-i)^2+x}}\right).
\label{mo_6_a} \eeq \ewt
 For
 a sufficiently
 small $\dzz$, the first term under the logarithm in
 Eq.~(\ref{mo_6_a}) can be neglected compared to the second one.
We then obtain the field-dependent part of $\Xi_{\log}$ as
 \beq \Xi_{\log} =- \frac{c^{3/2}z}{
 8 \pi^2 v^2_F }~ \frac{|\tilde \Delta|^{7/2}}{\omega^{1/2}_0},
\label{mo_7} \eeq where $z$ is the universal, i.e.,
cutoff-independent, part  of the integral \beq
\mathrm{Re}\int_0^\infty dx \int_0 ^\infty dy \ln\left[(y^{2/3}
-i)^2+x\right]. \label{mo_8} \eeq This integral can be evaluated
exactly and its universal part
 is equal to \beq z = \frac{8
 \sqrt{2} \pi}{35} \label{mo_9}, \eeq so that \beq \Xi_{\log} =-
\frac{\sqrt{2}c^{3/2}}{35 \pi }~ \frac{|\tilde
\Delta|^{7/2}}{v_F^2\omega^{1/2}_0}. \label{mo_10} \eeq Using Eq.
(\ref{n_23}) for $\omega_0$, expressing ${\tilde \Delta}$ via $M$,
and adding the $|M|^2$ term,
 we obtain for the thermodynamic potential
  as a function of magnetization $M$
\beq \Xi (T=0,M) =\frac{M^2}{4 \nu} (1 + g_{s,0}) -
\frac{|M|^{7/2}}{\nu^{5/2} E_c^{3/2}} + b M^4, \label{mo_11} \eeq
where
  the energy scale $E_c$ is defined as~\cite{comm_mo}
 \beq E_c =
A\left(\frac{{\tilde g}}{ak_F}\right)^{4/3}\epsilon_F
\label{mo_12} \eeq with \beq
A=c^{-1}\left[\frac{35}{4}\left(6\sqrt{3}\right)^{1/2}\right]^{2/3}\approx
5.22.
 \eeq
 We see that $\Xi (T=0, M)$ is
still nonanalytic
  at QCP,
 but
  the leading nonanalyticity
  becomes $|M|^{7/2}$ instead of $|M|^3$ in the FL-regime. Still, $|M|^{7/2}$
  is larger then the next-to-leading analytic term ($M^4$).
  Correspondingly, the non-linear susceptibility scales with the
 magnetic field as $\chi\propto +|H|^{3/2}$.
 This scaling is dual to the $|q|^{3/2}$ form of the susceptibility at
 finite $q$. \cite{pepin_prl,pepin_prb}

 The crossover between the FL and non-FL forms of the
 thermodynamic potential [Eqs.~(\ref{mo_3}) and (\ref{mo_11})]
 occurs at the same energy where
   $\Sigma^F$
 crosses over  between
 the FL and non-FL forms, i.e.,
 $\Xi$ is given by Eq.~(\ref{mo_11}) for $|M|\gg \omega_0\nu/\lambda^3$
 and by Eq.~(\ref{mo_3}) otherwise.
 In both cases,
 the nonanalytic terms are negative, which implies that a ferromagnetic quantum-critical point
  {\it is intrinsically
 unstable  against a first-order transition}.\\

\paragraph{Finite T.}

The form of the thermodynamic potential at finite temperatures
depends on whether the temperature is above or below the scale
$T_{\mathrm{QC}}$, separating the regimes where the contributions
from either finite or zero boson Matsubara frequencies  dominate
($T>T_{\mathrm{QC}}$ and $T <T_{\mathrm{QC}}$, respectively).
  In both cases, the
nonanalytic, $|M|^{7/2}$ term is replaced by a regular, $M^4$ one;
however, the temperature dependence of the prefactor of the $M^4$
is different in the two regimes. To see this, one can neglect the
$(aq)^2$ term in Eq.~(\ref{mo_6}), integrate over $q$, expand the
resulting expression to order $\dzz^4$, and convert the Matsubara
sum into a contour integral. Following the same steps that led us
to Eq.~(\ref{Xihigh}) away from the QCP, we expand
$\coth(\Omega/2T)$ as $2T/\Omega+\Omega/6T$ and keep only the
second term in this expansion, which determines the coefficient of
the $\dzz^4$ term
 \beq \Xi_{\log} =  \frac{\dzz^4}{T}\int
^T_0 d\Omega\frac{\Omega}{{\tilde \Omega}^2}.
 \label{s1}\eeq

For $T>T_{\mathrm{QC}}$, ${\tilde \Omega}\propto \Omega^{2/3}$ and
the integral in Eq.~(\ref{s1}) scales as $T^{2/3}$.
Correspondingly, the $M^4$ term in $\Xi$ is $-M^4/T^{1/3}$. For
$T<T_{\mathrm{QC}}$, ${\tilde\Omega}\propto \left(T/|\ln
T|\right)^{1/2}$, and the $M^4$ term behaves as $-M^4|\ln T|$.

In addition,  the $T$-dependence of the $M^2$ term, which
determines the $T$-dependence of the (inverse) spin
susceptibility, changes from $-T|\ln\left(1+g_{s,0}\right)|$ at
  small $T$ to $-T|\ln T|$ at higher $T$.
The crossover occurs at $T\sim T_1$, where $T_1 \propto
\epsilon_F/\lambda^2$ (Ref.~\onlinecite{pepin_prb}). Since the
fermion self-energy enters the $\dzz^2$ term only under the
logarithm, the difference between the regimes $T >
T_{\mathrm{QC}}$ and $T_1 < T < T_{\mathrm{QC}}$ is only in the
numerical prefactor of the $T|\ln T|$ term. As it was pointed out
in Ref.~\onlinecite{pepin_prb}, the negative $T|\ln T|$ dependence
dominates over the $T$-dependence of $\chi$ within the HMM theory,
which is given by the square of the thermal correlation length
$\chi^{-1}_{\mathrm{HMM}}\propto \xi^{-2}(T)\propto T/|\ln T|$ and
is weaker by a factor of $(\ln T)^2$.

\subsection{Phase diagram of a ferromagnetic quantum phase transition}
\label{sec:pt}

In this Section, we analyze two possible scenarios for a
ferromagnetic quantum phase transition in 2D, namely, the
breakdown of a continuous transition and the spiral instability of
a uniform magnetic state.

\subsubsection{First-order phase transition}

First, we assume that finite-$q$ fluctuations of the order
parameter are negligible, and analyze a potential instability of a
continuous second-order phase transition.
 Usually, the effect of nonanalyticities in the free energy
on phase transitions is described in terms of the Landau-Ginzburg
functional of the type given by Eq. (\ref{tp_gen}), where the
prefactors of regular (quartic and higher order) terms are assumed
to be determined by fermions with energies of order of the
bandwidth.\cite{belitz_rmp} Phenomenological nature of these
prefactors makes it difficult to make specific predictions in this
approach.  Here, we will follow a different approach and calculate the \emph{%
entire} thermodynamic potential in a model with a long-range
exchange interaction of radius $a\gg k_{F}^{-1}.$ The downside of
this approach is the choice of a particular model. The upside is
that not only nonanalytic but also
 analytic ($M^2, M^4$, etc)
  terms can be
found explicitly, and therefore--at least within this model--one
can make certain predictions about the nature of the phase
transition.

We will show that, in this particular model, the characteristic
energy scale corresponding to the first-order phase transition is
larger than the scale
 for the spiral instability, and falls into the regime where the
 mass renormalization is small by a factor $1/(ak_F) \ll 1$.  For a
 case when the interaction is short-ranged
 ($ak_F \lesssim 1$),  both the
first order transition and the spiral instability occur in the
strong coupling, critical regime, and which one occurs first
depends on the (unknown) prefactor of a regular $M^4$ term.

For the case $ak_F \gg 1$, we
  first assume and then verify that the mass renormalization factor
$\lambda $ in Eq.~(\ref{mo_11}) can be
 set
  to unity. We will
also see
  that the jump of spin polarization at the first-order
phase transition, while still small compared to its maximum value
(unity),
  is large enough
  so that  one should analyze the full thermodynamic potential $\Xi (M,T)$
 rather than  its expansion up to order $M^4$.
  Keeping in mind these two points, we write the
$T=0$ thermodynamic potential as a function of $M$ as \bwt
\begin{equation}
\Xi \left( M\right) =\frac{\delta M^{2}}{4\nu }+\int \frac{d\Omega _{m}}{%
2\pi }\int \frac{d^{2}q}{\left( 2\pi \right) ^{2}} \ln \left(
\left( aq\right) ^{2}+\frac{\left| \Omega _{m}\right|
}{\sqrt{\left( \Omega _{m}-iM/\nu \right) ^{2}+\left(
v_{F}q\right) ^{2}}}\right) \label{xim}
\end{equation}
\ewt
 In Eq.~({\ref{xim}),
 we set the coupling constant $\bar{g}$ to unity and
  neglected  $\delta$
   under the logarithm; we will see that
   typical values of  $(aq)^2$ are much larger than $\delta$.
  Equation (\ref{xim})
can be reduced to a dimensionless form by rescaling $x=\left(
aq\right) ^{2},$ $y=\Omega _{m}a/v_{F},$ $\zeta =Ma/\nu v_{F},$
and $E\left( \zeta \right) =4a^{2}\Xi \left( M\right) /\nu
v_{F}^{2}$. An expansion of the logarithmic part starts with the
term of order $\left(M^2/\nu\right) ak_F$. We absorb this term
into a renormalization of
 $\delta$ in $\delta M^2/(4\nu)$,
 so in all formulas
  beyond this point
  $\delta$ is already a renormalized parameter. Subtracting off
the
  $M$-independent part, we obtain for the dimensionless
thermodynamic potential in these variables
\begin{equation}
E\left( \zeta \right) =\zeta ^{2}\left[ \delta
-\frac{1}{ak_{F}}V\left( \zeta \right) \right] ,  \label{Ezeta}
\end{equation}
where
\begin{equation}
V\left( \zeta \right) =-\frac{1}{\pi \zeta ^{2}}\int_{-\infty
}^{\infty }dy\int_{0}^{\infty }dx\ln \left( \frac{x+\frac{\left|
y\right| }{\sqrt{\left( y-i\zeta \right)
^{2}+x}}}{x+\frac{|y|}{\sqrt{y^2+x}}}\right) -d
 \label{q_v}\end{equation} and \bwt
\begin{equation}
d=\frac{1}{\pi}\int_{0}^{\infty }dyy\int_{0}^{\infty }dx\frac{%
\left( 2y^{2}x\sqrt{y^{2}+x}+y^{3}-x^{2}\sqrt{y^{2}+x}%
-xy\right) }{\pi \left( y^{2}+x\right) ^{2}\left( x\sqrt{%
y^{2}+x}+y \right) ^{2}}\approx 0.63.
\end{equation}\ewt
 For small $\zeta$, the expansion of $V\left( \zeta \right) $
starts with a nonanalytic term: $V\left( \zeta \ll 1\right)
=\left| \zeta \right| /3+O\left( \zeta ^{2}\right)$. Substituting
the first, $\left| \zeta \right|/3 $, term into (\ref{Ezeta}), we
reproduce the nonanalytic, cubic term in Eq. (\ref{mo_3}).  At
larger $\zeta ,$ function $V\left( \zeta \right) $ goes through a
maximum and falls off at $\zeta \gg 1$ (cf. Fig.~\ref{fig:max}).

The first-order phase transition occurs when the minimum in the
thermodynamic potential  at finite magnetization approaches zero
(cf. Fig. (\ref{fig:meta}), i.e, when the following two conditions
are satisfied simultaneously
\begin{equation}
E\left( \zeta _{\text{cr}}\right) =0\text{ and }E^{\prime }\left( \zeta _{%
\text{cr}}\right) =0.
\end{equation}
This yields
\begin{subequations}
\begin{eqnarray}
&&\delta _{\text{cr}} =V\left( \zeta _{\text{cr}}\right) /ak_{F} \label{delta_a}\\
&&2\delta _{\text{cr}}-\frac{2}{ak_{F}}V\left( \zeta _{\text{cr}}\right) -%
\frac{\zeta V^{\prime }\left( \zeta _{\text{cr}}\right) }{ak_{F}}
=0.\label{delta_b}
\end{eqnarray}
\end{subequations}
Substituting (\ref{delta_a}) into (\ref{delta_b}), we see that the
jump of magnetization corresponds to an extremum of $V\left( \zeta
\right) $
\begin{equation}
V^{\prime }\left( \zeta _{\text{cr}}\right) =0.
\end{equation}
In our case, this extremum is a maximum.  The first-order phase
transition occurs when the line $\delta _{\text{cr}}ak_{F}$
touches the maximum of $V\left( \zeta \right)$, as shown in
Fig.~\ref{fig:max}. Numerical calculation gives \beq
 \zeta _{\text{cr}}\approx 1.11\;\text{and}\; \delta _{%
\text{cr}}\approx 0.21/ak_{F}.\label{q_4}\eeq Since
$\zeta_{\text{cr}}\sim 1$, a critical point cannot be determined
by expanding $E(\zeta)$ in $\zeta$ and keeping only a few first
terms. Coming back to dimensional variables, we see that the jump
of spin polarization at the transition $M_{\text{cr}}/n\sim
1/ak_F\ll 1$. The effective Zeeman splitting at the transition is
$\dzz_{\text{cr}}\sim\epsilon_F/ak_F\ll \epsilon_{F}$.

We now verify
 whether
  neglecting of the mass renormalization
was permissible. The critical distance to the QCP
$\delta_{\text{cr}}\sim 1/ak_F$ corresponds to
 $\lambda
\sim 1/\sqrt{ak_F}
 \ll 1$. Therefore, mass renormalization
is, indeed, irrelevant. Another way to see this is to notice that
the Zeeman splitting at the
  transition
$\dzz_{\text{cr}}\sim \epsilon_F/ak_F$ is parametrically larger
than the characteristic energy $\omega_0\sim \epsilon_F/(ak_F)^4$,
separating the regimes of weak and strong mass renormalization.

\begin{figure}[tbp]
\caption{(color online) Schematic behavior of the thermodynamic
potential $\Xi$ as a function of magnetization $M$. Upper curve:
the first-order phase transition has not been reached yet but the
system exhibits a metamagnetic transition in finite field. Lower
curve: the first-order phase transition occurs.} \label{fig:meta}
\begin{center}
\epsfxsize=0.6\columnwidth\epsffile{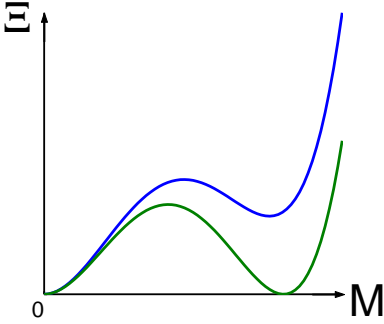}
\end{center}
\end{figure}

\begin{figure}[tbp]
\caption{(color online) Function $V(\zeta)$ defined by
Eq.~(\ref{q_v}). The first-order phase transition occurs when the
maximum of $V$ reaches the value of
 $\delta \times (ak_F)$.} \label{fig:max}
\begin{center}
\epsfxsize=0.8\columnwidth\epsffile{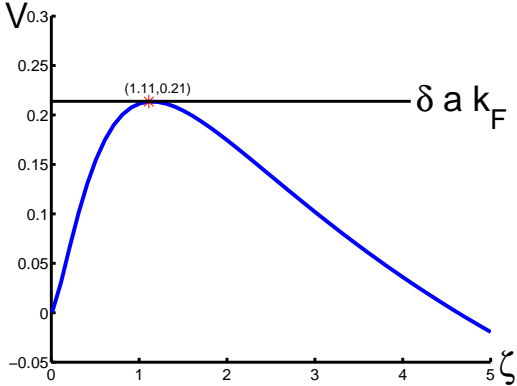}
\end{center}
\end{figure}

At finite temperatures, the first-order transition occurs along
the dashed line in
 Fig.~\ref{fig:phased}, which starts at $\delta=\delta_{\text{cr}}$
at terminates at a tricritical point
$(\delta_{\text{tc}},T=T_{\mathrm{tc}})$ (Ref.~\onlinecite
{griffiths,belitz05}). For $T>T_{\mathrm{tc}}$, the transition is
second-order (solid line in Fig.~\ref{fig:phased}). The
tricritical temperature can be estimated from the condition that
 the $M^4/(\nu^4 v^2_F T)$ term, which replaces the $|M|^3$ one at finite $T$,
 becomes comparable to the regular $bM^4$ term.
In our model,
   $b\sim a/\nu^4 v_F^3$.
This gives $T_{\text{tc}}\sim v_F/a\sim \dzz_{\text{cr}}$. To find
the numerical prefactor, one needs to evaluate the entire
thermodynamic potential at finite $T$ but we are not going to
dwell on it here.

\begin{figure}[tbp]
\caption{(color online) Phase diagram of a ferromagnetic phase
transition. The lines of
 second- and first-order
 transitions (solid and dashed lines,
respectively)
 are separated by a tricritical point.}
\label{fig:phased}
\begin{center}
\epsfxsize=1.0\columnwidth\epsffile{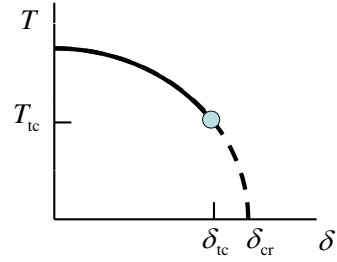}
\end{center}
\end{figure}

\subsubsection{Spiral magnetic phase}

We now turn to another scenario of a phase transition, which is
possible due to a nonanalytic behavior of the spin susceptibility
as a function of the wavenumber.
\cite{belitz,chm03,pepin_prl,pepin_prb} Away from the QCP,
$\chi(q)$ scales as $|q|$; near the QCP, the non-FL mass
renormalization changes this scaling to $|q|^{3/2}$. As in the
previous Section, it will turn out that
  for $ak_F \gg 1$, the instability occurs before the system enters into
 the $|q|^{3/2}$ regime.
  Therefore, we start with a FL form of $\chi(q)$
derived in Ref.~\onlinecite{pepin_prb}. In the long-range
interaction model, \beq
\chi^{-1}(q)\propto\delta+\left(aq\right)^2-\frac{4}{3\pi}\left(1+\lambda\right)\frac{|q|}{k_F}.
\label{q_1}\eeq Equation (\ref{q_1}) is valid for a moderate mass
renormalization: $(1+\lambda) \ll 1/\sqrt{\delta}$. A spiral
instability as $q=q_{\text{cr}}$ occurs when conditions
$\chi^{-1}=0$ and $d\chi^{-1}/d|q|=0$ are satisfied
simultaneously. This gives \bea
&&q_{\text{cr}}=\frac{2}{3\pi}\frac{1+\lambda}{k_Fa^2}\label{q_2}\\
&&\delta^{\text{s}}_{\text{cr}}=\frac{4}{9\pi^2}\frac{(1+\lambda)^2}{k_F^2a^2}.\label{q_3}
\eea Since $\lambda$ depends on $\delta$ itself, as specified by
Eq.~(\ref{n_23}), Eq.~(\ref{q_3}) represents an equation for
$\delta^{\text{s}}_{\text{cr}}$. Solving this equation, we find
that the spiral instability occurs at \beq
\delta^{\text{s}}_{\text{cr}}=\left(\frac{1}{3\pi}+\sqrt{\frac{1}{\left(3\pi\right)^2}+\frac{1}{2\pi}}\right)\frac{1}{\left(ak_F\right)^2}
\approx 0.52/\left(ak_F\right)^2. \eeq At this $\delta$,
$\lambda\sim 1$ and $(1+\lambda)^2\delta\sim\delta\ll 1$, so that
Eq.~(\ref{q_1}) is, indeed, applicable.

 Comparing the critical
values of $\delta$ for the spiral instability and the first-order
phase transition [cf. Eq.~(\ref{q_4})], we see that \beq
\delta_{\text{cr}}/\delta^{\text{s}}_{\text{cr}}\sim ak_F\gg 1.
\eeq This means that the first-order transition occurs before the
spiral instability.
\subsubsection{Re-entrant second-order phase transition}

Another interesting consequence of the nonanalytic behavior of the
susceptibility is a re-entrant second-order phase transition. This
effect occurs due to a nonanalytic temperature dependence of
$\chi$. Suppose that we are still far away from the tricritical
point, so that the nonanalyticity of $\Xi$ as a function of $M$
does not play a role. Taking into account the regular $T$
dependence of $\chi$, which arises from energies of order of the
bandwidth, we have  from (\ref{mo_5})
\begin{equation}
\chi^{-1}(T)=\chi_0\left[\delta+(T/T_0)^2-(T/\epsilon_F)\ln ( |\delta|^{-1})%
\right],  \label{reen}
\end{equation}
where $T_0\sim W$. The critical temperature of the second-order
transition is determined from the condition $\chi^{-1}(T_c)=0$. In
the absence of the nonanalytic term, the
 solution exists only for $\delta<0$ at $T_{c}=T_0\sqrt{-\delta}$
 and the transition line $T_c(\delta)$ has a negative curvature for negative $%
\delta$.  Due to the nonanalytic term,  there exist two solutions of $%
\chi^{-1} (T_c) =0$ for $\delta >0$
 (see Fig.~\ref{fig:reentrant}). One of them, $T_{c1}$, vanishes as $%
\delta/\ln\left(\delta^{-1}\right)$ at $\delta\to 0$, while the
second behaves as $T_{c2}\sim\ln\left(\delta^{-1}\right)$. The two
branches match at some (positive) value of $\delta=\delta_{R}$. As
a result, the phase transition  occurs at $\delta>0$, and the
phase diagram exhibits a re-entrant behavior. Depending on whether
the tricritical point is to the right or to the left of the
reentrant point, the phase diagram has the forms shown in Fig.
\ref{fig:reentrant}.

\begin{figure}[tbp]
\caption{(color online) Re-entrant phase diagram of a second-order
ferromagnetic phase transition. The two curves correspond to two
solutions of Eq.~(\ref{reen}).} \label{fig:reentrant}
\begin{center}
\epsfxsize=0.8\columnwidth\epsffile{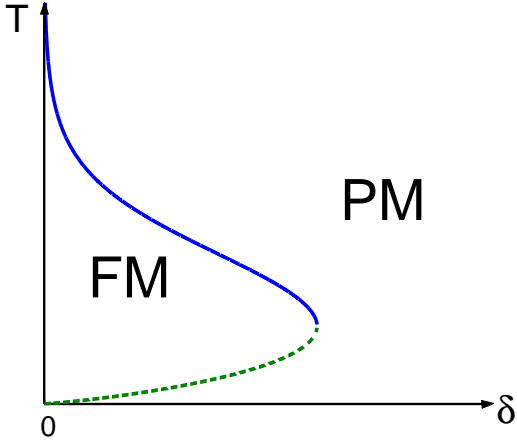}
\end{center}
\end{figure}

\subsection{Cooper channel near the quantum critical point}
\label{sec:cooper_qcp} We now address the issue
 of Cooper renormalization near a ferromagnetic QCP. We recall that
away from the QCP, the
 backscattering amplitude
  vanishes at
  $T\to 0$ as
$1/\ln [\max\{T,\dzz\}]$ due to singular renormalizations in the
Cooper channel. Consequently, the backscattering contribution to
the spin susceptibility is reduced by a factor of
 $1/\ln^2[\max\{T,\dzz\}]$
as compared to non-backscattering contributions.
 The question is to what extent the
Cooper renormalization affects
 the susceptibility near the QCP.
 This is
  an important issue because the sign of $\chi (T,H)$ determines whether
  the transition becomes first order or remain continuous.  In
 the
   preceding
     discussion of the phase transition,
    we approximated the scattering amplitude by  a single component
$f_{s,0}$, which diverges at the QCP, and neglected all other
components. This is certainly inconsistent with the vanishing of
$f_s (\pi)$ at $T=0$, as it is the sum of all components that must
vanish. We now include Cooper renormalization into consideration.
  For definiteness, we consider $\delta \chi (T, H=0)$.

First, we
  show
  that the logarithmical renormalization
of $f_{s} (\pi)$
 starts
 below some temperature which becomes
  progressively smaller as the
system approaches the QCP
 and $|f_{s,0}|$ increases.
To estimate
  this scale,
  we
  consider a vertex $\Gamma(k,p;p,k)$.
When ${\bf k}$ and ${\bf p}$ are projected onto the Fermi surface
and corresponding frequencies are set to zero, $\Gamma$ coincides
with the scattering amplitude $f_s(\theta)$, which depends on the
angle $\theta$ between ${\bf k}$ and ${\bf p}$. The backscattering
amplitude
 corresponds to $\theta=\pi$.
 \begin{figure}[tbp]
\caption{Renormalization of vertex $\Gamma(k,p;p,k)$ (hatched) in
the Cooper channel. Dashed lines are irreducible Cooper
amplitudes.}\label{fig:cooper}
\par
\begin{center}
\epsfxsize=1.0\columnwidth\epsffile{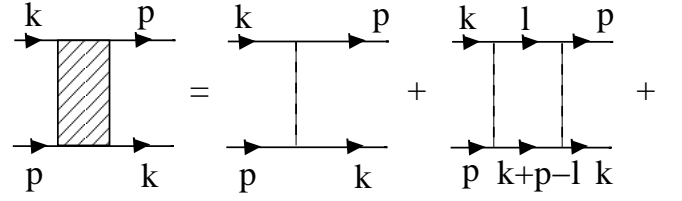}
\end{center}
\end{figure}
This vertex can be expressed
 as
a
  series of diagrams in the Cooper channel  (cf. Fig.\ref{fig:cooper}).
 The dashed lines
are irreducible amplitudes
 $ \gamma_C (\theta)$
 which,
 in general, depend on the scattering angle, e.g., on the angle
$\theta'$ between ${\bf k}$ and ${\bf l}$ in second term of the
series. Integrating over the magnitude of ${\bf l}$ and summing
over the internal fermion frequencies, we obtain
 for the Cooper bubble
 \beq
\Pi_{c}(\theta) \propto
\frac{1}{\lambda}\ln\frac{1}{\cos\theta/2},\label{Cooper}\eeq
 where,
 we remind,
 $\lambda\gg 1$ is the mass renormalization factor.
 As a
result, the series for $f_{s}(\theta)$ reads \beq
f_s(\theta)=\gamma_C(\theta)+\frac{\gamma^{(2)}_C(\theta)}{\lambda}\ln\frac{1}{\cos(\theta/2)}+
\frac{\gamma^{(3)}_C(\theta)}{\lambda^2}\ln^2\frac{1}{\cos(\theta/2)}+\ldots
\eeq where $\gamma_C^{(j)}$ are the angular convolutions of two
and more
 irreducible amplitudes; for example,
$\gamma^{(2)}_{C}(\theta)=\int d\phi
\gamma_C(\phi)\gamma_C(\theta-\phi)$, etc.
 Since
 irreducible amplitudes
 $\gamma_C(\theta)$ and $\gamma^{(j)}(\theta)$ are regular
functions of $\theta$ near $\pi$,
 they can be replaced by constants $\gamma_C(\pi)$ and
$\gamma^{(j)}(\pi)$ for $\theta\approx \pi$. Near $\theta = \pi$,
$f_s(\theta)$ then becomes
 \bea f_s (\theta) & \approx& c_1 + \frac{c_2}{\lambda} \ln\frac{1}{\cos
(\theta/2)} + \frac{c_3}{\lambda^2} \ln^2\frac{1}{\cos (\theta/2)}
+... \nonumber\\
&&=f_{s,0}+\sum_{n>0}f_{s,n}\cos(n\theta). \label{yepp5} \eea The
Cooper bubbles in Eq.~(\ref{yepp5}) have regular expansions over
angular harmonics; for example $\ln\left(\cos\theta/2\right)=- \ln
2+\sum_{n>0}(-)^{n+1}\cos(n\theta)/2n$, etc.  Equating the
prefactors of the $\cos(n\theta)$ terms in the first and second
lines of Eq.~(\ref{yepp5}), we obtain \bea
f_{s,0}&=&c_1+\frac{c_2}{\lambda}\ln 2+\ldots\nonumber\\
f_{s,n}&=&\frac{c_2}{\lambda}\frac{(-)^n}{2n}+\ldots \eea
According to our assumption
 of a Pomeranchuk-like instability, $|f_{s,0}|\gg 1$ while
$|f_{s,n>0}|\lesssim 1$. Therefore, $c_1=f_{s,0}+O(1)$.
 Dropping terms of order unity, the first line of Eq.~(\ref{yepp5})
can be rewritten as \beq f_s (\theta) = f_{s,0} +
\frac{c_2}{\lambda} \ln\frac{1}{\cos(\theta/2)} +
\frac{c_3}{\lambda^2}
\ln^2\frac{1}{\cos(\theta/2)}+\ldots\label{911a}\eeq The
angular-dependent terms in Eq.~(\ref{911a}) are of order unity for
$\theta\sim 1$. As $\theta$ approaches $\pi$, they increase and
cancel the large, angular-independent term, $f_{s,0}$. This is how
the vanishing of $f_s(\pi)$ happens. However, the cancellation
occurs only for angles {\it exponentially} close to $\pi$:
$\pi-\theta\sim \exp\left(-|f_{s,0}|\lambda/c_1\right).$
  Associating typical $\pi - \phi$ with typical
$T/\epsilon_F$, we conclude that
 the characteristic temperature,
 below which  Coooper renormalization becomes crucial,
  is
exponentially small in $|f_{s,0}| \lambda \sim (1 +
g_{s,0})^{-3/2}$.

 These exponentially small
  temperatures are of no relevance for the preceding discussion, as
 they are
 lower that the temperature of the tricrical point, at
which the second-order phase transition turns into a first-order
one.

Still, the
 very fact that Cooper
renormalizations bring about higher harmonics of $f_s(\phi)$,
which scale as $\ln{\epsilon_F/T}$
 requires
  some
 attention
 because the divergent harmonic
 of the scattering amplitude
 , $f_{s,0}$,
  enters
$\delta \chi (T)$ only logarithmically
 [see Eq.~(\ref{fs0T}],
 and the presence of extra $\ln T$ terms
  may affect the temperature dependence of $\chi$.

To address this issue, we use
 a simplified model, in which partial components of $f_s(\theta)$ with
$n>0$ are absorbed
 into an
   effective  temperature-dependent component $f_{s,1}$
  modelled as
 \beq f_{s,1} = \frac{\frac{t}{\lambda}
\ln{\frac{\epsilon_F}{T}}}{1 + \frac{t }{\lambda |f_{s,0}|}
\ln{\frac{\epsilon_F}{T}}}, \label{yepp6} \eeq where $t \sim 1 $
is a constant. This interpolation formula reproduces correctly
 the limiting forms of $f_{s,1}$. Indeed,
 if  $\ln\left(\epsilon_F/T\right)/\lambda |f_{s,0}|$ is small,
$f_{s,1}$ scales as $(1/\lambda) \ln{\epsilon_F/T}$, consistent
with the perturbation theory. If
$\ln\left(\epsilon_F/T\right)/\lambda |f_{s,0}|$ is large,
$f_{s,1}$ approaches $f_{s,0}$, consistent with the vanishing of
$f_s (\pi) = f_{s,0} -f_{s,1}$ at $T=0$.

We now use Eqs. (\ref{yepp1})  and (\ref{yepp2}) for the
susceptibility and compare
 the contributions from $f_{s,0}$ and $f_{s,1}$.
Before substituting $f_{s,1}$ from Eq.~(\ref{yepp6}) into Eq.
(\ref{yepp1})
 for the susceptibility, we need to
 account for
the mass renormalization
  in the expressions for
 the functions  $F_1 (f_{s,0})$
 and $F_2 (f_{s,0})$,
 which amounts to replacing $f_{s,0}$ by $f_{s,0}/\lambda$ in the
arguments of $F_1$ and $F_2$.
 Since
$F_1 (f_{s,0})$ scales as $\ln |f_{s,0}|$, it is not affected by
mass renormalization. On the other hand, the asymptotic behavior
of $F_2 (f_{s,0}) \approx 2 f_{s,0}$ in the absence of mass
renormalization
 must be replaced  by $2 f_{s,0}/\lambda$.

Using the modified expressions for $F_1$ and $F_2$ in
Eq.~(\ref{yepp1}) and
 the
  perturbative form
$f_{c,1} \approx  (c/\lambda) \ln{\epsilon_F/T}$, valid for all
but
 exponentially small $T$, we find
\beq \delta\chi(T, H=0) = \chi^{2D}_0
\left(\frac{{\tilde\mu}_B}{\mu_B}\right)^2 \frac{T}{{\tilde
\epsilon}_F} \left[ \ln{|f_{s,0}|} - \frac{4 c}{\lambda^2}
\ln{\frac{\epsilon_F}{T}} \right]. \label{yepp7} \eeq This is the
key result.  We see that Cooper renormalization generates a $ \ln
T$
 correction to our  earlier result for $\chi (T)$ in the
 isotropic scattering model.
 However,
   at all but exponentially small $T$
 the correction
 is
    small in $1/\lambda^2$ and can be safely neglected.
We conclude, therefore, that the sign and magnitude of $\chi (T,
H=0)$ near the QCP is not affected by
 renormalizations in the
Cooper channel.

The same consideration and conclusion hold also for $\chi (H,
T=0)$.

\section{Magnetic response in 3D}
\label{sec:3D}

In this Section, we discuss the nonanalytic behavior of the spin
susceptibility in 3D. The behavior of $\delta\chi(T,q, H=0)$ has
been considered in
 Refs.~\onlinecite{bealmonod68,rasolt,pethick,belitz,chm03,chmm,betouras,dassarma}.
 As the Kohn anomaly is logarithmic in 3D, the
corresponding nonanalyticities are also only logarithmic:
$\delta\chi(T,q, H=0)\propto q^2\ln {\cal E}$, where ${\cal E}
=\max \{
 v_F|q|,T
\}$, and $\delta\chi(T=0,q=0, H)\propto H^2\ln |H|$. It was argued
in Refs.~\onlinecite{belitz_rmp,belitz05} that
 these
  non-analyticities give rise to tricritical points,
observed
 in 3D
  itinerant ferromagnets. \cite{belitz_rmp} However, the
 relative
 weakness of nonanalytic terms in 3D raises a concern whether
 the tricritical behavior is indeed caused by
 the many-body physics
  or is
   due to
some single-particle effects, such as
  features in the density of states
   \cite{schofield} or magnetoelasticity \cite{larkin}.
In this Section,  we
 show that logarithmic nonanalyticities
 are weakened even further by
 the mass renormalization
 in the vicinity of a 3D FM QCP, to $q^2\ln|\ln {\cal E}|$ and $H^2 \ln |\ln |H||$ forms, respectively .

 Another peculiarity of the 3D case is that
 the $T$-dependence of $\chi (T, H=0, q=0)$ is analytic:
$\delta\chi(T)\propto c_{3D}T^{2}$
(Refs.~\onlinecite{bealmonod68,pethick}). For a Galilean-invariant
system (with  a quadratic fermion dispersion), such as He$^3$, the
coefficient $c_{3D}$ was shown to be negative in
Ref.~\onlinecite{bealmonod68}, i.e., the $T$-dependence of $\chi$
has the same sign as in a Fermi gas. We reanalyze this result in
Sec.~\ref{sec:3DT}
  and show that the magnitude and {\it sign} of $c_{3D}$
are nonuniversal, i.e.,  they depends on details of the fermion
dispersion. This may explain  while $c_{3D}$ is positive in
 He$^3$ (Ref.~\onlinecite{He3chi}) but negative in some
exchange-enhanced paramagnets. \cite{co}

\subsubsection{
 Magnetic-field dependence of the susceptibility} \label{sec:3DH}
We first discuss the nonanalytic behavior of the susceptibility at
$T=0$ in the perturbation theory.\\

\paragraph{Perturbation theory}

To obtain the
 $H^2 \ln |H|$ field dependence of the spin susceptibility in 3D at
$T=0$, it
 suffices
  to expand the integral for  $\Xi $ to order
$H^{4}$ and
 verify that the prefactor diverges logarithmically.
 Cutting off the singularity at the scale set by  $H$, one arrives at the $H^4\ln |H|$ behavior of $\Xi$ and, therefore, the $H^2\ln |H|$ behavior of
$\delta\chi$.

 At second order,
$\Xi$ is still given by Eq.~(\ref{second}), where
$\Pi_{\uparrow\downarrow}$
 is now the 3D  polarization bubble. For simplicity, we assume that the interaction is
  local, i.e.,
   $U(q) =\mathrm {const}\equiv U$.

 Similarly to the 2D
case, $\Pi_{\uparrow\downarrow}$  can be separated into the static
and dynamic parts as
 \begin{equation}
\Pi_{\uparrow \downarrow }\left( \Omega _{m},q\right) =-\nu\left(1-\frac{i\Omega _{m}}{%
2v_{F}q}\ln \frac{i\Omega _{m}+v_{F}q+\dz}{i\Omega
_{m}-v_{F}q+\dz}\right). \label{bubbleud3D}\end{equation} Here and
till the end of this Section
 $\nu=mk_F/2\pi^2$ is
 density of states at the Fermi energy
  in 3D.
  Expanding the square
of $\Pi_{\up\down}$ to order $\dz^{4}$, we obtain
\begin{equation}
\Pi _{\uparrow \downarrow }^{2}\left( \Omega _{m},q\right)
=\ldots+
\nu^{2} \left( \frac{\dz}{%
v_{F}q}\right) ^{4}Q^{(2)} \left( \frac{\Omega
_{m}}{v_{F}q}\right)+ \ldots \label{911_b}\end{equation} where
dots stand for $\Delta$-independent and regular $\Delta^2$ terms,
and
\begin{equation} Q^{(2)}
  \left(
x\right)=x^{2}\frac{2\arctan (x^{-1})x\left[ x^{2}-1\right]
+x^{2}+4/3}{(x^{2}+1)^{4}}. \label{Q2}\end{equation} Substituting
  Eq.~(\ref{911_b}) into Eq.~(\ref{second}) and integrating
 over $\Omega _{m}$,
   we find that the $q$-integral is indeed
 logarithmic.
  Cutting
    off the logarithmic singularity
    by the field, we obtain
\begin{eqnarray}
\Xi _{2}(T=0,H) &=&-\frac{u^{2}\dzz^{4}}{(2\pi )^{3}v_{F}{}^{3}}%
\int_{|\dzz|/\vf}^{k_F}\frac{dq}{q}
\int_{-\infty }^{\infty } dy \text{ }%
Q\left(y\right)\nn \\
&=&-\frac{u^2}{48\pi ^{2}}\frac{\dzz^4}{v_{F}^{3}}\ln
\frac{\epsilon_F}{\left| \dzz\right| },
 \label{xi23D}\end{eqnarray}
where $u = U \nu$ and we used that $\int_{-\infty }^{\infty }dx
Q^{(2)}\left( x\right) =\pi /6$. Consequently,
\begin{equation}
\delta \chi^{\left( 2\right) }=u^2 \frac{\dz^2}{4\epsilon_F^2} \ln
\frac{\epsilon_F}{\left| \dzz\right| }\chi_{0}^{3D},
\end{equation}
where $\chi_0^{3D}=mk_F\mu_B^2/\pi^2$
  is the spin susceptibility of a 3D Fermi gas. We see that that the sign of the
field-dependence of the second-order contribution is metamagnetic,
i.e., $\chi$ increases with $H$. This conclusion contradicts to
Ref.~\onlinecite{bealmonodH}, where $\chi$ was found to decrease
with $H$ as $H^2$ (without a logarithmic factor). Notice also that
typical $\Omega_m$ and $q$ in Eq.~(\ref{xi23D}) are such that
$\om\sim \vf q\gg \dz$. Therefore,
 even the second-order contribution
 in 3D
 does not arise entirely from
backscattering processes.

Higher-order terms modify the second-order result by
 renormalizing static vertices in the skeleton diagram with
 two dynamical bubbles,
 and  add new
processes involving larger number of dynamic bubbles. For example,
the nonanalytic part of diagram $e$ in Fig.~\ref{fig:fig2}
contains the cube of the dynamic bubble. Expanding
$\Pi^3_{\up\down}$
 to order $\dz^4$, we obtain
\begin{equation}
\Pi _{\uparrow \downarrow }^{3}\left( \Omega _{m},q\right) =\ldots+ \text{ }%
\nu^{2} \left( \frac{\dzz}{v_{F}q}\right) ^{4} Q^{(3)}\left(
\frac{\Omega _{m}}{v_{F}q}\right)+ \ldots
\end{equation}
where
\begin{widetext}
\begin{equation}
Q^{(3)}\left( x\right)=x^{2}\frac{3x^{2}\left( x^{2}+1\right)
\arctan
^{2}(x^{-1})+x\left( 4+3x^{2}\right) \arctan (x^{-1})-3x^{2}-1}{%
(x^{2}+1)^{4}%
}.
\end{equation}
\ewt Substituting this expansion into Eq.~(\ref{xi3e}), and
performing the integrations over $\om$ and $q$ in the same way as
in Eq.~(\ref{xi23D}), we obtain
\begin{equation}
\delta \chi^{\left( 3e\right) }=-\frac{\pi
^{2}-8}{8}u^{3}\frac{\dzz^2}{4\epsilon_F^2} \ln
\frac{\epsilon_F}{\left| \dzz\right| }\chi_{0}^{3D},
\end{equation}
where we used that $\int_{-\infty }^{\infty }dxZ\left( x\right)
=\pi (\pi ^{2}-8)/16$.
 As it also the case in 2D,
 the signs of the second- and
third-order contributions are
  opposite.\\

\paragraph{Quantum-critical behavior in 3D}

 The vicinity of a ferromagnetic QCP in 3D can be analyzed
 within the Eliashberg theory, in the same way it was done in
Sec.~\ref{sec:qcp} for the 2D case. Since our goal here is to
obtain only the qualitative behavior of $\chi$, we will not repeat
the manipulations within the spin-fermion model but
 just consider the RPA for $\Xi$, which reproduces correctly both
the weak-coupling and quantum-critical limits. Summing up the
RPA-series
  and replacing $u$
by $-g_{s,0}$ we obtain
\begin{widetext}
\begin{equation}
\Xi (T=0,H) = \frac{1}{4 \pi^3} \int_{-\infty}^\infty d \Omega_m
\int_0^\infty q^2 dq \ln \left[\delta + \frac{i
  \Omega_m}{2
v_F q} \ln \frac{i\Omega _{m}+v_{F}q+\dzz}{i\Omega
_{m}-v_{F}q+\dzz}\right] + \ldots \label{yep_9}
\end{equation}
\end{widetext}
where $\Gamma = -g_{s,0}/(1+ g_{s,0})$
 and dots stand for regular terms. Expanding
 the integrand to fourth order in $\dzz$, and integrating over
$\Omega_m$ and $q$, we obtain
 \begin{equation}
\Xi (T=0,H) =
 -
 \frac{1}
  {2\pi ^{3}}\frac{\dzz^4}{v_{F}^{3}}
  Z(\Gamma)\ln \frac{\epsilon_F}{\left| \dzz\right|},
 \label{yep_6}
\end{equation}
where
\begin{widetext}
\begin{equation}
Z(\Gamma) = \int_0^{\pi/2} d\phi \sin^2\phi \cos^2 \phi
\left[\frac{\cos 2\phi}{S} +
  \frac{11 \cos^2 \phi -2}{6
S^2} +
  \frac{\cos^2\phi}{S^3} +
 \frac{cos^2\phi}{4S^4}\right] \label{yep_7}
\end{equation}
\end{widetext}
and $S = \Gamma^{-1}- \phi \cot \phi$. For small $\Gamma$,
$Z(\Gamma) \approx \pi \Gamma^2/24$, and Eq.~(\ref{yep_6}) reduces
to Eq.~(\ref{xi23D}). At the QCP, $\Gamma \rightarrow \infty$, and
$K(\Gamma) \approx 0.032
$. Substituting this into (\ref{yep_6}) and differentiating twice
over $H$, we obtain at the QCP
\begin{equation}
\delta \chi \approx 0.24 \frac{\dzz^2}{4\epsilon_F^2} \ln
\frac{\epsilon_F}{\left| \dzz\right| }\chi_{0}^{3D}. \label{yep_8}
\end{equation}

Equation (\ref{yep_8}) is, however, incomplete as the RPA neglects
mass renormalization which becomes singular near the QCP.
 Near a 3D QCP, the fermion self-energy behaves
  as $\Sigma
(\omega_m) \propto \omega_m |\ln \omega_m|$, i.e., at small
frequencies it is parametrically larger than a bare $\omega_m$
term in the fermion propagator. Reevaluating the polarization
bubble for dressed fermions, we obtain the same expression as in
Eq.~(\ref{bubbleud3D}), except for $\Omega_m$ under the logarithm
is replaced by $\Omega_m \ln{|\Omega_m|}$. Substituting this into
Eq.~(\ref{yep_9}), setting $\Gamma = \infty$, expanding again to
order $\dzz^4$ and integrating first over $\Omega_m$ and then over
$q$, we obtain
\begin{equation}
\Xi _{2}(T=0,H) \sim \dzz^{4} \int_{|\dzz|/\vf}^{k_F}\frac{dq}{q
\ln q} \sim \dzz^4 \ln |\ln |\dzz||.
 \label{yep_10}
\end{equation}
 Consequently,
\begin{equation}
\delta \chi\sim \chi_{0}^{3D} {\dzz^2}{\epsilon_F^2} \ln \ln
\frac{\epsilon_F}{\left| \dzz\right|}. \label{yep_11}
\end{equation}
This is a very weak nonanalytic dependence.

\subsubsection{Temperature dependence of the spin susceptibility}
\label{sec:3DT}
  The temperature dependence of the spin
susceptibility in 3D, $\delta \chi (T) \propto T^2$,
 was found perturbatively
  in Refs.~\onlinecite{pethick,chm03}
and in the paramagnon model by Beal-Monod et al.
 (Ref.\onlinecite{bealmonod68}).   Here, we
consider the temperature dependence of the spin susceptibility of
a 3D FL near a ferromagnetic QCP.
 As in Ref.~\onlinecite{bealmonod68}, we consider fermions
 with local interaction $U$ near a ferromagnetic QCP, so that
 $U\nu= -g_{s,0} \approx 1$.  However, in contrast with Ref.~\onlinecite{bealmonod68}, we assume an isotropic but
otherwise arbitrary energy spectrum $\varepsilon_k$. Correspondingly, the density of states, $%
\nu \left(\varepsilon_{k}\right)$, is an arbitrary function of the
energy.
 As it will be shown in this Section, this generalization leads to
a
 possibility of reversing the sign of the
$T$-dependence of $\chi$.

\begin{figure}[tbp]
\caption{Diagrams for the thermodynamic potential that give rise
to the temperature dependence of the spin susceptibility in 3D.
Top row: ladder diagrams; bottom row: ring diagrams.
}\label{fig:ladder}
\par
\begin{center}
\epsfxsize=1.0\columnwidth\epsffile{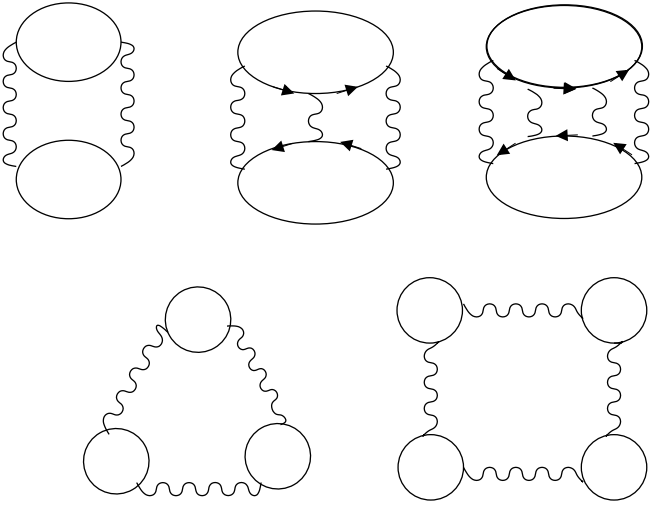}
\end{center}
\end{figure}
The thermodynamic potential is given
  by Eq.~(\ref{bm}); however,
  in contrast to the 2D case,
 the $T$-dependent part of the susceptibility comes not only from the ladder but
 also from the ring diagrams for $\Xi$. Consequently,
 the $\delta \Xi (T)$ term in Eq.~(\ref{bm})
  contains both types of diagrams
(cf. Fig.~\ref{fig:ladder}):
 $\delta\Xi=\delta\Xi_{L}+\delta\Xi_{R}$, where
\begin{subequations}
\begin{eqnarray}
\delta\Xi_{L} &=&T\sum_q \left[ \ln \left( 1+U\Pi _{\uparrow
\downarrow }\right)\right]  \label{Fladder} \\
\delta\Xi_{R} &=&\frac{1}{2}T\sum_q\left[ \ln \left( 1-U^{2}\Pi
_{\uparrow }\Pi _{\downarrow }\right) +U^{2}\Pi _{\uparrow }\Pi
_{\downarrow }\right], \label{Fring}
\end{eqnarray}
\end{subequations}
 and
\beq\Pi_{\up,\down}(\Omega_m,q)=T\sum_{k}G_{\up,\down}(k+q)G_{\up,\down}(k)\label{same}\eeq
is the polarization bubble composed of fermions with the same
spin.
  The
second-order diagram can be considered as a part of either the
ladder or the ring series; we associate it with the ladder series
and subtract the second-order term from the ring sequence.

The details of the calculation are presented in Appendix
\ref{sec:app3D}. Here, we present the results for
  two different regimes near
 a QCP.
  The first one is
 the FL regime of small $T$ and finite $1 + g_{s,0}$:
 $\left(1+g_{s,0}\right)^{3/2}\gg T/\epsilon_F$
 . In this regime,
$\delta \chi (T)$ scales as $T^2$,
 and the prefactor is
 the sum of the universal ($u$) and nonuniversal ($n$)
contributions:
\begin{equation}
\chi^{-1}(T)=(\chi(0))^{-1}\left[1+\left( A_u+A_n\right)
\left(\frac{T}{\epsilon_F}\right)^2\right],
\label{T3D}\end{equation} where
  $\chi(0)=2\mu_B^2\nu/\left(1+g_{s,0}\right)$. The universal
contribution has the same origin as in 2D:
 it arises due to a long-range dynamic interaction between fermions
 mediated by particle-hole pairs.
 The prefactor of this contribution, $A_u$, is given by
 \beq
A_u=-\frac{1}{3}~\frac{1}{\left(1+g_{s,0}\right)^2}.\label{univ}
\eeq
 The nonuniversal part comes from the regular terms in the expansion of the polarization bubble
  in the momentum and
    frequency.
The magnitude and the \emph{sign} of the nonuniversal part depend
on details of the fermion dispersion:
 \beq A_n=\frac{\pi^2}{4}\frac{\left( \nu ^{\prime }\right)
^{2}-(5/3)\nu \nu ^{\prime \prime }}{\alpha\nu
^{2}}\frac{\epsilon_F^2}{\left(1+g_{s,0}\right)^2},
\label{nonuniv}\eeq where $v_k=d\varepsilon_k /dk$,
$m_k^{-1}=d^{2}\varepsilon_k /dk^{2}$,
$\nu'=d\nu(\varepsilon_k)/d\varepsilon_k|_{\varepsilon_k=\epsilon_F}$,
$\nu''=d^2
\nu(\varepsilon_k)/d\varepsilon_k^2|_{\varepsilon_k=\epsilon_F}$.
 The coefficient
  $\alpha$ determines the $q$ dependence of the
  static bubble for free fermions in 3D
 \beq\Pi(\Omega_m=0,q)=-\nu\left[1-\frac{\alpha}{12}\left(\frac{q}{k_F}\right)^2\right]
 \eeq
and is given by
\begin{equation}
\alpha =\frac{k_{F}^{2}}{\nu }\frac{d}{d\varepsilon_k}\left\{ \nu \left[ 2%
\frac{v_k}{k}+\frac{1}{m_k}\right]
-\frac{2}{3}\frac{d}{d\varepsilon_k }\left[ \nu v_k^{2}\right]
\right\} \Big|_{\varepsilon_k =\epsilon_F}.
\label{alpha}\end{equation} For a power-law dispersion,
$\varepsilon_k =Ak^{\gamma }$, we have $\alpha =(\gamma +1)/3$.
For a
 quadratic dispersion,  $\gamma=2,\alpha =1$ and \beq
A_n = \frac{\pi^2}{6} \frac{1}{(1 + g_{s,0})^2}. \label{yep12}
\eeq
 Comparing (\ref{univ}) and (\ref{yep12}), we see that
$A_n$ is positive and larger in magnitude than $%
A_u$: $A_{u}/A_{n}=-\pi^2/2\approx -4.93$.
 This is the result found in Ref.~\onlinecite{bealmonod68}.
 In this
case, $\chi$ decreases with $T$: \beq \chi (T) = \frac{
  \chi(0)}{1 + g_{s,0} + \frac{\pi^2-2}{6}\frac{1}{1
+ g_{s,0}} \left(\frac{T}{\epsilon_F}\right)^2} \label{yep14} \eeq
 This case is perhaps most relevant for
He$^{3}$. However, for
 more complex dispersions (relevant for anisotropic FLs in metals),
the universal term can win over the nonuniversal one. Indeed,
because
 $\alpha$ must be positive regardless of the
dispersion (otherwise, a system of free fermions on a lattice
would have a magnetic instability at finite $q$), the sign of
$A_n$ is
  determined by the the sign of the combination
 $(\nu')^2-(5/3)\nu\nu''$, entering Eq.~(\ref{nonuniv}).
We see that $A_n>0$ as long as $\nu''<0$. On the other hand, if
the Fermi energy is near the minimum in the density of states,
where $\nu'=0$ and $\nu''>0$, then $A_n<0$ and $\chi$ increases
with $T$. Therefore, both types of the $T$-dependences are
possible in metals. An increase of $\chi(T)$ with $T$ was observed
in a number of strongly paramagnetic metals \cite{co}.

Next, we consider the quantum critical regime: $T/\epsilon_F\gg
\left(1 + g_{s,0}\right)^{3/2}$.
 Here  both universal and non-universal terms scale as
 $T^{4/3}$ (Ref.~\onlinecite{3/4}). A detailed calculation shows
that
  \beq \chi^{-1} (T) =
 \left(2 \mu_B^2 \nu\right)^{-1}
b\left(B_u+B_n\right) \left(T/\epsilon_F\right)^{4/3}
 \label{yep_16} \eeq where $B_u=-8/\pi^2$ is the universal,
low-energy contribution, $B_n$ is the nonuniversal contribution
whose form depends on the details of the fermion dispersion,
 and $b =2^{2/3} (6\pi)^{1/3} \Gamma(4/3) \zeta
(4/3)/3^{5/2}\approx 0.871$, with $\Gamma(x)$ and $\zeta(x)$ being
the $\Gamma$- and $\zeta$-functions, respectively. For the
quadratic dispersion, $B_n=1>8/\pi^2$ and the susceptibility is
positive, i.e., the system is stable. It is
 possible, however, that $B$ is smaller than $8\pi^2$ for a more
complex dispersion, in which case the susceptibility is negative
 at the QCP.
 This
 implies
 that the transition line $T_c (g_{s,0})$
  may have
 ``wrong''
 sign of the slope at small $T$, as in Fig.~\ref{fig:reentrant}, and
the system
 exhibits a reentrant ferromagnetic transition before reaching the
QCP at $T=0$.

There is also a more fundamental reason for the re-entrant
behavior. In the calculation that led us to Eq.~(\ref{yep_16}), we
neglected the fermion self-energy, which scales as $\omega_m \ln
\omega_m$ at the QCP.
 Including the self-energy, we find that the universal, negative
contribution to the inverse susceptibility [the $-8/\pi^2$ term in
Eq.~(\ref{yep_16})] acquires an extra factor of $(\ln T)^2$, while
the non-universal contribution remains the same.
 As a
result, the universal
 term wins at low enough $T$, and the system displays a re-entrant
behavior even for the quadratic dispersion.

 \section{Conclusions}
\label{sec:concl}

In this paper, we have considered nonanalytic spin response of an
interacting Fermi system, both away from and near a ferromagnetic
quantum critical point. Our two primary goals were i) to establish
the sign of the magnetic-field and temperature dependences of the
spin susceptibility in the Fermi-liquid regime and ii) analyze the
stability of the continuous ferromagnetic quantum critical point
in 2D. We found that higher-order processes, involving more than
one particle-hole pair, may reverse the anomalous (positive) sign
of the single particle-hole pair contribution in a 2D Fermi
liquid. A controllable calculation within a large-$N$ model shows
that this effect is more important than Cooper renormalizations of
the backscattering amplitude considered in
Refs.~\onlinecite{finn,finn_new}. For a 3D Fermi liquid, we showed
that the sign of the $T$-dependence of $\chi$ is determined by the
balance between universal and non-universal contributions; while
the former depends only on the density of states and Fermi
velocity near the Fermi level, the latter is sensitive to the
actual form of the fermion spectrum away from the Fermi energy. We
found that whereas the sign of $T$ and $H$ dependences of $\chi$
in the Fermi liquid regime depends on the detailed form of the
interaction, the anomalous (positive) sign of these dependences is
restored near a ferromagnetic quantum critical point. At the same
time, the role of Cooper renormalizations is diminished even
further near criticality. Analyzing different mechanisms for a
breakdown of second-order phase transition, we showed, within the
model of a long-range exchange interaction, that the first-order
phase transition preempts an instability towards a spiral magnetic
phase. \acknowledgments We acknowledge helpful discussions with
  I. L. Aleiner, J. Betouras, S. Chesi,
 A. M. Finkelstein, L. I. Glazman,
K. B. Efetov, D. Efremov, D. Loss,  C. P{\'e}pin, M. Reznikov, R.
Saha, G. Schwiete,  P. Simon, M. Shayegan, and R. Zak, and support
from NSF-DMR 0604406 (A. V. Ch.)
 D. L. M.
 acknowledges the hospitality of the
Basel Center for Quantum Computing and Quantum Coherence
(Switzerland), Laboratoire de Physique des Solides, Universit{\'e}
Paris-Sud (France), where parts of this work were done, and thanks
the RTRA Triangle de la Physique for financial support.
 We thank C. Wei and P. Marlin for their help in preparing this
manuscript.
 \\

\appendix
\section{Spin susceptibility of a 2D Coulomb system in the large $N$ limit} \label{sec:largeN}
In this Appendix, we calculate the nonanalytic magnetic-field
dependence of the spin susceptibility for a valley-degenerate 2D
electron gas with Coulomb interaction. We assume that there are
$N_v$ degenerate orbital valleys, so that the total
(spin$\times$valley) degeneracy is $N=2N_v \gg 1$.
 For simplicity, we consider only the $T=0$ case.

\paragraph{\textbf{Ring diagrams}}

 Since each polarization bubble comes with a large factor
of $N$, the leading contribution to the thermodynamic potential is
 given by a series of ring diagrams [diagrams {\it a,b}... in
Fig.~\ref{fig:fig2}], which contain a maximum number of bubbles at
each order.
 The sum of ring diagrams gives
\begin{eqnarray}
\Xi _{N=\infty}
=\frac{1}{2}%
T\sum_{q}\ln \left[ 1-\frac{N}{2}U\left( q\right) \left( \Pi
_{\uparrow }+\Pi _{\downarrow }\right) \right],
\label{xiRPA}\end{eqnarray} where $U(q)=2\pi e^2/q$ is the bare
Coulomb potential. The nonanalytic dependence of $\Xi$ is
determined by $q$ near $2k_F$.
  For $q \approx 2k_F$,  the
second term under the logarithm in Eq.~(\ref{xiRPA}) is of order
$N(e^{2}/k_{F})m\sim g_N$, where $g_N$ is the dimensionless
coupling constant defined in Eq.~(\ref{gc}). For $g_N\ll 1$, one
can expand the logarithm to second order in $U(q)$. This
reproduces a weak-coupling result of Ref.~\onlinecite{betouras}.
In the opposite limit of $g_N\gg 1$ (strong coupling), the second
term under the logarithm dominates, and the field-dependent part
of $\Xi_{N=\infty}$ reduces to
\begin{equation}
\Xi _{N=\infty}=\frac{1}{2}T\sum_{q}\ln \left[ 1-\left(
P_{\uparrow }+P_{\downarrow }\right) /2\right] ,
\label{xiRPA_2}\end{equation}%
where
 $P_{\up,\down}$ are the dynamic parts of the bubbles with the same
spins, defined as
 \begin{equation}
\Pi_{\up,\down}(\om,q)=-\frac{m}{2\pi}\left(1-P_{\up,\down}\right).
\label{yep_5}
\end{equation}

To evaluate $\Xi_{N=\infty}$, we expand the logarithm in
Eq.~(\ref{xiRPA_2}) in Taylor series\bwt
\begin{equation}
\Xi_{N=\infty} =-\frac{1}{2}T\sum_{q}\left[ \frac{1}{2}\left(
P_{\uparrow }+P_{\downarrow }\right) +\frac{1}{2}\left[
\frac{1}{2}\left( P_{\uparrow
}+P_{\downarrow }\right) \right] ^{2}+\frac{1%
}{3}\left[ \frac{1}{2}\left( P_{\uparrow }+P_{\downarrow }\right)
\right]
^{3}+\frac{1}{4}\left[ \frac{1}{%
2}\left( P_{\uparrow }+P_{\downarrow }\right) \right] ^{4}+\ldots
\right]. \label{LN2}
\end{equation}
\ewt This expansion can be viewed as a series of fictitious ring
diagrams with each bubble being either $P_{\uparrow }$ or
$P_{\downarrow }$ and the effective interaction being equal to
unity.

At $T,H=0$ and $q\approx 2k_F$,
  \bea P_{\up}=P_{\down}=\left(\frac{{\bar
q}}{2k_F}+ \sqrt{\left(\frac{{\bar
q}}{2k_F}\right)^2+\left(\frac{\om}{4\epsilon_F}\right)^2}
\right)^{1/2}, \label{pi2kf}\eea where ${\bar q}=q-2k_F$ and
$|{\bar q}|\ll k_F$ (Ref.~\onlinecite{chm03}).
 For $|\om|\ll |{\bar q}|\vf$,
Eq.~(\ref{pi2kf}) reduces to \bea P_{\up}=P_{\down}=\left\{
\begin{array}{cl}
&\left({\bar q}/k_F\right)^{1/2};\;{\rm for }\;{\bar q}>0;\nn\\
&\sqrt{2}|\om|/2\vf\left(k_F|{\bar q}|\right)^{1/2},\;{\rm for}\;
{\bar q}<0.
\end{array}
\right. \eea One can easily verify that the field dependence of
the $\Xi_{N=\infty}$ comes only from the products of
 $P_{\up,\down}$ with opposite spins. At second order, there is
only one such a term: $P_{\up}P_{\down}$. The $q$-integral of
$P_{\up}P_{\down}$ diverges logarithmically for ${\bar q}<0$.
Finite magnetic field splits the Fermi momenta of spin-up and
spin-down fermions.
  This, along with with finite $\om$,
regularizes the integral: \bea\int d{\bar q}P{\up}P_{\down}&\sim&
\om^2\int^{-\max\{|\om|,|\dz|\}}_{-k_F}d{\bar q}/|{\bar
q}|\nn\\&&=-\om^2\ln \max\{|\om|,|\dz|\}.\eea This is the same
logarithmic behavior that we have already encountered in the
perturbation theory [cf. Eq.~ (\ref{Xi2H2D_a})]--it leads to a
$|\dz|^3$ term in $\Xi$.  At third order, the nonanalytic
 dependence on $\Delta$
comes from  $P_{\up}^2P_{\down}$. The $q$ integral of
$P_{\up}^2P_{\down}$ yields $|\om|^3\int d{\tilde q}/|{\tilde
q}^{3/2}\sim |\om|^{5/2}$. This leads to a $|\dz|^{7/2}$
nonanalyticity in $\Xi$, which is weaker than the $|\dz|^3$ one.
All higher order contributions produce either analytic terms or
nonanalytic terms with higher powers of $\dz$. Therefore, as long
as only ring diagrams are considered, the only source of a
$|\dz|^3$ nonanalyticity is
the second order term in  (\ref{%
LN2}). Hence,
 \begin{equation}
\Xi _{N=\infty }=-\frac{1}{8}T\sum_{q}P_{\uparrow }P_{\downarrow
}. \label{LN3}
\end{equation}
This situation is to be contrasted with the regular perturbation
theory [cf.~Sec.~\ref{sec:delta}], where ring diagrams are built
from the full bubbles $\Pi_{\up,\down}$ (see \ref{yep_5}).
 There, diagrams to all orders yield
 $|\dz|^3$ terms. For example, at third order
 one generates a $|\dz|^3$ term by keeping
 $P_{\up,\down}$ parts in two out of three bubbles and replacing
the third
 one by its constant part ($-m/2\pi$). A strong-coupling expansion
that we consider here involves
 ring diagrams built from $P_{\up,\down}$
instead of the full bubbles. Since $P_{\up,\down}$
 do not contain constant terms, diagrams containing
more than two $P_{\up,\down}$ do not give rise to a $|\dz|^3$
nonanalyticity.

An explicit computation of $\Xi_{N=\infty}$ from Eq.~(\ref{LN3})
is quite involved, because one needs to know the
 expressions for $2k_F$ bubbles in a finite field. Fortunately,
there is no need to perform such a computation because the
nonanalytic part of the second-order diagram for a short-range
interaction
\begin{eqnarray}
\Xi _{2} &=&-\frac{%
1}{2}U^{2}T\sum_{q}\Pi _{\uparrow }\Pi _{\downarrow }
\end{eqnarray}
contains the same combination of $P_{\up}P_{\down}$ as
Eq.~(\ref{LN3}).
 Therefore,
\begin{equation}
\Xi _{N=\infty }=\frac{1}{4}\Xi _{2}|_{u=1},
\label{map}\end{equation} where $u=\nu U$. Using
  Eq.~(\ref{Xi2H2D_a})
  $\Xi_{2}$
  and differentiating twice
 with respect to
 the field, we obtain
\begin{equation}
\delta\chi_{N=\infty }=\frac{|\dzz|}{8\epsilon_F}
 \chi^{2D}_0. \label{chiRPA}\end{equation}

Notice that this result
 could be obtained from the
second-order ring diagram by replacing the wavy line by the
screened Coulomb potential for $N\to\infty$
\begin{equation}
\tilde{U}=\frac{%
2\pi e^{2}}{q+2\pi e^{2}\frac{m}{2\pi }N}\approx \frac{2\pi }{mN}.
\label{screen}\end{equation}

The $2k_F$ ring diagrams  can be viewed
 as series of backscattering events,
 hence Eq.~(\ref{chiRPA}) is
 the backscattering contribution to the susceptibility  for a
Coulomb interaction in the large-$N$ limit.

\paragraph{
\textbf{Cooper channel}} \label{sec:cooper_coulomb}

 Out of the
subleading, $1/N$ diagrams that modify
  the backscattering
 contribution to  $\delta\chi$, the most important ones are the Cooper
diagrams which
 come with an additional logarithm, $L_C=\ln( W/|\Delta|)$. At low
energies, when $L_C>N$, the Cooper diagrams become comparable to
the ring ones.

 To third order,
  a Cooper diagram is
 diagram {\it c} in Fig.~\ref{fig:fig2}
 with wavy line
 replaced by the screened Coulomb interaction from
Eq.~(\ref{screen}).
 This diagram reads
  \begin{eqnarray}
\Xi _{C,3} &=&\frac{1}{3}\left( \frac{N}{%
2}\right) ^{2}\tilde{U}%
^{3}T\sum_{q}\left( \Pi _{\uparrow \downarrow }^{C}\right) ^{3},
\label{cooper3}
\end{eqnarray}
where $\Pi_{\uparrow \downarrow }^{C}$ is a Cooper bubble composed
of fermions with opposite spins \beq \Pi^{C}_{\up\down}=T\sum_k
G_{\up}(k+q)G_{\down} (-k). \eeq
 At $T=0$,
\bea \Pi _{\uparrow
\downarrow }^{C}&=&\frac{m}{2\pi }%
\text{Re}\ln \frac{W}{\om +i\dzz+\sqrt{(\om +i\dzz)^{2}+\left( v_{F}q\right) ^{2}}}\nn\\
&&=\frac{m}{2\pi }%
\left[ L_{C}+P^C_{\up\down}\right], \label{cooperlog} \eea where
$P^C_{\up\down}$ is a dimensionless function of two variables, $x
=\Omega_m /|\dzz|$ and $y=\left( v_{F}q/\dzz\right) ^{2}$:
\begin{equation}
P^C_{\up\down}=-\text{Re}%
\ln \left[ x+i\text{sgn}\dzz+\sqrt{(x+i\text{sgn}\dzz)^{2}+y}%
\right].
\end{equation}
 Substituting Eq.~(\ref{cooperlog}) into
Eq.~(\ref{cooper3}), we obtain
\begin{eqnarray}
\Xi _{C,3} &=&\frac{T}{12N}\sum_{q}\Big(
L_{C}^{3}+3L_{C}^{2}P^C_{\up\down}\nn\\
&&+3L_{C}\left[P^C_{\up\down}\right]^{2}+
\left[P^C_{\up\down}\right]^{3}\Big)\label{161}.
\end{eqnarray}
 Cooper renormalization of the backscattering amplitude
 comes from the third term
in Eq.~(\ref{161})
\begin{equation}
\Xi _{C,3}=\frac{1}{4N}%
L_{C}^{{}}T\sum_{q}\left[P^C_{\up\down}\right]^{2},
\label{yep_6_1}
\end{equation}
while other terms either give field-independent contributions
  or
  generate $1/N$ corrections to the prefactor
  of the $\left|\dz\right| ^{3}$
 term.
 We now recall that the
second-order diagram (diagram {\it a} in Fig.~\ref{fig:fig2}) can
be represented equivalently either in the particle-hole or the
particle-particle form.
 The latter reads
\begin{equation}
\Xi _{2}=-%
\frac{1}{2}u^2T\sum_{q}\left[P^C_{\up\down}\right]^{2}\label{911_c}.
\end{equation}
 Comparing Eqs.~(\ref{yep_6_1}) and (\ref{911_c}), we find that
\begin{equation}
\Xi _{C,3}=-\frac{L_{C}}{2N}\Xi _{2}|_{u=1} = - \frac{2 L_C}{N}
\Xi_{N=\infty}.
\end{equation}
As a result,
\begin{equation}
\delta\chi_{C,3}=-\frac{2L_{C}^{{}}}{N}\delta\chi_{N=\infty},
\end{equation}
where $\delta\chi_{N=\infty}$ is given by Eq.~(\ref{chiRPA}).

Higher order Cooper diagrams (diagram {\it d} in
Fig.~\ref{fig:fig2} and similar diagrams at higher orders) form a
ladder. At any order $n$, we need to select only the
$L_{C}^{n-2}\left[P_{\up\down}^c\right]^{2}$ term from the
$n^{\text {th}}$ power of the Cooper bubble. The $n^{\text{th}}$
order Cooper diagram gives \bea
\Xi _{C,n}&=&\frac{\left( -\right) ^{n-1}}{%
n}\left( \frac{N}{2}\right)\frac{n(n-1)}{2}
^{2}L_{C}^{n-2}C_{2}^{n}\left( \frac{m\tilde{U}%
}{2\pi }\right) ^{n}\nonumber\\
&&\times T\sum_{q}\left[P^c_{\up\down}\right]^{2}. \eea
 Re-expressing again $T\sum_{q}\left[P^C_{\up\down}\right]^{2}$ via
$\Xi _{2},$ we obtain for the $n^{\text {th}}$ order Cooper
contribution to the susceptibility
\begin{equation}
\delta\chi_{C,n}=\left( -\right) ^{n}(n-1) \left( \frac{L_{C}}{N}%
\right) ^{n-2}\delta\chi_{N=\infty},\text{ for }n\geq 3.
\end{equation}
Summing up all orders, we obtain
\begin{equation}
\delta\chi_{C}=\sum_{n=3}^{\infty}\delta\chi_{C,n}=\delta\chi
^{N=\infty }\left( \frac{1}{(1+L_{C}/N)^{2}}%
-1\right).   \label{LN4}
\end{equation}
Combining this result with Eq.~(\ref{chiRPA}), we find for the
total backscattering contribution $\delta\chi_{\mathrm{BS}}
=\delta\chi
_{N=\infty }+\delta\chi_{C}$%
\begin{equation}
 \delta\chi_{\mathrm{BS}}
=\frac{|\Delta|}{8\epsilon_F}\frac{\chi^{2D}_0}{\left(1+L_{C}/N
\right)^2}. \label{LN5}
\end{equation}

\paragraph{\textbf{$1/N$ corrections}}

The $1/N$ corrections to Eq.~(\ref{LN5}) come from diagrams of
third and higher orders in the screened interaction
(\ref{screen}),
 excluding the main logarithmic parts of Cooper diagrams.
 Some of these diagrams, e.g., diagrams {\it b,d}
and {\it f} in Fig.~\ref{fig:fig2}, belong to backscattering
 type
  and
  contribute either non-logarithmic corrections to the
 backscattering amplitude or subleading logarithmic corrections in the Cooper channel.
For example, fourth-order Cooper diagram {\it d} contains not only
the $L^2_C$ term, already taken into account when summing up the
leading logarithmic terms, but also a subleading, $L_C$, term.
These diagrams change the two terms in
 the denominator
 of (\ref{LN5}) as $1 \to 1+ O(1/N)$ and $L_C/N \to L_C (1/N + O(1/N^2))$.
 At low energies, i.e., for $L_C\gg N$, the entire
backscattering contribution is then
  given by an asymptotic
limit of Eq.~(\ref{LN5}): \beq \delta\chi_{\mathrm{BS}}
=\frac{|\Delta|}{8\epsilon_F}\frac{\chi^{2D}_0N^2}{L_{C}^2}~\left(1
+ O\left(\frac{1}{N}\right)\right). \label{174}\eeq

In addition,  there are diagrams of  different type, which do {\it
not} participate in the renormalization of the  backscattering
amplitude.
 To leading order in $1/N$, this is
  diagram {\it e} in Fig. \ref{fig:fig2}.
 This diagram has already been calculated
in Sec.~\ref{sec:other}: we just need to substitute $u=1/N$ into
Eq.~(\ref{chi3e}) and multiply the result by $(N/2)^2$. Combining
this result with the backscattering contribution, we obtain the
 low-energy form of $\delta\chi$ in the large-$N$ model given by
Eq.~(\ref{175a}).

\section{Equivalence between the thermodynamic and Kubo-formula approaches for the susceptibility.
} \label{sec:equiv}

In this work, we were using a thermodynamic approach to spin
susceptibility, in which one first evaluates the thermodynamic
potential
 in finite magnetic field and  then finds the susceptibility by
 differentiating
 the potential with respect to the field. Alternatively, one can
find the susceptibility via the Kubo formula for the spin-spin
correlation function. Certainly, the results of the two approaches
must be equivalent in the linear-response regime. However, when
going beyond the lowest-order linear-response diagrams, it is
quite easy to miss certain diagrams which are generated
automatically in the thermodynamic approach. Moreover, the
diagrams for $\chi$ generated by differentiating the thermodynamic
potential work even beyond the linear-response regime, i.e, for an
arbitrary ratio of the temperature and the Zeeman energy (but for
Zeeman energies still smaller than the Fermi one, when the
magnetic-field dependence of the bare interaction can be
neglected).

We now show how the diagrams for $\chi$ are generated within the
RPA in the spin channel for a local bare interaction:
$U(q)=\mathrm{const}=u/\nu$.
 In this case, the field-dependent part of the thermodynamic
potential is given by Eqs.~(\ref{bm}) and (\ref{n_5a}).
  The
polarization operator is given by Eq. (\ref{pi})
 with the Green's functions
 from
 Eq.~(\ref{n_1}).
 Differentiating Eq.~(\ref{n_5a}) and using the identities \beq
\frac{\partial
 G_{\uparrow,\down}}{\partial H} =
 \mp {\tilde \mu}_B
  G^2_{\uparrow,\down}
 \label{e_4} \eeq
we obtain \beq \delta \chi (T,H) = \delta \chi_1 + \delta \chi_2,
\label{e_5}\eeq where
\begin{widetext}
\bea &&  \delta \chi_1 = 2{\tilde \mu}_B^2
 T^2
 \sum_q \sum_{k}
\frac{U}{1 + U \Pi_{\uparrow \downarrow}(q)}
 \left[
 G^2_{\uparrow}(k) G^2_{ \downarrow} (k +q)
  -
  G^3_{\uparrow}(k) G_{ \downarrow} (k +q)  -
  G_{\uparrow}(k) G^3_{ \downarrow} (k +q)\right] \nonumber \\
 && \delta \chi_2 = {\tilde \mu}_B^2 T^3\sum_q \sum_{k_1}
\sum_{k_2} \frac{U^2}{(1 +U \Pi_{\uparrow \downarrow} (q))^2}
\left[G^2_{\uparrow}(k_1) G_{ \downarrow} (k_1 +q) -
G_{\uparrow}(k_1) G^2_{ \downarrow} (k_1 +q)\right]\nonumber \\
&&\times
 \left[G^2_{\uparrow}(k_2) G_{ \downarrow} (k_2 +q) - G_{\uparrow}(k_2) G^2_{ \downarrow} (k_2 +q)\right].
\label{e_6} \eea
\end{widetext}
 Algebraic expressions for $\delta\chi_1$ and $\delta\chi_2$ are
equivalent to diagrams
 in Fig.~\ref{fig:lr}.
 \begin{figure}[tbp]
\caption{Diagrams for the spin susceptibility generated by
differentiating the thermodynamic potential with respect to the
magnetic field.}\label{fig:lr}
\par
\begin{center}
\epsfxsize=1.0\columnwidth\epsffile{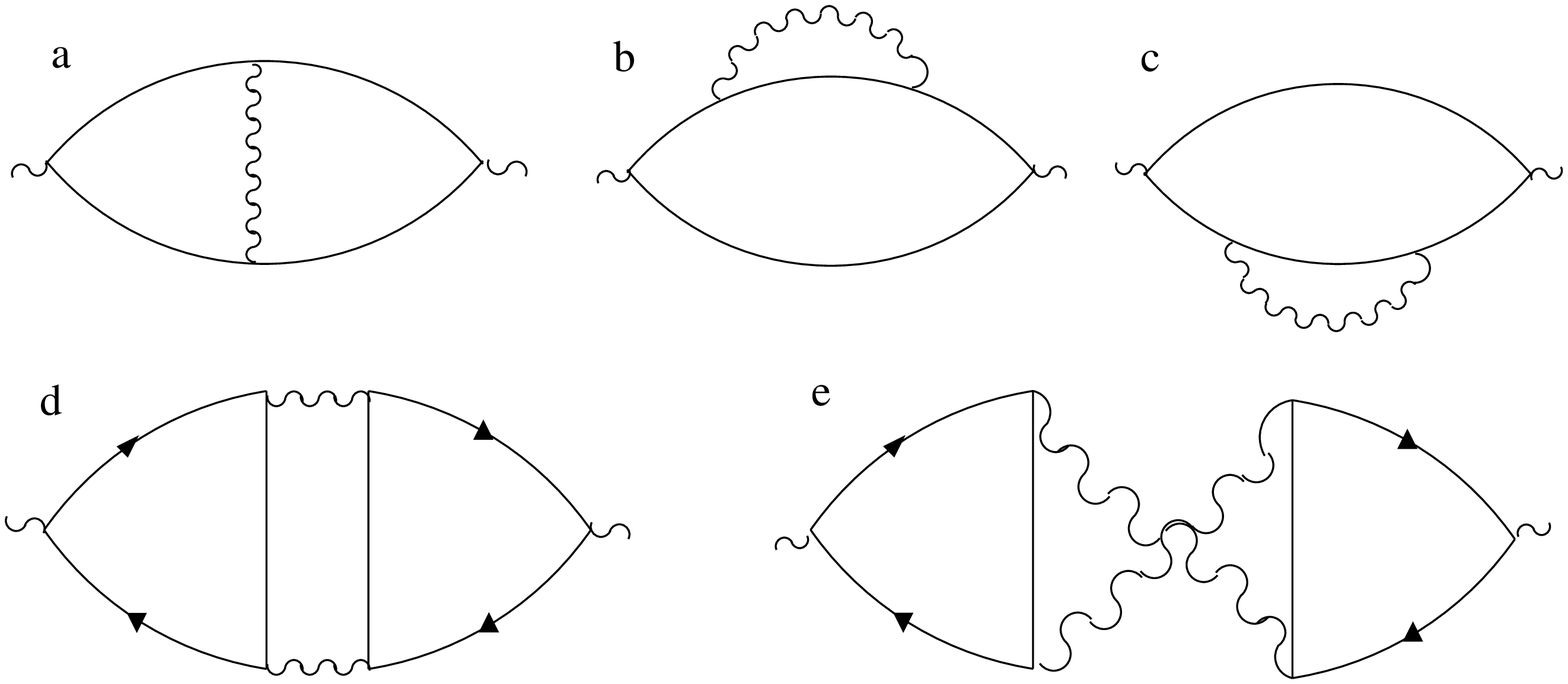}
\end{center}
\end{figure}
  The first term in $\delta \chi_1$
is equivalent to diagram {\it a} for the  vertex correction
insertions into the bare particle-hole bubble; the second and
third terms are equivalent to diagrams and {\it b} and {\it c},
which are self-energy corrections.
  The contribution $\delta \chi_2$ is equivalent to
 "Aslamazov-Larkin"
 diagrams {\it d} and {\it e},
which contain  two triads of fermion propagators, connected by the
interaction lines.
  The wavy lines
 in these diagrams are spin propagators $1/(1 + U
\Pi_{\uparrow \downarrow} (q))$, while the vertices contain
 Pauli matrices
 ${\bf
\sigma}_{\alpha \beta}$. The same diagrams have been used in
Ref.~\onlinecite{pepin_prb} to analyze diagrammatically the
momentum dependence of the spin susceptibility.

\section{Temperature dependence of the spin susceptibility in 3D}
\label{sec:app3D}

In this Appendix, we derive Eqs.~(\ref{T3D}-\ref{alpha}),
describing the temperature dependence of the spin susceptibility
in a 3D FL near a ferromagnetic QCP.

As we said in the main text, the field-dependent part of the
thermodynamic potential is represented by a sum of the ladder and
ring diagrams, which contain polarization bubbles made of fermions
both with
 the same and opposite spins:
 $\Pi_{\uparrow,\uparrow}$ ($\Pi_{\downarrow,\downarrow}$)) and
 $\Pi_{\uparrow\downarrow}$, respectively [cf.
Eqs.~(\ref{Fladder},\ref{Fring})]. In 2D, the nonanalytic, $O(T)$
 behavior of the spin susceptibility is associated with
 the field dependence of the
  the dynamic part of $\Pi_{\uparrow\downarrow}$.
 In 3D, the $T$ dependence of $\chi$ turns out to be analytic, and one needs to consider
 the field dependence of
 both static and dynamic parts of
 $\Pi_{\uparrow\downarrow}$,
  as well as the field dependence
  of
  $\Pi_{\uparrow}$ and $\Pi_{\downarrow}$
  (cf. Eq.~(\ref{same})).

We begin with $\Pi_{\uparrow\downarrow}$. Expanding
$\Pi_{\uparrow\downarrow}(0,0)$ to order $\dzz^2$, we obtain \bwt
\begin{eqnarray}
\Pi _{\uparrow \downarrow }\left( 0,0\right)=\int
\frac{d^{3}k}{\left( 2\pi \right) ^{3}}\int \frac{d\omega_m}{2\pi
}G_{\uparrow }\left( \omega_m ,k\right) G_{\downarrow }\left(
\omega_m ,k\right)
=\frac{1}{\dzz}\int_{-\dzz /2<\varepsilon _{k}-\epsilon_F<\dzz/2}\frac{d^{3}k}{%
\left( 2\pi \right) ^{3}}=-\nu -\frac{\nu ^{\prime \prime }\dzz
^{2}}{24}. \label{delta2}\end{eqnarray} \ewt We will also need a
$q^{2}$ term in the expansion of a static bubble in zero magnetic
field.  To obtain this term for an arbitrary but isotropic
dispersion, we expand the difference
$\delta\Pi(q)=\Pi(\Omega=0,q)-\Pi(\Omega_m=0,q\to 0)$ to order
$q^2$: \bwt\beq\delta\Pi(q)=\int \frac{d^{3}k}{\left( 2\pi \right)
^{3}}\left[\frac{f_{0}\left( \varepsilon _{\mathbf{k+q}}\right)
-f\left(
\varepsilon _{\mathbf{k}}\right) }{\varepsilon _{\mathbf{k+q}}-\varepsilon _{%
\mathbf{k}}}-f'_0(\varepsilon_{\mathbf{k}})\right]
=\int \frac{d^{3}k}{\left( 2\pi \right) ^{3}}\left[ \frac{1}{2}%
f_{0}^{\prime \prime }\delta \varepsilon +\frac{1}{6}f_{0}^{\prime
\prime \prime }\left(\delta\varepsilon\right)^2 \right] , \eeq
\ewt where $f_0(\varepsilon_{\bf k})$ is the Fermi function and
\begin{equation}
\delta \varepsilon =\varepsilon _{\mathbf{k+q}}-\varepsilon
_{\mathbf{k}}.
\end{equation}
For an isotropic system $\varepsilon_\mathbf{k}=\varepsilon_{k}$
and \bwt
\begin{equation}
\delta\varepsilon =\varepsilon \left( \sqrt{k^{2}+2\mathbf{k\cdot q+q}^{2}}%
\right) -\varepsilon _{k}=\varepsilon_k ^{\prime }\left( qx+\frac{%
q^{2}}{2k}\right) +\frac{1}{2}q^{2}x^{2}\left( \varepsilon_k
^{\prime \prime }-\varepsilon_k ^{\prime }/k\right) .
\end{equation}
\ewt For the quadratic spectrum, the last term is equal to zero.
Integrating by parts, we find
\begin{eqnarray}
\int \frac{d^{3}k}{\left( 2\pi \right) ^{3}}f_{0}^{\prime \prime
}\delta \varepsilon_k &=&\frac{q^{2}}{6}\frac{d}{d\varepsilon_k
}\left[\nu \left( \varepsilon_k \right) \left(
2\frac{\varepsilon_k ^{\prime }}{k\left( \varepsilon_k \right)
}+\varepsilon_k ^{\prime \prime }\right)
\right]\Big|_{\varepsilon_k=
 \epsilon _{F}}\nn\\
\int \frac{d^{3}k}{\left( 2\pi \right) ^{3}}f_{0}^{\prime \prime
\prime }\left(\delta\varepsilon_k\right)^2
&=&-\frac{q^{2}}{3}\frac{d^{2}}{d\varepsilon_k
^{2}}\left[ \nu \left( \varepsilon_k \right) (\varepsilon_k ^{\prime })^{2}%
\right]\Big|_{\varepsilon_k =
 \epsilon _{F}}. \eea Adding up these two results, we obtain
\begin{equation}
\Pi \left( 0,q\right) =-\nu +\nu \alpha \frac{q^{2}}{12k_{F}^{2}},
\label{q2}\end{equation} where $\alpha$ is given by
Eq.~(\ref{alpha}).

 Combining
Eqs.~(\ref{delta2}) and (\ref{q2}) with the dynamic part
 of $\Pi_{\up\down}$
 from
Eq.~(\ref{bubbleud3D}), we obtain
 a
  complete result for $\Pi_{\uparrow\downarrow}$:
 \bea \Pi _{\uparrow \downarrow }\left( \Omega _{m},q\right)
&=&-\nu -\nu''\frac{\dzz ^{2}}{24}+\nu \alpha \frac{%
q^{2}}{12k_{F}^{2}}\nn\\
&&+\frac{i\nu \Omega _{m}}{2v_{F}q}\ln \frac{%
i\Omega _{m}+v_{F}q+\dzz}{i\Omega _{m}-v_{F}q+\dzz
}.\label{bubble3DDelta} \eea

Next, we consider $\Pi_{\uparrow}$ and $\Pi_{\downarrow}$, which
depend on $\dzz $ only via the static parts
\begin{equation}
\Pi _{\uparrow,\downarrow}\left( 0,0\right) =-\nu \left(\epsilon_F
\pm \dzz /2\right) =-\left( \nu \pm \nu^{\prime }\frac{\dzz
}{2}+\nu^{\prime \prime }\frac{\dzz ^{2}}{8}\right). \label{up}
\end{equation}

Using Eq. (\ref{Fladder}) for $\delta \Xi_L (T, H)$, we obtain
 \bwt
\begin{equation}
\delta \chi _{L}^{-1}=\frac{1}{\left( \mu _{B}\nu \right)
^{2}}\frac{\partial ^{2}\delta\Xi_{L}}{\partial \dzz
^{2}}\Big|_{\dzz =0}=\frac{1}{\left( \mu _{B}\nu \right)
^{2}}T\sum_q \left[ \frac{U}{1+U\Pi _{\uparrow \downarrow
}}\frac{\partial ^{2}\Pi _{\uparrow \downarrow }}{\partial \dzz
^{2}}-\frac{U^{2}}{\left( 1+U\Pi _{\uparrow \downarrow }\right)
^{2}}\left( \frac{\partial \Pi _{\uparrow \downarrow }}{\partial
\dzz }\right) ^{2}\right] \Bigg|_{\dzz =0}. \label{ladder}
\end{equation}
\ewt
 The derivatives of $\Pi_{\up\down}$ in  Eq.~(\ref{ladder}) are found with the help of Eq.~(\ref{bubble3DDelta}):
 \begin{subequations}
\begin{eqnarray}
\frac{\partial ^{2}\Pi _{\uparrow \downarrow }}{\partial \dzz ^{2}}%
\Bigg|_{\dzz =0} &=&-\frac{\nu ^{\prime \prime }}{12}
 +2 \nu \frac{%
\Omega _{m}^{2}}{\left( \Omega _{m}^{2}+v_{F}^{2}q^{2}\right)
^{2}}
\label{pipp} \\
\left( \frac{\partial \Pi _{\uparrow \downarrow }}{\partial \dzz }\right) ^{2}%
\Bigg|_{\dzz =0} &=&-\nu ^{2}\frac{\Omega _{m}^{2}}{\left( \Omega
_{m}^{2}+v_{F}^{2}q^{2}\right) ^{2}} \label{pip2}
\end{eqnarray}
\end{subequations}
The first term in Eq. (
 \ref{ladder}) gives a nonuniversal
 $T^2$ contribution,
   which can be obtained by
   expanding the prefactor
$\left(1+U\Pi _{\uparrow \downarrow }\right)^{-1}$ to first order
in $\Omega _{m}/v_{F}q$ and also expanding $\Pi_{\up\down}$ itself
to order $\left(q/k_{F}\right)^2$. Doing so, we find
 \bwt \bea \left(\delta \chi
_{L}^{(n)}\right)^{-1}=-\frac{1}{12\left( \mu _{B}\nu \right)
^{2}}
\frac{\nu ^{\prime \prime }}{\nu }T\sum_q\left({1+g_{s,0}+\frac{\pi \left| \Omega _{m}\right| }{2v_{F}q}%
+\alpha \frac{q^{2}}{12k_{F}^{2}}}\right)^{-1} = -
\frac{1}{12\chi(0)}\frac{k_{F}^{2}\vf^2%
}{\alpha}\frac{\nu ^{\prime \prime
}}{\nu}\frac{T^{2}}{\left(1+g_{s,0}\right)^ 2 \nu \vf^3 },
\label{nl_res} \eea \ewt where $\chi(0)=2
\mu_B^2\nu/\left(1+g_{s,0}\right)$.
 The Matsubara summation was performed using
\begin{equation}
T\sum_{\Omega _{m}}|\om|= \mathrm {const} -\frac{\pi }{3}T^{2}.
\end{equation}
The second term in Eq.~
 (\ref{ladder})
 gives a
 contribution of order  $T^2 \ln {|1 + g_{s,0}|}/(1 + g_{s,0})$,
which diverges weaker upon approaching the QCP than the $T^2/(1 +
g_{s,0})^2$
 contribution in Eq.~(\ref{nl_res}). Therefore, this contribution
  can be
 neglected.

The universal part of the ladder contribution
  comes from the second term of
Eq.~(\ref{ladder}). We have
 \bwt \bea \left(\delta \chi _{L}^{\left(
u\right)}\right)^{-1}=\frac{1}{\left(\mu
_{B}\nu \right) ^{2}}T\sum_q\frac{1}{\left( 1+g_{s,0}+%
\frac{\pi \left| \Omega _{m}\right| }{2qv_{F}}\right)
^{2}}\frac{\Omega _{m}^{2}}{\left( \Omega
_{m}^{2}+v_{F}^{2}q^{2}\right) ^{2}} =-\frac{2}{3\pi
^{2}\chi(0)}\frac{T^{2}}{\left(1+g_{s,0}\right)^2\nu v_{F}^{3}}.
\label{ul_res} \eea \ewt The total ladder contribution is the sum
of Eqs. (\ref{nl_res}) and (\ref {ul_res}).

The contribution to $\chi$ from ring diagrams is obtained in the
same way, and the result is
 \bwt
\begin{equation}
\delta \chi _{R}^{-1}=\frac{1}{2\left( \mu _{B}\nu \right)
^{2}}T\sum_q\left\{ -U^{2}\frac{\frac{%
\partial ^{2}}{\partial
 \dzz
  ^{2}}\left( \Pi _{\uparrow}\Pi _{\downarrow
}\right) }{1-U^{2}\Pi _{\uparrow}\Pi _{\downarrow
}}+U^{4}\frac{\left( \frac{\partial }{\partial
 \dzz }\left( \Pi _{\uparrow}\Pi _{\downarrow}\right) \right)
^{2}}{\left( 1-U^{2}\Pi _{\uparrow}\Pi _{\downarrow}\right)
^{2}}\right\} \Big|_{
 \dzz =0}.
 \label{ring}
\end{equation}
\ewt Using Eq.~(\ref{up}), we find that the second term in
Eq.~(\ref{ring}) vanishes, while the first one  reduces to
\begin{equation}
\delta \chi _{R}^{-1}=\frac{1}{4 (\mu_B \nu)^2}\left( \left( \nu
^{\prime }\right) ^{2}-\nu \nu ^{\prime \prime }\right)
T\sum_{\Omega _{m}}\int \frac{d^{3}q}{\left( 2\pi \right)
^{3}}\left[ \frac{U^{2}}{1-U^{2}\Pi ^{2}}\right],
\label{ring1}\end{equation} where $\Pi$ is the polarization
operator in the absence of the field. Obviously,
 the entire ring contribution is non-universal.

 Near a ferromagnetic QCP,
i.e, when $g_{s,0}\approx -1$,
  Eq.~(\ref{ring1})
 can be simplified
  to
\bwt
\begin{eqnarray}
\delta \chi _{R}^{-1}=\frac{1}{4\mu _{B}^{2}\nu} \frac{\left(\nu
^{\prime }\right)^{2}-\nu \nu''_F}{\nu ^{2}}
T\sum_q\left(1+g_{s,0}+\frac{\pi \left|\Omega_{m}\right|}{2v_{F}q}
+\alpha \frac{q^{2}}{12k_{F}^{2}} \right)^{-1}
=\frac{1}{8\chi(0)}\frac{k_{F}^{2}\vf^2}{\alpha}\frac{\left( \nu
^{\prime }\right) ^{2}-\nu \nu ^{\prime \prime }}{\nu
^{2}}\frac{T^{2}}{\left(1+g_{s,0}\right)^2\nu v_F ^3}.
\end{eqnarray}
\ewt Combining $\delta\chi _{R}^{-1}$ with
 Eq. (\ref{nl_res}) for the nonuniversal part of the ladder
contribution, we obtain Eq.~(\ref{nonuniv}), while the universal
part of $\delta\chi_{L}^{-1}$ gives Eq.~(\ref{univ}).


\begin{thebibliography}{99}
\bibitem{comm_mo}
The numerical prefactor in $E$ differs
  from the one
obtained in \protect\cite{we_short}, because the thermodynamic
potential was defined in a different way. It agrees with the one
obtained recently~\protect\cite{bec}
 in the analysis of $\Xi (T,M)$ in the ferromagnetically ordered phase.
\end{thebibliography}
\end{document}